\documentclass[journal]{IEEEtran}

\ifCLASSOPTIONcompsoc
\usepackage[caption=false,font=normalsize,labelfont=sf,textfont=sf]{subfig}
\else
\usepackage[caption=false,font=footnotesize]{subfig}
\fi

\usepackage{hyperref}
\hypersetup{hidelinks}
\usepackage{xcolor}
\definecolor{color}{rgb}{0,0,0}

\usepackage{multirow}
\usepackage{stfloats}
\usepackage{float}
\usepackage{cite}
\usepackage{amsmath}
\usepackage{amssymb}
\usepackage{algorithm}
\usepackage{algorithmic}
\usepackage{url}
\usepackage{epsfig}
\usepackage{graphicx}
\usepackage{makecell}
\usepackage{footmisc}
\usepackage{sepfootnotes}
\usepackage{caption}
\usepackage[para]{threeparttable}
\newcommand{\tabincell}[2]{\begin{tabular}{@{}#1@{}}#2\end{tabular}} 

\begin{document}
	%
	
	\title{A Novel Length-Flexible Lightweight Cancelable Fingerprint Template for Privacy-Preserving Authentication Systems in Resource-Constrained IoT Applications}
	
	\author{Xuefei Yin, 
		Song Wang,
		Yanming Zhu,
		and Jiankun Hu$^{*}$, \IEEEmembership{Senior Memeber,~IEEE}
		\thanks{Xuefei Yin is with the School of Engineering and Information Technology, University of New South Wales, Canberra, ACT 2600, Australia (e-mail: xuefei.yin@unsw.edu.au).} 
		\thanks{Song Wang is with the Department of Engineering, La Trobe University, Australia, VIC 3086, Australia (e-mail: song.wang@latrobe.edu.au).}
		\thanks{Yanming Zhu is with the School of Computer Science and Engineering, University of New South Wales, Sydney, NSW 2052, Australia (e-mail: yanming.zhu@unsw.edu.au).} 
		\thanks{($\ast$ Corresponding author) Jiankun Hu is with the School of Engineering and Information Technology, University of New South Wales, Canberra, ACT 2600, Australia (e-mail: j.hu@adfa.edu.au).} 
	}
	
	\maketitle
	
	\begin{abstract}
		 Fingerprint authentication techniques have been employed in various Internet of Things (IoT) applications for access control to protect private data, but raw fingerprint template leakage in unprotected IoT applications may render the authentication system insecure. 
		 Cancelable fingerprint templates can effectively prevent privacy breaches and provide strong protection to the original templates. 
		 However, to suit resource-constrained IoT devices, oversimplified templates would compromise authentication performance significantly.
		 In addition, the length of existing cancelable fingerprint templates is usually fixed, making them difficult to be deployed in various memory-limited IoT devices.		 
		To address these issues, we propose a novel length-flexible lightweight cancelable fingerprint template for privacy-preserving authentication systems in various resource-constrained IoT applications.  
		The proposed cancelable template design primarily consists of two components: 1) length-flexible partial-cancelable feature generation based on the designed re-indexing scheme; and 2) lightweight cancelable feature generation based on the designed encoding-nested-difference-XOR scheme. 
		Comprehensive experimental results on public databases~FVC2002 DB1-DB4 and FVC2004 DB1-DB4 demonstrate that the proposed cancelable fingerprint template achieves equivalent authentication performance to state-of-the-art methods in IoT environments, but our design substantially reduces template storage space and computational cost.
		More importantly, the proposed length-flexible lightweight cancelable template is suitable for a variety of commercial smart cards (e.g., C5-M.O.S.T. Card Contact Microprocessor Smart Cards CLXSU064KC5). 
		To the best of our knowledge, the proposed method is the first length-flexible lightweight, high-performing cancelable fingerprint template design for resource-constrained IoT applications.
		
	\end{abstract}
	
	
	%
	\IEEEpeerreviewmaketitle
	
	\section{Introduction}
	
	\IEEEPARstart{W}{ith} rapid technological advancements, the Internet of Things (IoT) has emerged as a network connecting various sensors or smart devices via the Internet or other communication channels \cite{Survey_2019}. 
	The `things' connected in the IoT may perform functions of data collection, data processing or information communication.  
	However, these functions are vulnerable to privacy leakage~\cite{RN2413, IoT-security-2018, Demysifying}, especially during data collection and data processing, if there is no protection to raw information.
	Therefore, identity authentication has been one of the promising options for access control in IoT applications~\cite{ riad2019sensitive, pinno2020controlchain}.
	An identity authentication system usually consists of two procedures: enrollment and verification. The enrollment procedure is aimed at registering a user by generating and storing the user's template, while the verification procedure attempts to match the template generated for a query against the enrolled template.
	
	Fingerprints have proven to be one of the most popular and efficient biometric traits and have been pervasively used for identity authentication~\cite{RN2422, RN1145}.  
    Compared to traditional token-based identity authentication~\cite{obaidat2019biometric, ren2016biometrics, jiang2021enhancing}, fingerprint-based authentication systems are user-friendly because, unlike passwords, fingerprints won't be forgotten.
    Along with convenience, however, fingerprint-based authentication systems without any protection also expose IoT applications to privacy breaches and security risks.
    First and foremost, raw fingerprint data leakage in unprotected IoT applications may render the authentication system insecure, because the raw fingerprint data can be repeatedly utilized to spoof the authentication system. Simultaneously, a finger would be `lost' forever once its raw fingerprint data is compromised. 
    Another issue is regarding legal regulations on data privacy around the world.
	For example, to protect personal private information, specific laws and legal regulations have been formulated in many regions and countries, such as the \textit{General Data Protection Regulation}~\footnote{https://gdpr-info.eu} in European Union,
	the \textit{Personal Information Protection Law of the People's Republic of China} in China, and the \textit{California Privacy Rights Act}~\footnote{https://oag.ca.gov/privacy/ccpa} 
	in the United States. 
	Therefore, it is essential to implement protection schemes to fingerprint authentication systems in IoT applications.
	
	Fingerprint protection can be typically divided into two categories: 1) cryptography-based approaches and 2) cancelable fingerprint template approaches.
	In cryptography-based approaches, cryptographic techniques (e.g., symmetric/asymmetric encryption and homomorphic encryption) are commonly utilized to encrypt fingerprint templates so as to avoid original template leakage ~\cite{RN2557, RN2558, RN2560}. The benefit is that encrypted templates tend to be very secure and cannot easily cracked.
	The downside is that the encryption and decryption processes are usually time-consuming. Therefore, the cryptography-based methods are unsuitable for resource-constrained IoT devices~\cite{RN2558}. 
	
	Cancelable biometrics is another template protection technique~\cite{RN1102, RN2561, punithavathi2019partial, kaur2020privacy, jiang2020cancelable, patel2015cancelable}. 
	The core idea of cancelable fingerprint templates is to irreversibly transform the raw fingerprint template into a new template to avoid privacy leakage. 
	Four objectives are demanded in the design of cancelable fingerprint templates~\cite{RN1102, RN2519}: 
	1) diversity: different unrelated fingerprint templates can be obtained with disparate distortions; 
	2) revocability: a new template can be issued to replace the compromised template; 
	3) non-invertibility: it should be computationally infeasible to retrieve the original fingerprint template from the transformed (cancelable) template; and 
	4) accuracy: cancelable fingerprint templates should not significantly deteriorate the accuracy of fingerprint recognition.
	Therefore, cancelable fingerprint templates can effectively avoid privacy leakage and provide strong protection to the original templates. 
	However, to suit resource-constrained IoT devices, oversimplified cancelable fingerprint templates deployed in resource-constrained IoT applications would compromise authentication performance significantly.
	In addition, the length of existing cancelable fingerprint templates is usually fixed, making them difficult to be implemented in various resource-constrained IoT devices.
	Moreover, designing cancelable fingerprint templates that meet the above four objectives is challenging, especially for IoT applications.

	To address these issues, we design a length-flexible lightweight, high-performing cancelable fingerprint template for privacy-preserving authentication systems with applications to resource-constrained IoT devices. 
	The proposed cancelable fingerprint template is based on the state-of-the-art minutia cylinder-code~(MCC)~\cite{RN159}, which is a robust minutia-based local descriptor with excellent authentication performance on public fingerprint databases. 
	However, the original MCC is not designed for resource-constrained IoT devices. 
	More importantly, the original MCC has no template protection function.
	The proposed cancelable fingerprint template design consists of two components: 1) length-flexible partial-cancelable feature generation; and 2) lightweight cancelable feature generation.
	For the first component, we propose a simple, efficient yet effective method to flexibly re-index the original MCC feature.
	For the second component, we develop an encoding-nested-difference-XOR scheme.
	The novel cancelable template possesses four advantages: 1) flexible length: the template length can be flexibly adjusted to suit various memory-limited IoT devices; 2) lightweight: this makes the proposed template further applicable to memory- and computation-constrained IoT devices; 3) cancelablility: this protects raw fingerprint data against privacy leakage; and 4) high performance: extensive experiments demonstrate the satisfactory performance of the proposed cancelable template on eight public fingerprint datasets.  
	
	The main contributions of this study are summarized as follows:
	\begin{itemize}
		\item To the best of our knowledge, this study proposes the first length-flexible, high-performing privacy-preserving fingerprint template suited to various memory-limited IoT devices. 
		As IoT devices are usually embedded with varying storage space, it is essential to provide length-flexible but high-performing fingerprint templates. 
		\item We propose an innovative lightweight cancelable fingerprint template based on the re-indexing operation and the encoding-nested-difference-XOR operation. 
		The template size is reduced by up to 85\% (around 64K bits) while achieving superior verification performance in the privacy-preserving IoT environment. 
		The cancelable characteristic can also protect the original fingerprint data against hill-climbing and pre-image attacks, thus making the proposed template appropriate for resource-constrained IoT applications.
		\item Comprehensive experimental results obtained on eight public benchmark datasets FVC2002 DB1-DB4~\footnote{http://bias.csr.unibo.it/fvc2002/}~\cite{RN2562} and FVC2004 DB1-DB4~\footnote{http://bias.csr.unibo.it/fvc2004/}~\cite{RN2022} demonstrate that the proposed template achieves equivalent authentication accuracy to the state-of-the-art cancelable fingerprint templates in IoT settings, but our design significantly reduces template storage space and computational cost. 
	\end{itemize}
	
	The rest of this paper is organized as follows. 
	We review state-of-the-art studies on privacy-preserving fingerprint templates and lightweight fingerprint authentication systems for IoT applications in Section~\ref{sec:RelatedWorks}.	We detail the proposed cancelable template in Section~\ref{sec:Framework}. 
	We present the experimental setting and analyze the experimental results in Section~\ref{sec:experiment}. 
	We conclude the paper in Section~\ref{sec:conclusion}.
	
	\section{Related Work}\label{sec:RelatedWorks}
	Cryptographic techniques (e.g., symmetric/asymmetric encryption and homomorphic encryption) have been used to protect original fingerprint templates by encrypting them~\cite{RN2557, RN1242, RN2559}. 
	Xi \textit{et al.}~\cite{RN1242} reviewed topical cryptographic techniques and fingerprint biometrics and discussed the applications of the cryptographic technique in fingerprint-based authentication systems. 
	Kim \textit{et al.}~\cite{RN2557} proposed using fully homomorphic encryption to protect the original fingerprint image by encrypting its features. This method can provide strong protection to the original template. However, this method is time-consuming and unsuitable for resource-constrained IoT devices.
	Yang \textit{et al.}~\cite{RN2559} introduced a similar homomorphic encryption-based fingerprint authentication method, in which minutiae pairs are used as original features. However, the authentication accuracy (EER=8.25\%) of this method is unsatisfactory.
	Azzaz \textit{et al.}~\cite{RN2768} proposed a symmetric encryption-based method to encrypt a fingerprint image instead of its features (e.g., minutiae) to avoid privacy leakage. The disadvantage is that fingerprints could be lost forever once the cipher key is leaked. Besides, the encryption and decryption would increase computational complexity. 
	Liu \textit{et al.}~\cite{RN2771} presented a fingerprint encryption-based online fingerprint authentication scheme, in which homomorphic addition is used to encrypt fingerprint data. However, this method is cloud-oriented and unsuitable for IoT applications.
	In summary, cryptography-based fingerprint authentication methods tend to be time-consuming and resource-intensive due to the encryption and decryption operations. Besides, original fingerprint information are still at risk to privacy breaches due to key-related hacking.
		
	Another popular protection scheme is cancelable fingerprint template techniques, which are aimed to irreversibly transform the raw fingerprint template into a new one to avoid privacy breaches~\cite{RN1102}. 
 	Kho \textit{et al.}~\cite{RN2582} proposed a cancelable fingerprint template design based on the local minutia descriptor and permutated randomized non-negative least square. 
	Wu \textit{et al.}~\cite{RN2772} designed a privacy-preserving cancelable pseudo-template based on a random distance transformation technique. 
	Kavati \textit{et al.}~\cite{RN2775} proposed a cancelable fingerprint template protection scheme using elliptical structures guided by fingerprint minutiae. Although this method provides strong protection to the raw fingerprint template, the authentication accuracy is poor with EER of 7.3\% and 5.13\% for FVC2002 DB1 and DB2, respectively.
	Tran \textit{et al.}~\cite{RN2519} proposed a multi-filter matching framework for cancelable fingerprint template design and achieved good authentication performance. 
	Bedari \textit{et al.}~\cite{RN2728} presented an alignment-free cancelable MCC-based fingerprint template design.
	Similarly, Yin \textit{et al.}~\cite{RN2761} proposed an IoT-oriented cancelable fingerprint template based on the MCC feature and achieved state-of-the-art authentication performance in an IoT environment.
	Unlike aforementioned methods, Lee \textit{et al.}~\cite{RN2774} developed a tokenless cancelable template for multi-modal biometric systems, where the real-valued face and fingerprint vectors are fused into a cancelable template.
	In summary, compared to cryptography-based fingerprint template protection methods, cancelable fingerprint templates can effectively protect raw fingerprint data because the cancelable template instead of the raw template is stored in the authentication system. 
	However, most of these approaches are designed for cloud applications or powerful devices rather than resource-constrained IoT applications. 
	Besides, most of these cancelable templates are usually of fixed length, making them unsuitable for resource-constrained IoT devices. 
	
	Fingerprint-based authentication systems in IoT environments have been explored in the literature~\cite{RN134, RN2551, RN2510}. 
	Habib \textit{et al. }~\cite{RN2563} introduced an authentication framework based on biometric and radio fingerprinting for the IoT in an eHealth application. Through the embedded authentication system, the framework can guarantee that the monitored private data is associated with the correct patient. 
	Punithavathi \textit{et al.}~\cite{RN2551} proposed a lightweight fingerprint authentication system based on machine learning for smart IoT devices in a cloud computing environment.
	However, the authentication accuracy evaluated on public datasets FVC2002 DB1-DB2 and FVC2004 DB1-DB2 is poor.
	Golec \textit{et al.}~\cite{RN2777} introduced a fingerprint-based authentication system in an IoT environment, where the fingerprint data in the communication channel and database is protected by the AES-128-bit key encryption method. 
	Sabri \textit{et al.}~\cite{RN2779} developed a fingerprint-based authentication framework for match-on-card and match-on-host applications, but the fingerprint template is unprotected.
	Kumar~\cite{RN2778} utilized a fingerprint authentication system in an IoT environment to defend communication channels against black hole attacks. However, the fingerprint template used in the authentication system is vulnerable to privacy breaches. 
	In summary, fingerprints or fingerprint features have been used for identity authentication in various IoT applications and even on resource-constrained IoT devices, such as smart cards. However, the original fingerprint data in these studies faces privacy leakage issues. 
	
	\section{The Proposed Lightweight Cancelable Fingerprint Template}\label{sec:Framework}
	A fingerprint authentication system typically consists of two procedures: 1) enrollment and 2) verification.
	The enrollment procedure is aimed at registering a user by generating and storing the user's template, while the verification procedure is aimed at generating a template for a query user and matching the template against the enrolled one.
	The enrollment procedure usually consists of fingerprint acquisition via a fingerprint sensor, template generation and template storage. 
	The verification procedure usually consists of fingerprint acquisition, template generation and template matching.
	{\color{color}In cloud-based applications, a fingerprint is captured on the end-user side and then transferred to the cloud for template generation, template storage and template matching. Thus, the end-user is responsible for capturing a fingerprint, transferring it to the cloud, and then receiving the verification result from the cloud. The security issue here is that the private fingerprint data is held by the cloud. This may cause privacy leakage due to security concerns in relation to cloud servers or attackers.
	In the IoT applications discussed in this work, the fingerprint data does not leave the IoT. The IoT application takes responsibility for fingerprint acquisition, template generation, template storage and template verification. As opposed to cloud-based applications where the raw fingerprint needs to be transferred to the cloud, in IoT applications, a cancelable template stays in the IoT. As an advantage, the raw fingerprint enrolled in the IoT application is securely protected because a compromised cancelable template would not reveal the raw fingerprint information.}

	The core step in both enrollment and verification is template generation. This work proposes a novel method for generating a lightweight cancelable fingerprint template for resource-constrained privacy-preserving IoT applications.
	In the rest of this section, we firstly introduce the preliminary procedure about minutia extraction and minutia-based MCC feature extraction in Section~\ref{sec:preliminary}. Then, we describe the details of partial-cancelable feature generation in Section~\ref{sec:partial} and lightweight cancelable feature generation in Section~\ref{sec:cancelable}. Finally, we present template matching in Section~\ref{sec:matching}.
	
	\subsection{Preliminary Procedure} \label{sec:preliminary}
	\subsubsection{Minutia Extraction} \label{sec:extraction}
	Minutiae as a popular feature starting point have been widely used in fingerprint biometrics. In this paper, minutiae are also utilized to generate the proposed IoT-oriented cancelable fingerprint template. 
	Given a fingerprint image captured by the embedding fingerprint sensor, $n$ minutiae are extracted to represent this fingerprint, denoted by $\mathbf{T} = \{\mathbf{m}_{\color{color}1}, \mathbf{m}_2, \cdots, \mathbf{m}_n\}$. Each minutia is in the format of ISO/IEC 19794-2,\footnote{https://www.iso.org/standard/50864.html} defined by $\mathbf{m}_i = \{x_i,y_i,\theta_i\}$ where $x_i$ and $y_i$ are the coordinates in pixels and $\theta_i \in [0,2\pi]$ stands for the minutia orientation. 
	In the proposed IoT-oriented fingerprint authentication system, minutia extraction is conducted upon the minutia extraction algorithm, Mindtct~\cite{RN675}, from the open-source NIST biometric image software.\footnote{https://www.nist.gov/services-resources/software/nist-biometric-image-software-nbis}
	
	\subsubsection{MCC Template}
	The MCC template~\cite{RN159} is a robust minutia-based local feature representation and has been proved successful in fingerprint authentication.
	As the MCC feature is defined by the relative relationship between a minutia and its neighboring minutiae, the MCC feature possesses some desirable properties, such as translation- and rotation-invariance, and fixed length. 
	The MCC feature is defined for each minutia and represented by a cylinder which is discretized into cube-like cells. 
	The value for each cell is used to measure the relative distance contribution between the cell and neighboring minutiae, as well as to measure the relative orientation contribution between the cell, the reference minutia and neighboring minutiae.
	
	The MCC feature for each minutia contains two vectors: 1) the cell value vector and 2) the cell validity vector. The cell value is calculated by the distance and orientation contributions, while the cell validity is used to indicate the cell status. An MCC feature is represented by 
	$$
	\mathbf{v} = [\mathbf{c}, \mathbf{b}],
	$$
	where $\mathbf{c}$ denotes the cell value vector and $\mathbf{b}$ the cell validity vector~\cite{RN159}. 
	According to the parameter settings for the MCC feature in~\cite{RN159}, the cylinder diameter is set to $N_S = 16$ cells and the height of the cylinder is set to $N_D=5$ cells. 
	Therefore, the length of the cell value vector $\mathbf{c}$ is represented by $L_{\mathbf{c}} = 1,280$ (i.e., $N_S\times N_S\times N_D$), while the length of the cell validity vector $\mathbf{b}$ is represented by $L_{\mathbf{b}} = 256$ (i.e., $N_S\times N_S$).
	
	\subsection{Length-Flexible Partial-cancelable Features}\label{sec:partial}
	A simple, efficient yet effective scheme is proposed to generate the partial-cancelable feature by re-indexing the original MCC feature. 
	The new feature contains two parts: 1) the cell value part and 2) the cell validity part.
	To design a lightweight feature, we assign a percentage value $p \in [50\%, 100\%]$ to control the length of the new cancelable feature.
	Given the MCC feature vector $\mathbf{v}$ with the length $L_{\mathbf{c}}$, 
	its index set $\mathbf{I}$ is defined by
	\begin{equation}
		\mathbf{I} = \{1,2,\cdots,L_{\mathbf{c}}\},
	\end{equation} 
	its cell value part $\mathbf{c}$ is represented by 
	\begin{equation}
		\mathbf{c} = (c_1, c_2, \cdots, c_{L_{\mathbf{c}}}),
	\end{equation}
	and its cell validity part can be easily obtained by
	{\color{color} replicating the base mask for each cell section in the cylinder because each section shares the same base mask,}
	without causing ambiguity, represented by
	\begin{equation}
		\mathbf{b} = (b_1, b_2, \cdots, b_{L_{\mathbf{c}}}),
	\end{equation}
	where the $i^{th}$ bit $b_i$ denotes the validity of the $i^{th}$ value in the cell value part $\mathbf{c}$.
	
	A re-indexing set $\mathbf{I}'$ is generated by randomly selecting $l$ unique integers from the set $\mathbf{I}$, represented by 
	\begin{equation} \label{eq:reindex}
		\mathbf{I}' =  \{t_i|t_i \in \mathbf{I}, 1 \leqslant i \leqslant l\}, 
	\end{equation}
	where $ l = \lfloor p*L_{\mathbf{c}} \rfloor - mod(\lfloor p*L_{\mathbf{c}} \rfloor, 8)$~\footnote{The operator $\lfloor x \rfloor$ rounds $x$ to the nearest integer less than or equal to $x$, and $mod$ is the modulo operation.}. 
	$l$ is set to a multiple of eight to facilitate the subsequent feature extraction. For convenience, we alternatively denote $l = 8k$.
	The new cell value vector is then obtained by collecting the corresponding values from $\mathbf{c}$ with the index in $\mathbf{I}'$, expressed as 
	\begin{equation} \label{eq:8k}
		\mathbf{c}' = (c_{t_1}, c_{t_2}, \cdots, c_{t_{8k}}),
	\end{equation}
	and the cell validity vector is similarly obtained from $\mathbf{b}$, given by 
	\begin{equation} \label{eq:8k_mask}
		\mathbf{b}' = (b_{t_1}, b_{t_2}, \cdots, b_{t_{8k}}).
	\end{equation}
	In summary, the partial-cancelable feature is formulated by \begin{equation} \label{v_prime}
		\mathbf{v}' = [\mathbf{c}', \mathbf{b}'].
	\end{equation}
	
	This is a partial-cancelable feature, because it satisfies three of the four objectives of cancelable templates: diversity, revocability and accuracy.
	The diversity is guaranteed by many re-indexing sets that exist, namely $ \frac{L_{c}!}{(L_{c}-l)!}$.\footnote{The $!$ is the factorial operator, which returns the product of all positive integers less than or equal to a positive integer.} 
	Regarding the revocability, as the re-indexing process is controlled by a random generator, a new template can be easily obtained by choosing a different random seed. 
	The accuracy is also not much affected by this new feature.
	Especially, setting $p = 100\%$ maintains the same accuracy, 
	because the similarity between two features {\color{color}defined in\cite{RN159}} is order-invariant to the feature elements. 
	At this stage, the feature in Eq.~(\ref{v_prime}) does not achieve non-invertibility, because the original template may be retrieved by gathering the features and the corresponding index sets. 
	In Section~\ref{sec:cancelable}, we will propose a scheme to attain non-invertibility and a lightweight design.
	
	\subsection{Lightweight Cancelable Features} \label{sec:cancelable}
	To achieve the non-invertibility objective as well as the lightweight design, we propose an encoding-nested-difference-XOR scheme, which contains three operations: 1) the nested-difference operation, 2) the encoding operation; and 3) the bitwise XOR Boolean operation.
	As a notable benefit to resource-constrained IoT devices, the new feature will save approximately $87.5\%$ storage space when $p = 50\%$ compared to the bit-MCC feature~\cite{RN159}. For example, for the partial-cancelable feature with $p = 50\%$ containing $8k$ cell values, the proposed lightweight cancelable feature will result in $2k$ bits. 
	
	\subsubsection{The Nested-difference Operation} \label{sec:nestedDifference}
	This operation is to calculate the nested difference of four neighboring cell values in the partial-cancelable vector.
	{\color{color}
	For clarity, we define the first-layer nested difference by vector $\mathbf{e^{L_1}}$, whose $i^{th}$ element $e^{L_1}_{i}$, formulated by Eq. \eqref{eq:firstDif}, is calculated upon the partial-cancelable vector $\mathbf{c}'$ in Eq. \eqref{eq:8k}.
	\begin{equation}\label{eq:firstDif}
		e^{L_1}_{i} = c_{t_{2i-1}} - c_{t_{2i}},
	\end{equation}
	where $1 \leqslant i \leqslant 4k$.
	The second-layer nested difference vector $\mathbf{e^{L_2}}$ is then calculated upon the first-layer nested difference, represented by 
	\begin{equation} \label{eq:2k}
		\mathbf{e^{L_2}} = (e^{L_2}_1, e^{L_2}_2, \cdots, e^{L_2}_{2k}),
	\end{equation}
	where the $i^{th}$ element 
	\begin{equation}\nonumber 
	\begin{split}
	    e^{L_2}_i & = e^{L_1}_{2i-1} - e^{L_1}_{2i} \\
	    & = (c_{t_{4i-3}} - c_{t_{4i-2}}) - (c_{t_{4i-1}} - c_{t_{4i}}),
	\end{split}
	\end{equation}
	and $1 \leqslant i \leqslant 2k$.
	For convenience and without causing ambiguity, we use $\mathbf{e}$ to represent $\mathbf{e^{L_2}}$ and use $e_i$ to represent the $i^{th}$ element in $\mathbf{e}$.}
	As $c_i$ is in the range [0, 1], $e_i$ is therefore in the range [-2, 2].
	For the cell validity part, we use the OR Boolean operator to concatenate four neighboring cell masks so that valid cells can remain. The new validity vector is formulated by
	\begin{equation} \label{eq:d}
		\mathbf{d} = (d_1, d_2, \cdots, d_{2k}),
	\end{equation} 
	where the $i^{th}$ element $d_i = b_{t_{4i-3}} ~|~ b_{t_{4i-2}} ~|~ b_{t_{4i-1}} ~|~ b_{t_{4i}}$, $1 \leqslant i \leqslant 2k$, and $|$ denotes the OR Boolean operator.
	
	This procedure has three advantages: 
	1) the nested difference can significantly reduce the number of elements because it can incorporate four values; 
	2) the proposed operation increases the difficulty to revert to the original feature; and 
	3) the simple relationship between four values can effectively identify the distinguishability of the original feature, which is also supported by the experimental results in Section \ref{sec:validation} and \ref{sec:simulation}.
	
	\subsubsection{The Encoding Operation}\label{sec:encoding}
	The encoding operation is using two bits to encode the relationship between the nested difference and a threshold. For a well-defined threshold, this relationship can effectively model the original feature information without significantly deteriorating the matching accuracy.
	Given a nested difference $e$ and a threshold $\tau$ ($\tau$ is optimally set to 0.2 in our experiments), the encoding table is shown in Table~\ref{tab:encoding}.
	By encoding the vector $\mathbf{e}$ (in Eq. \eqref{eq:2k}) according to Table \ref{tab:encoding}, a new vector $\bar{\mathbf{e}}$ in bits is obtained as
	\begin{equation} \label{eq:encoding2k}
		\bar{\mathbf{e}} = (\bar{e}_1, \bar{e}_2, \cdots, \bar{e}_{2k}),
	\end{equation} 
	where each unit $\bar{e}_i$ contains two bits. Its validity vector is the same as $\mathbf{d}$ in Eq. \eqref{eq:d}.
	\begin{table}[htbp]
		\setlength\tabcolsep{15pt}
		\renewcommand{\arraystretch}{1.5}
		\caption{The encoding table}
		\label{tab:encoding}
		\centering
		\begin{tabular}{c|c} 
			\hline
			\hline
			\textbf{Relationship} &\textbf{Encoding bits} \\ 
			\hline 
			$\frac{e_i}{2} \geqslant \tau$&10 \\ 
			\hline
			
			$\frac{e_i}{2} \leqslant -\tau$ &01	\\ 
			\hline
			
			$otherwise$ &00	\\ 
			\hline
			\hline
		\end{tabular}
	\end{table}
	
	The encoding procedure has two key advantages: 1) the threshold in the encoding operation can enhance the privacy of the original feature, thus making it impossible to revert to the original MCC feature; and 2) the encoding that converts float values into bits can significantly reduce the storage space.
	
	\subsubsection{The Bitwise XOR Boolean Operation}\label{sec:xor}
	The XOR Boolean operation conducts the bitwise XOR between two neighboring units $\bar{e}_i$ and $\bar{e}_{i+1}$. 
	Given the encoded vector $\bar{\mathbf{e}} = (\bar{e}_1, \bar{e}_2, \cdots, \bar{e}_{2k})$, the new feature vector $\hat{\mathbf{e}}$ in bits is formulated by 
	\begin{equation}\label{eq:xorVector}
		\hat{\mathbf{e}} = (\hat{e}_1,\hat{e}_2,\cdots,\hat{e}_k),
	\end{equation}
	where 
	$\hat{e}_i = \bar{e}_{2i-1} \oplus \bar{e}_{2i}, 1 \leqslant i \leqslant k,$ and $\oplus$ denotes the bitwise XOR Boolean operator. For example, given $\bar{e}_{1} = 10 $ and $\bar{e}_{2} = 00 $, we obtain $\hat{e}_{1} = 10 \oplus 00 = 10.$ The corresponding validity vector is obtained by 
	\begin{equation}
		\hat{\mathbf{d}} = (\hat{d}_1, \hat{d}_2,\cdots,\hat{d}_k),
	\end{equation}
	where the $i^{th}$ element $\hat{d}_i = d_{2i-1} ~|~ d_{2i}$, $1 \leqslant i \leqslant k,$ and $|$ denotes the OR Boolean operator. 
	The proposed lightweight cancelable feature vector is then represented by $\hat{\mathbf{v}} = [\hat{\mathbf{d}}, \hat{\mathbf{e}}].$ The comparison of the feature length between the original MCC feature and the proposed lightweight cancelable feature is summarized in Table~\ref{tab:comparisonLength}. When $p = 50\%$, the length of the proposed cell value vector is approximately $\frac{L_c}{8}$. When $N_D < \frac{8}{p} \in [8, 16]$, the length of the proposed cell validity vector is less than that of the original cell validity vector; otherwise, we can alternatively use the base mask to easily obtain the cell validity vector without increasing extra storage costs. 
	\begin{table}[htbp]
		\setlength\tabcolsep{3pt}
		\renewcommand{\arraystretch}{1.5}
		\caption{Comparison of the feature length in the case of $N_S=16, N_D=5$, and $p=50\%$}
		\label{tab:comparisonLength}
		\centering
		\begin{tabular}{l|c|c} 
			\hline
			\hline
			&\textbf{Cell value vector} & \textbf{Cell validity vector} \\ \hline 
			The original MCC feature&$L_c = 1,280$ 					 & $L_b=256$ \\ \hline
			The proposed feature      &	$\frac{p}{4}L_c = 160$ 	 &$\frac{p}{8}N_DL_b = 80$\\ 
			\hline
			\hline
		\end{tabular}
	\end{table}

	\subsection{Template Matching}\label{sec:matching}
	Template matching is to decide whether two templates are matched, which is an essential process in biometric authentication. This procedure comprises two steps: 1) computation of the similarity between two feature vectors; and 2) computation of the decision score.
	
	\subsubsection{Computation of the Similarity between Two Feature Vectors}
	
	Given two feature vectors $\mathbf{v}_q = [ \hat{\mathbf{e}}_q, \hat{\mathbf{d}}_q]$ and $\mathbf{v}_p = [ \hat{\mathbf{e}}_p, \hat{\mathbf{d}}_p]$ coming from the query template and the enrolled template, respectively, 
	the intersection between the two cell validity vectors is defined by 
	\begin{equation}
		\mathbf{d}_{qp} = \hat{\mathbf{d}}_q \otimes \hat{\mathbf{d}}_p,
	\end{equation}
	where $\otimes$ denotes the bitwise AND Boolean operator. 
	To facilitate the subsequent computation, we must align the intersected validity vector with the cell value vectors. The aligned validity vector $\hat{\mathbf{d}}_{qp}$ is obtained by duplicating each bit of $\mathbf{d}_{qp}$, represented by
	\begin{equation}
		\hat{\mathbf{d}}_{qp} = (d_{qp,1},d_{qp,1},\cdots,d_{qp,k},d_{qp,k}).
	\end{equation}
	The similarity between two features is calculated by
	\begin{equation} \label{eq:sqp}
		s_{qp} = 1 - \dfrac{\parallel \hat{\mathbf{e}}_{q|p} \oplus \hat{\mathbf{e}}_{p|q} \parallel}{\parallel \hat{\mathbf{e}}_{q|p} \parallel+\parallel \hat{\mathbf{e}}_{p|q} \parallel},
	\end{equation}
	where
	\begin{equation}\nonumber 
		\begin{cases}
			\hat{\mathbf{e}}_{q|p} &=  \hat{\mathbf{e}}_{q} \otimes \hat{\mathbf{d}}_{qp},\\
			\hat{\mathbf{e}}_{p|q} &=  \hat{\mathbf{e}}_{p} \otimes \hat{\mathbf{d}}_{qp}.
		\end{cases}
	\end{equation}
	The similarity is in the range [0,1], where the higher the value, the more similar the two features are.
	
	\subsubsection{Computation of the Decision Score}
	The decision score is used to measure the matching probability between a query template and an enrolled template. 
	Given a query template containing $n$ feature vectors and an enrolled template containing $m$ feature vectors, a score matrix~$\mathbf{s}$ of size $n \times m$ is obtained by calculating the similarity of each pair of feature vectors from the query and enrolled templates. The element $s_{qp}$ of the score matrix~$\mathbf{s}$ is given by Eq.~\eqref{eq:sqp}. 
	The decision score is then calculated upon the score matrix~$\mathbf{s}$ using the local greedy similarity~(LGS) algorithm in~\cite{RN2538}.
	The decision score is in the range [0, 1], with a larger value indicating a higher matching probability between the query and enrolled templates.
	
	\section{Experiments}\label{sec:experiment}
	In this section, we evaluate the proposed template in an IoT environment in terms of matching accuracy and efficiency. First,
	We present the experimental setting in Section \ref{sec:setting}, including the benchmark datasets, the evaluation protocol, and the measurement metrics. 
	Next, we evaluate the effect of the feature length using different values of $p$ in Section~\ref{sec:emcc_p}.
	Then, we comprehensively compare the proposed lightweight cancelable template with state-of-the-art methods in Section~\ref{sec:validation}, and implement an IoT prototype system to evaluate the authentication performance on eight benchmark datasets in Section \ref{sec:simulation}.
	Finally, security analysis is conducted in Section~\ref{sec:security}.
	
	\subsection{Experimental Setting} \label{sec:setting}
	
	\subsubsection{Benchmark Datasets} \label{sec:dataset}
	Eight benchmark datasets are used in the experiments, including four from FVC2002~\cite{RN2562} and four from FVC2004~\cite{RN2022}.
	Each dataset is composed of eight hundred fingerprint images collected from one hundred fingers, with eight images per finger. 
	Details about the FVC2002 datasets and FVC2004 datasets are shown in Table~\ref{tab:FVC2002} and Table~\ref{tab:FVC2004}, respectively.
	\begin{table}[htbp]
		\setlength\tabcolsep{4pt}
		\renewcommand{\arraystretch}{1.5}
		\caption{Details about datasets FVC2002 DB1-DB4}
		\label{tab:FVC2002}
		\centering
		\begin{tabular}{c|c|c|c|c}
			\hline\hline
			&Fingers&Images per finger &Image size            & DPI \\ \hline
			\textbf{FVC2002 DB1} &100     &8                       &$388\times 374$& 500 \\ \hline
			\textbf{FVC2002 DB2} &100     &8                       &$296\times 560$& 569 \\ \hline
			\textbf{FVC2002 DB3} &100     &8                       &$300\times 300$& 500 \\ \hline
			\textbf{FVC2002 DB4} &100     &8                       &$288\times 384$& 500 \\ 
			\hline \hline
		\end{tabular}
	\end{table}
	
	\begin{table}[htbp]
		\setlength\tabcolsep{4pt}
		\renewcommand{\arraystretch}{1.5}
		\caption{Information about datasets FVC2004 DB1-DB4}
		\label{tab:FVC2004}
		\centering
		\begin{tabular}{c|c|c|c|c} 
			\hline \hline
			&Fingers&Images per finger &Image size            & DPI \\ \hline
			\textbf{FVC2004 DB1} &100     &8                       &$640\times 480$& 500 \\ \hline
			\textbf{FVC2004 DB2} &100     &8                       &$328\times 364$& 500 \\ \hline
			\textbf{FVC2004 DB3} &100     &8                       &$300\times 480$& 512 \\ \hline
			\textbf{FVC2004 DB4} &100     &8                       &$288\times 384$& 500 \\ 
			\hline \hline
		\end{tabular}
	\end{table}
	
	\subsubsection{Evaluation Protocol} \label{sec:protocols}
	The widely used FVC evaluation protocol is adopted to assess the performance of the proposed template. 
	In this protocol, genuine scores and imposter scores are calculated to evaluate the performance. 
	The genuine scores are obtained by matching each fingerprint image of a finger against the remaining ones of the same finger. 
	If the matching of $P$ against $Q$ is performed, the symmetric one (i.e., $Q$ against $P$) is not tested to avoid correlation. 
	For each dataset, the total number of genuine scores is therefore 2,800 (i.e., $(8\times7)/2\times100$).
	The imposter scores are obtained by matching the first fingerprint image of each finger against the first one of remaining fingers. Similarly, repeating tests are not performed. 
	For each database, the total number of imposter scores is therefore 4,950 (i.e., ($100\times99)/2$).
	
	\subsubsection{Measurement Metrics} \label{sec:metrics}
	The following metrics, which are commonly used in biometric authentication, are adopted to evaluate the authentication accuracy of the proposed template:
	\begin{itemize}
		\item False Matching Rate (FMR): the rate of a pair of fingerprints not from the same finger falsely decided as a match.
		\item False Non-Matching Rate (FNMR): the rate of a pair of fingerprints from the same finger falsely decided as a non-match.
		\item FMR$_{1000}$: the lowest FNMR for a threshold at which the FMR $ \leqslant 1 \text{\textperthousand} $.
		\item Equal-Error Rate (EER): the value at which the FNMR is equal to the FMR. The lower the EER, the better.
		\item Detection Error Tradeoff (DET) curve: the DET curve plots the FNMR against the FMR for a series of varying thresholds.
	\end{itemize}
	
	\subsection{Authentication Accuracy with Different Values of $p$} \label{sec:emcc_p}
	In this experiment, we evaluate the effect of the feature length, controlled by $p$ in Eq. \eqref{eq:reindex}, on the authentication accuracy in terms of the DET curve, the EER, and the FMR$_{1000}$. 
	To avoid redundant computation, we evaluate three feature lengths, namely $\frac{1}{4}L_c$, $\frac{1}{6}L_c$ and $\frac{1}{8}L_c$, with $p=1$, $p=2/3$ and $p=1/2$, respectively. 
	For convenience, we use `eMCC$_{1}$', `eMCC$_{2/3}$' and `eMCC$_{1/2}$' to indicate these three features, respectively.
	Table~\ref{tab:comparisonLengthP} summarizes the relationship between the feature length and the parameter $p$ as well as the comparison of the length between these three features and the original MCC feature. 
	\begin{table}[htbp]
		\setlength\tabcolsep{3pt}
		\renewcommand{\arraystretch}{1.5}
		\caption{Comparison of the feature length with different values of $p$ in the case of $N_S=16$ and $N_D=5$}
		\label{tab:comparisonLengthP}
		\centering
		\begin{tabular}{l|c|c} 
			\hline
			\hline
			&\textbf{Cell value vector} & \textbf{Cell validity vector} \\ \hline 
			The original MCC feature					 &$L_c = 1,280$ 					 & $L_b=256$ \\ \hline
			eMCC$_{1}$ with $p=1$      				   &$\frac{p}{4}L_c = 320$     &$\frac{p}{8}N_DL_b = 160$\\  \hline
			eMCC$_{2/3}$ with $p=\frac{2}{3}$   &$\frac{p}{4}L_c = 212$ 	   &$\frac{p}{8}N_DL_b = 106$\\ \hline 
			eMCC$_{1/2}$ with $p=\frac{1}{2}$   &$\frac{p}{4}L_c = 160$ 	   &$\frac{p}{8}N_DL_b = 80$\\ 
			\hline
			\hline
		\end{tabular}
	\end{table}
	To minimize the side effects caused by missing and spurious minutiae, the commercial software Verifinger 12.1~\footnote{https://www.neurotechnology.com/verifinger.html} is employed in this experiment to extract minutiae. 
	The LGS algorithm mentioned in Section~\ref{sec:matching} is used to perform the template matching.
	
	\subsubsection{Comparison of DET Curves for Different Values of $p$}
\begin{figure}[htbp]
		\centering
		\subfloat[DET curves on FVC2002 DB1]
		{
			\includegraphics[width=0.48\linewidth]{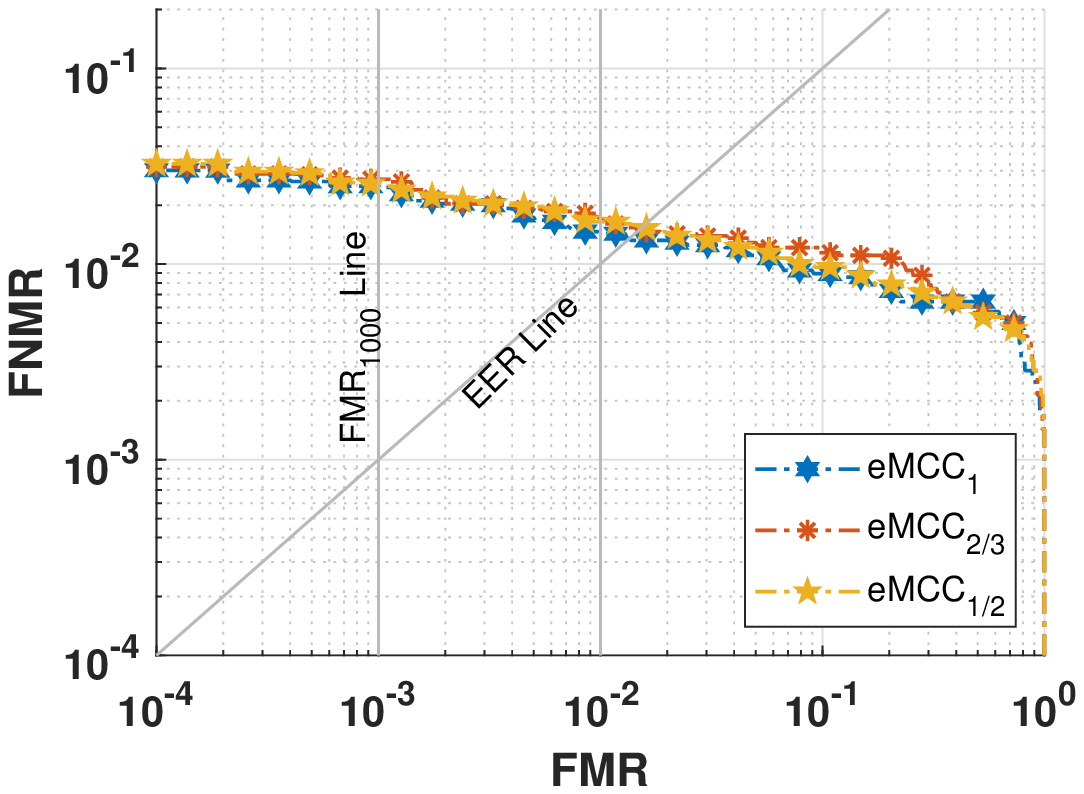}
		}
		\subfloat[DET curves on FVC2002 DB2]
		{
			\includegraphics[width=0.48\linewidth]{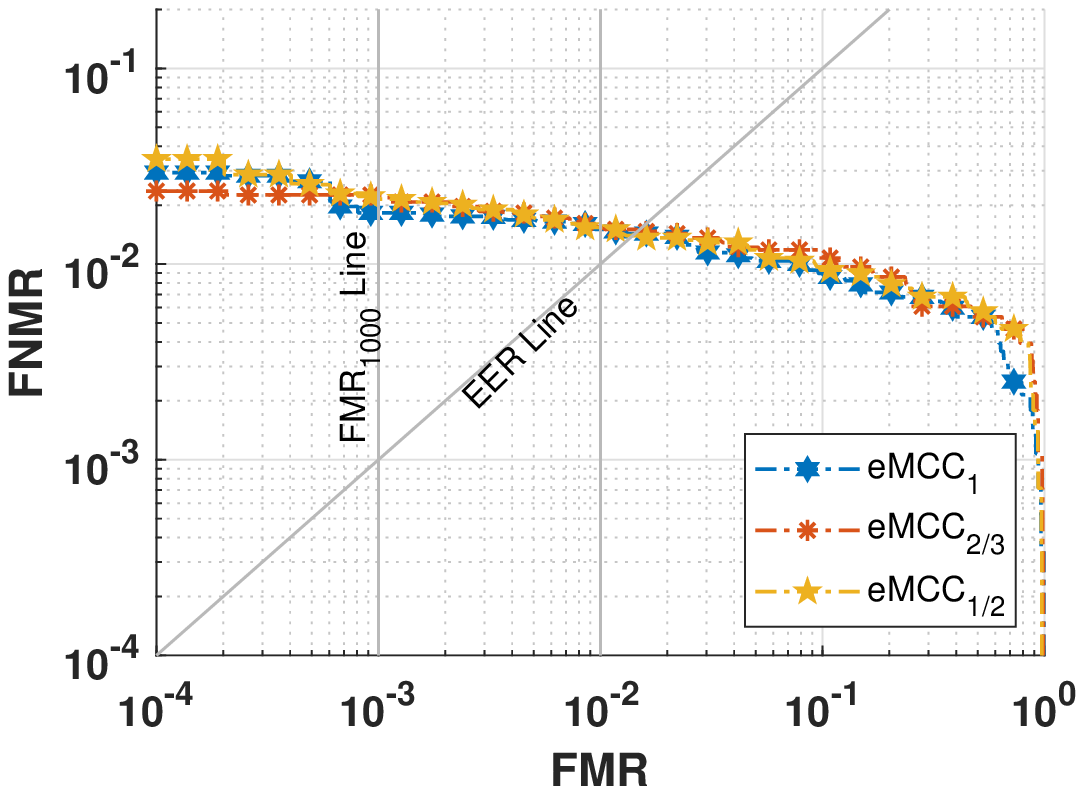}
		}
		\vfil
		\subfloat[DET curves on FVC2002 DB3]
		{
			\includegraphics[width=0.48\linewidth]{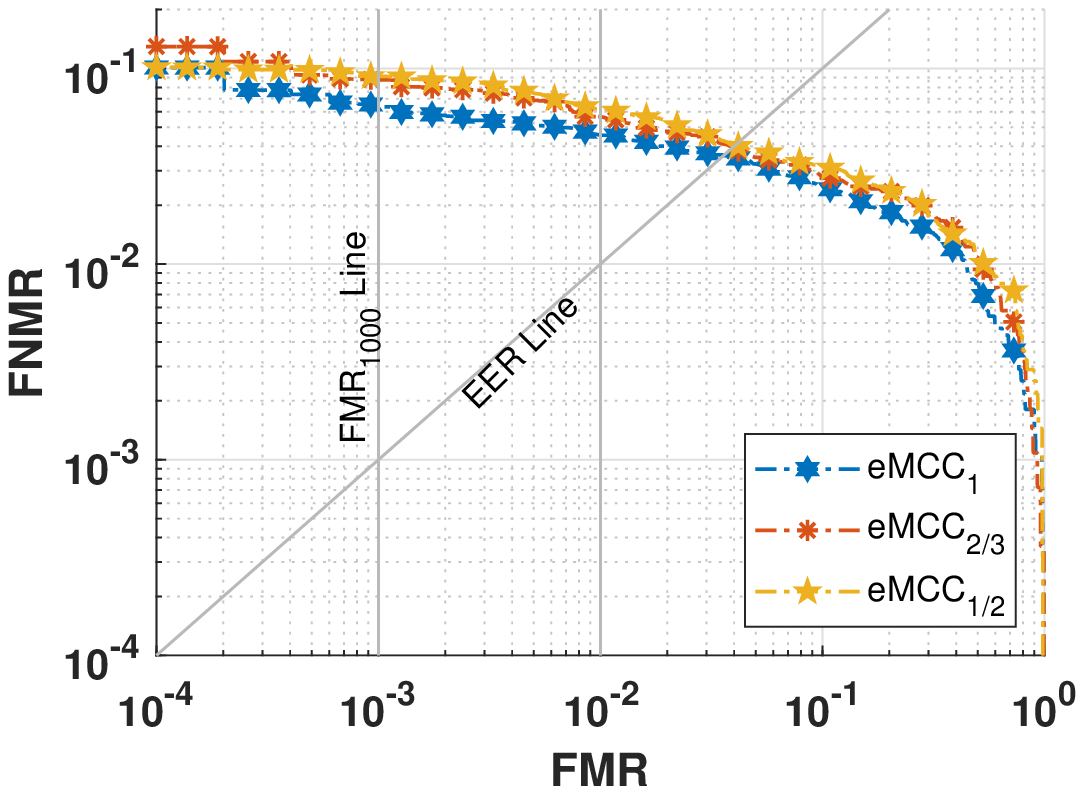}
		}
		\subfloat[DET curves on FVC2002 DB4]
		{
			\includegraphics[width=0.48\linewidth]{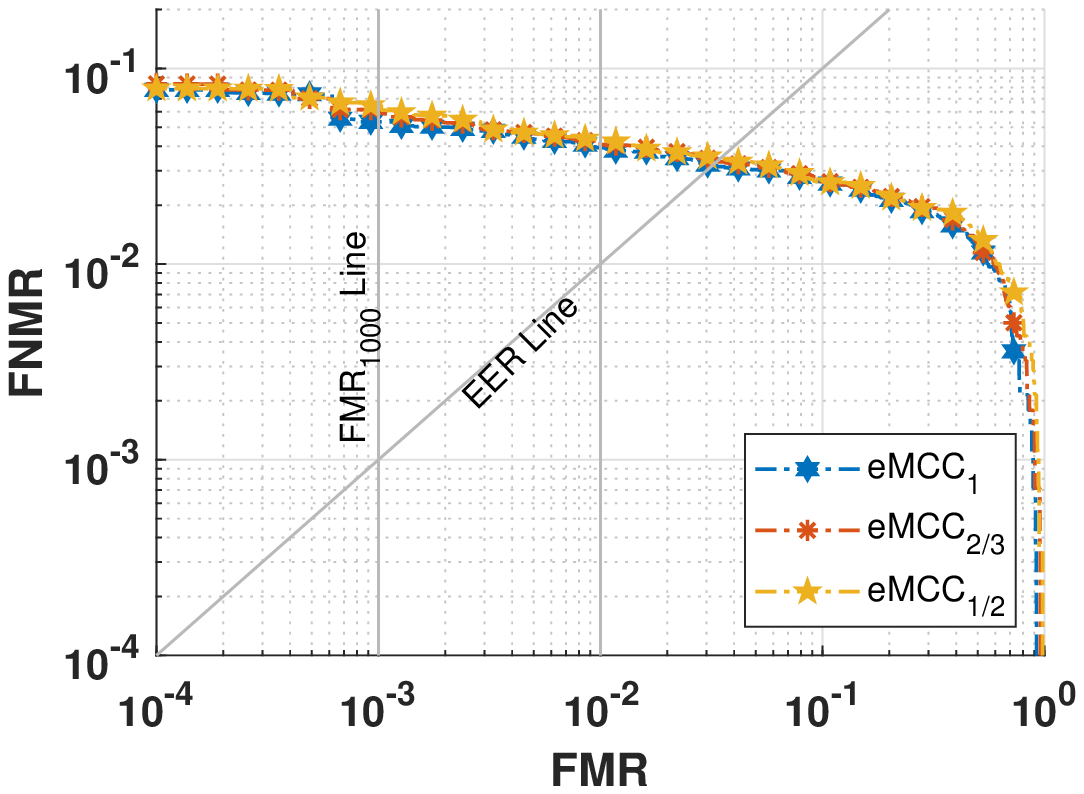}
		}
		\vfil
		\subfloat[DET curves on FVC2004 DB1]
		{
			\includegraphics[width=0.48\linewidth]{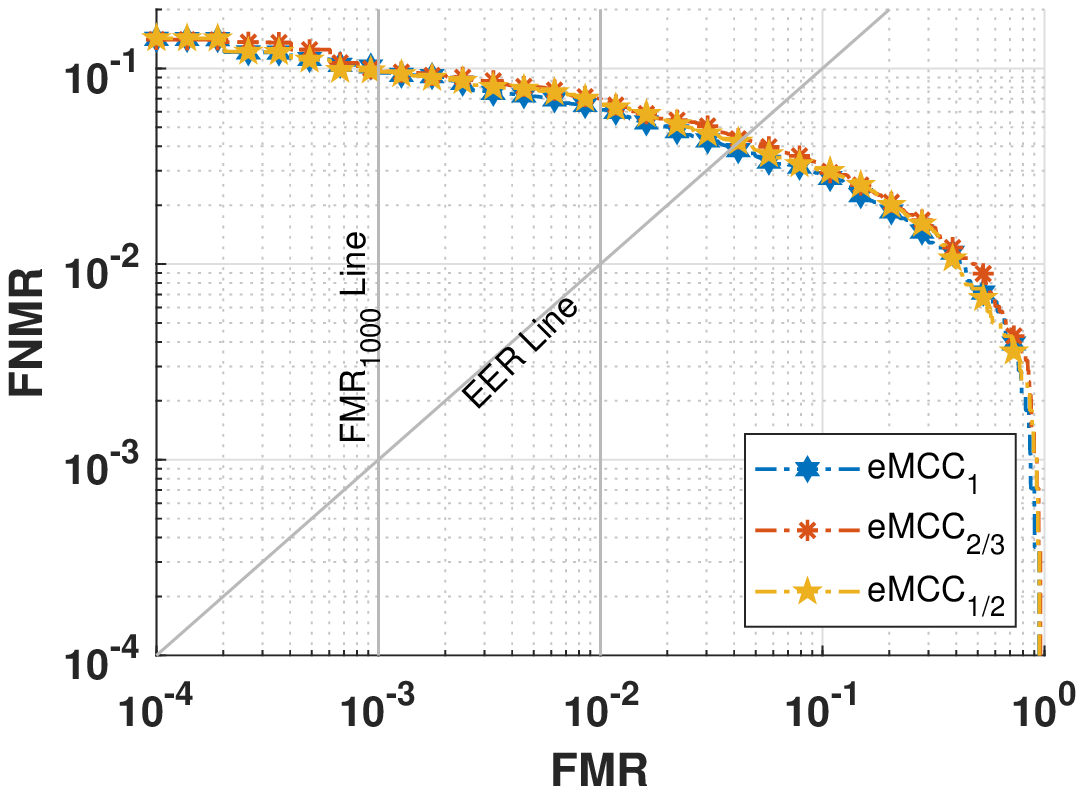}
		}
		\subfloat[DET curves on FVC2004 DB2]
		{
			\includegraphics[width=0.48\linewidth]{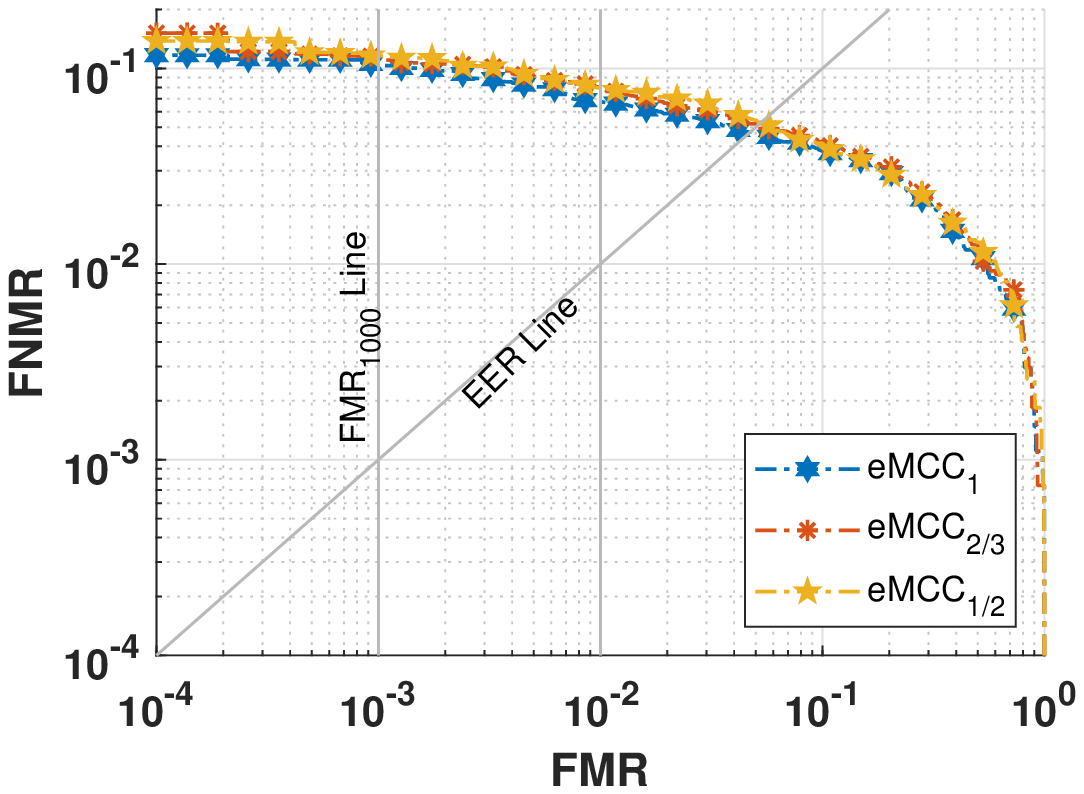}
		}
		\vfil
		\subfloat[DET curves on FVC2004 DB3]
		{
			\includegraphics[width=0.48\linewidth]{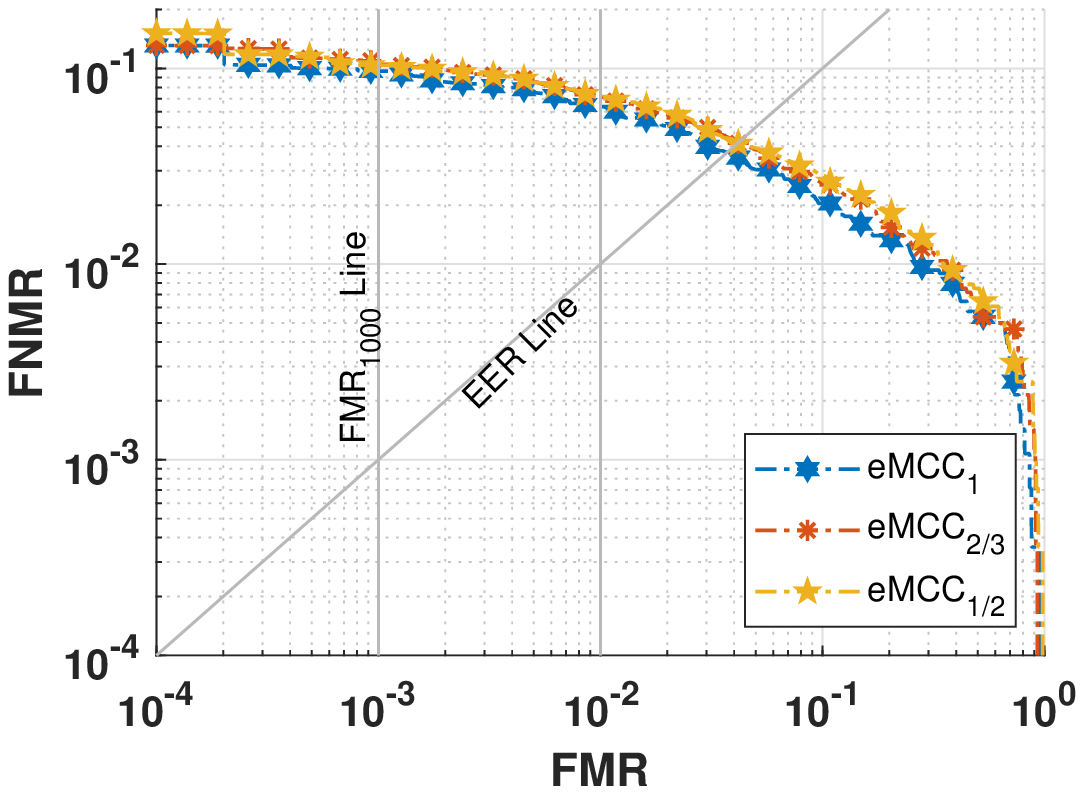}
		}
		\subfloat[DET curves on FVC2004 DB4]
		{
			\includegraphics[width=0.48\linewidth]{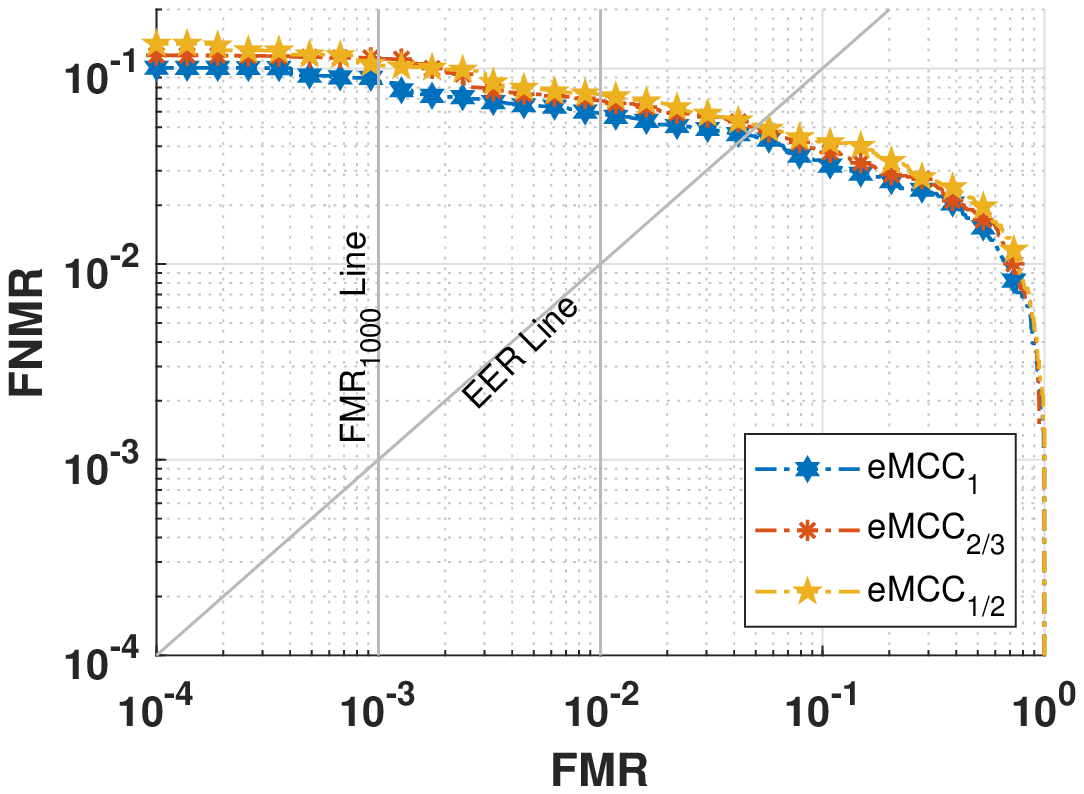}
		}
		\caption{Comparison of DET curves by eMCC$_1$, eMCC$_{2/3}$ and eMCC$_{1/2}$ evaluated on datasets FVC2002 DB1-DB4 and FVC2004 DB1-DB4.}
		\label{fig:FVC2002_DET_emcc_all}
	\end{figure}

	Fig. \ref{fig:FVC2002_DET_emcc_all} shows the comparison of DET curves evaluated by eMCC$_1$, eMCC$_{2/3}$, and eMCC$_{1/2}$ on datasets FVC2002 DB1-DB4 and FVC2004 DB1-DB4.
	It is clearly shown that similar DET curves are obtained by eMCC$_1$, eMCC$_{2/3}$, and eMCC$_{1/2}$ on these eight datasets, especially on datasets FVC2002 DB1, FVC2002 DB2, FVC2002 DB4, FVC2004 DB1, FVC2004 DB2, and FVC2004 DB3. 
	We can also observe that there are no significant differences at the intersections between the DET curves and the FMR$_{1000}$ line and the EER line.
	The DET curves obtained by eMCC$_1$ on these eight datasets are slightly better than those obtained by eMCC$_{2/3}$, and eMCC$_{1/2}$, which is because eMCC$_1$ incorporates the whole information of the original MCC feature, while eMCC$_{2/3}$, and eMCC$_{1/2}$ only utilize two thirds and half of the original MCC feature, respectively. It is worth noting that there are fewer differences between the DET curves obtained by eMCC$_{2/3}$ and those obtained by eMCC$_{1/2}$.
	In summary, eMCC$_{1}$ performs marginally better than eMCC$_{2/3}$ and eMCC$_{1/2}$, while eMCC$_{2/3}$ and eMCC$_{1/2}$ achieve much similar performance.
	
	\subsubsection{Comparison of the EER and FMR$_{1000}$ Evaluated with Different Values of $p$}
	Table~\ref{tab:EER_FMR1000_emcc_all} demonstrates the comparison of verification accuracy in terms of the EER and FMR$_{1000}$ evaluated by eMCC$_1$, eMCC$_{2/3}$, and eMCC$_{1/2}$ on FVC2002 DB1-DB4 and FVC2004 DB1-DB4.
	As shown in Table~\ref{tab:EER_FMR1000_emcc_all}, eMCC$_1$ achieves slightly better EER on most of these eight datasets than eMCC$_{2/3}$ and eMCC$_{1/2}$, except on FVC2002 DB2 where eMCC$_{1/2}$ achieves a slightly better EER than eMCC$_{1}$ and eMCC$_{2/3}$. 
	On these eight datasets, eMCC$_{2/3}$ and eMCC$_{1/2}$ achieve comparable accuracy in terms of the EER, evidenced by eMCC$_{2/3}$ performing better on five of these eight datasets than eMCC$_{1/2}$, while eMCC$_{1/2}$ obtains better EER on the other three datasets than eMCC$_{2/3}$.
	Regarding FMR$_{1000}$, eMCC$_{1}$ performs better on seven of these eight datasets than eMCC$_{2/3}$ and eMCC$_{1/2}$, while on FVC2004 DB1, eMCC$_{1/2}$ achieves a slightly better result.
	eMCC$_{2/3}$ and eMCC$_{1/2}$ achieve comparable accuracy of FMR$_{1000}$, evidenced by eMCC$_{1/2}$ achieving better results on five of these eight datasets than eMCC$_{2/3}$, while eMCC$_{2/3}$ performs better on the other three datasets than eMCC$_{1/2}$.
	In summary, eMCC$_{1}$ performs better on most of these eight datasets than eMCC$_{2/3}$ and eMCC$_{1/2}$, while eMCC$_{2/3}$ and eMCC$_{1/2}$ achieve much similar performance.
	\begin{table}[htbp]
		\setlength\tabcolsep{2pt}
		\renewcommand{\arraystretch}{1.5}
		\caption{Comparison of verification accuracy in terms of the EER and FMR$_{1000}$ obtained by eMCC$_1$, eMCC$_{2/3}$, and eMCC$_{1/2}$ on FVC2002 DB1-DB4 and FVC2004 DB1-DB4.}
		\label{tab:EER_FMR1000_emcc_all}
		\centering
		\resizebox{\linewidth}{!}
		{
			\begin{tabular}{c |c||c|c|c||c|c|c} 
				\hline		\hline
				\multicolumn{2}{c||}{\multirow{2}{*}{Dateset}}	&\multicolumn{3}{c||}{\textbf{EER (\%)}}  	          &  \multicolumn{3}{c}{\textbf{FMR$_{1000}$ (\%)}}		\\ \cline{3-8} 
				\multicolumn{2}{c||}{}                                             & eMCC$_1$ & eMCC$_{2/3}$ & eMCC$_{1/2}$  & eMCC$_1$ & eMCC$_{2/3}$ & eMCC$_{1/2}$	\\ \hline \hline
				
				\multirow{4}{*}{\textbf{FVC2002}} & DB1		        & 1.35             &1.46                   &1.55		             & 2.50            & 2.71                   &2.57		\\		\cline{2-8}
				& DB2		       & 1.43	          &1.47                   &1.40		                &1.83             &2.26                    &2.22       \\      \cline{2-8}
				& DB3			   & 3.61	          &3.97                   &4.04              		&6.53             &8.79                     &9.12		\\ 		\cline{2-8}
				& DB4		       & 3.18	          &3.33                   &3.50		                &5.33             &6.15                     &6.48		\\ 		\hline		\hline
				
				\multirow{4}{*}{\textbf{FVC2004}} & DB1       & 3.89 &4.32 &4.21	 & 10.21 & 9.97&9.75 \\ 	\cline{2-8}
				& DB2		& 4.72	&5.20&5.28		&10.52 &11.53&11.61		\\ 		\cline{2-8}
				& DB3		& 3.71	&4.15&4.11		&9.71 &10.92&10.38		\\ 		\cline{2-8}
				& DB4		& 4.54	&5.01&5.12		&8.95 &11.32&10.55		\\ 		\hline		\hline
			\end{tabular}
		}
	\end{table}
	
	{\color{color}
		\subsection{Performance  Against the Number of Nesting Layers }
		
		The main idea of our nested-difference operation is to extract discriminative features exhibiting the difference of neighboring feature cell pairs. So, in the first nesting layer $\mathbf{e^{L_1}}$ in Eq. \eqref{eq:firstDif}, a nested difference $e^{L_1}_{i}$ involves two cells of the cell vector. Four cell values contribute to the nested difference in the second layer $\mathbf{e^{L_2}}$ in Eq. \eqref{eq:2k}. Eight cells of the cell vector will contribute to the nested difference in the third layer $\mathbf{e^{L_3}}$, where the $i^{th}$ element $e^{L_3}_{i} $is formulated by Eq. \eqref{eq:newDif_8}.
    	\begin{equation}\label{eq:newDif_8}
    	\begin{split}
    	    e^{L_3}_i = & ~ e^{L_2}_{2i-1} - e^{L_2}_{2i} \\
    	    = & ~ ((c_{t_{8i-7}} - c_{t_{8i-6}}) - (c_{t_{8i-5}} - c_{t_{8i-4}})) \\
    	    & ~ - ((c_{t_{8i-3}} - c_{t_{8i-2}}) - (c_{t_{8i-1}} - c_{t_{8i}})).
    	\end{split}
    	\end{equation}
		We use dMCC$_1$ to denote the new template defined by Eq. \eqref{eq:newDif_8} with $p=1$.
	 Experiments are conducted to show the performance against the number of nesting layers (i.e., number of cell vector values involved in the nested difference). 

		
		As shown in Table \ref{tab:EER_FMR1000_emcc_all_4}, Fig. \ref{fig:FVC2002_DET_all_4} and Table \ref{tab:templateLength_8}, compared to eMCC$_1$, dMCC$_1$ obtains much worse accuracy in terms of EER and FNMR$_{1000}$, although it saves half storage space. 
		Compared to eMCC$_{1/2}$ which has the same storage space, dMCC$_1$ still achieves much worse accuracy in terms of EER and FNMR$_{1000}$ for all four datasets.
		Apparently, two layers of nesting can strike the best balance between the template size and authentication accuracy. 
		\begin{table}[htbp]
			\setlength\tabcolsep{2pt}
			\renewcommand{\arraystretch}{1.5}
			\caption{\color{color}Comparison of verification accuracy in terms of the EER and FMR$_{1000}$ obtained by bMCC, eMCC$_1$, eMCC$_{1/2}$, and dMCC$_1$ on FVC2002 DB1-DB4.}
			\label{tab:EER_FMR1000_emcc_all_4}
			\centering
			\resizebox{\linewidth}{!}
			{\color{color}
				\begin{tabular}{c||c|c|c|c||c|c|c|c} 
					\hline		\hline
					\multirow{2}{*}{Dateset}	&\multicolumn{4}{c||}{\textbf{EER (\%)}}  	          &  \multicolumn{4}{c}{\textbf{FMR$_{1000}$ (\%)}}		\\ \cline{2-9} 
					& bMCC & eMCC$_1$ & eMCC$_{1/2}$  & dMCC$_1$ &bMCC &  eMCC$_1$ &  eMCC$_{1/2}$&dMCC$_1$	\\ 	\hline 
					
					DB1& 1.04 &1.35 &1.55 &1.75 &2.04 &2.50 &2.57&3.22		\\		\hline
					DB2& 1.15 &1.43 &1.40 &2.11 &1.77 &1.83 &2.22&3.11       \\      \hline
					DB3& 3.00 &3.61 &4.04 &6.99 &6.69 &6.53 &9.12&13.24		\\ 		\hline
					DB4& 2.86 &3.18 &3.50 &4.23 &4.92 &5.33 &6.48&7.75		\\ 		\hline		\hline
					
				\end{tabular}
			}
		\end{table}
		
		\begin{figure}[htbp]
			\color{color}
			\centering
			\subfloat[DET curves evaluated on DB1]
			{
				\includegraphics[width=0.45\linewidth]{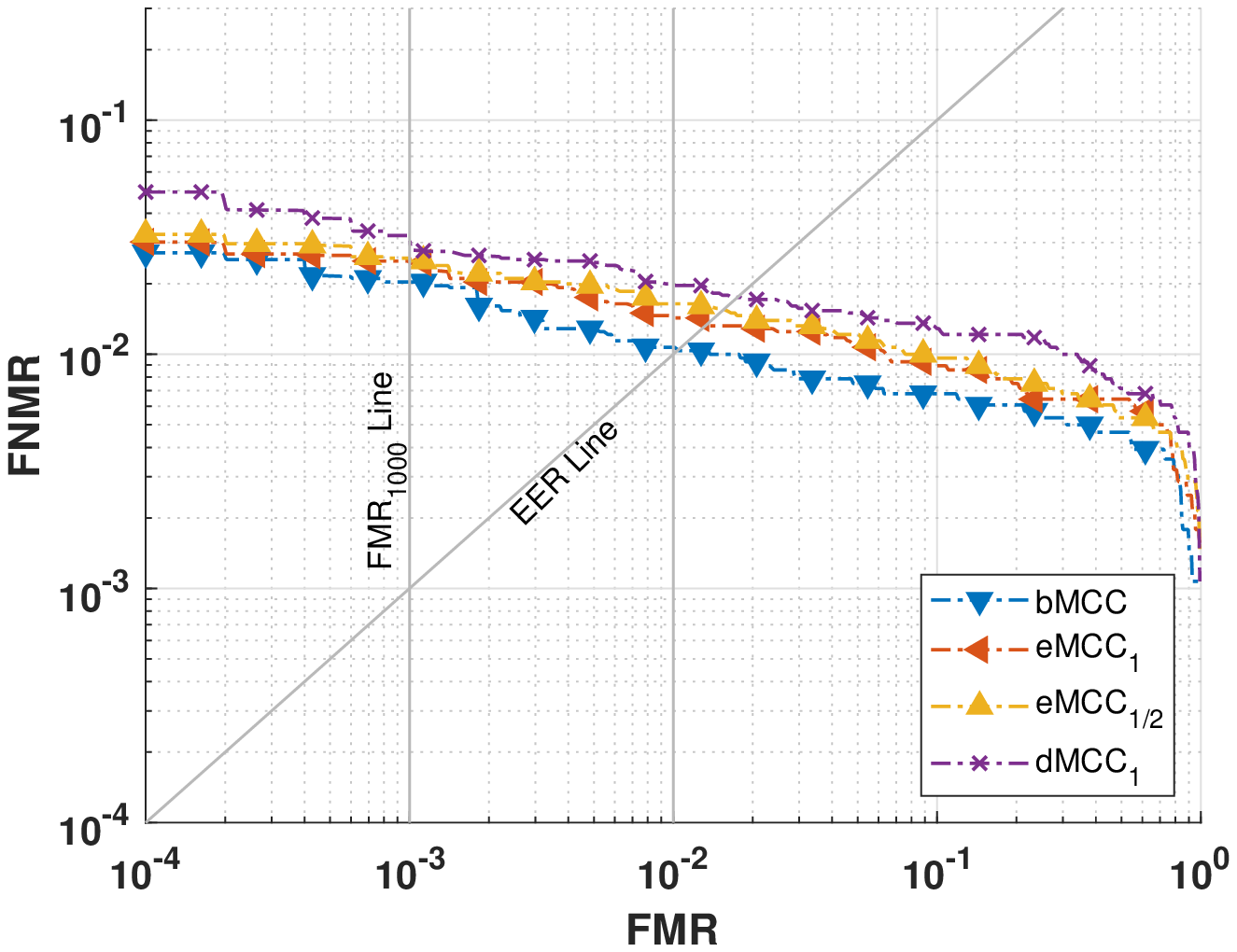}
			}
			\subfloat[DET curves evaluated on DB2]
			{
				\includegraphics[width=0.45\linewidth]{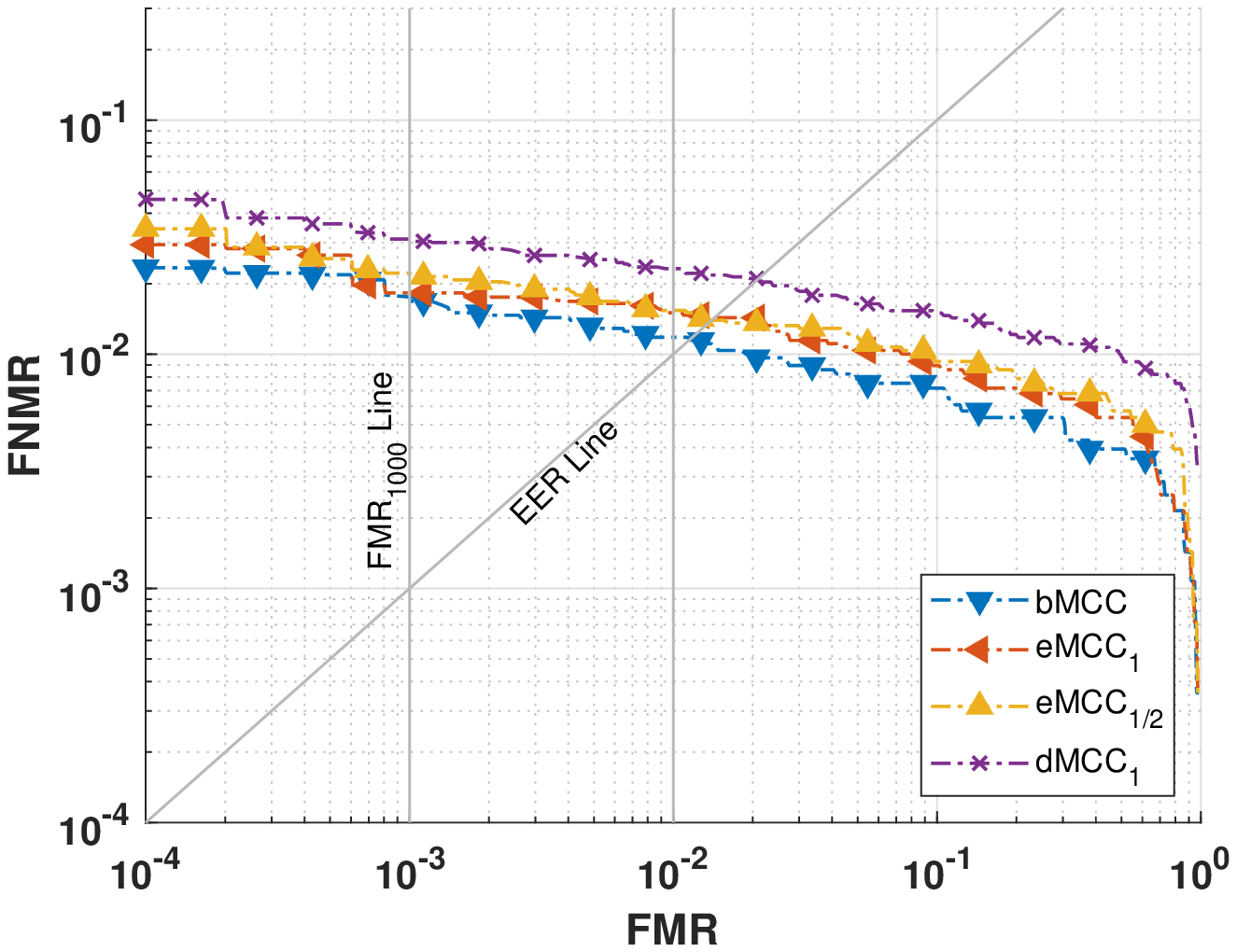}
			}
			\vfil
			\subfloat[DET curves evaluated on DB3]
			{
				\includegraphics[width=0.45\linewidth]{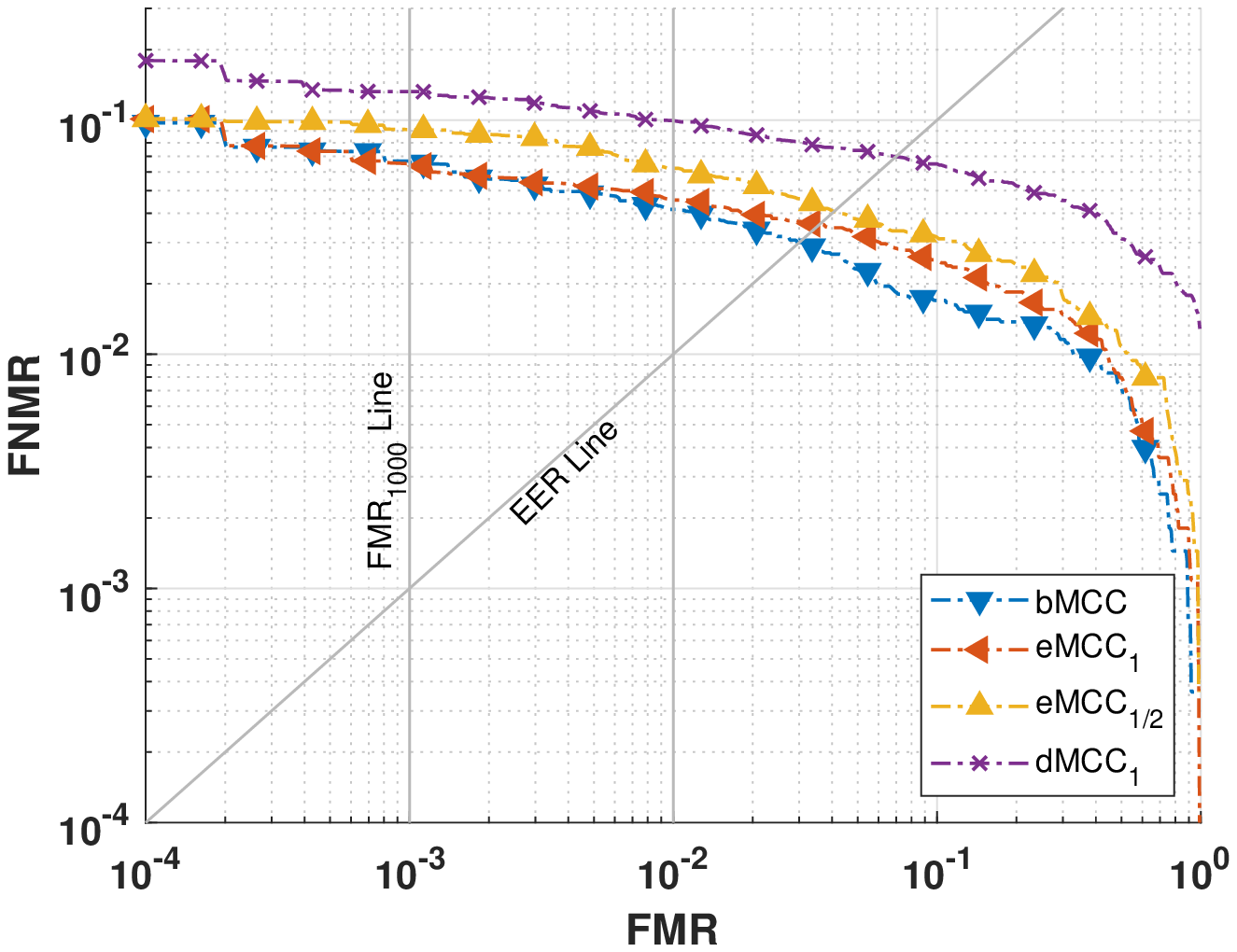}
			}
			\subfloat[DET curves evaluated on DB4]
			{
				\includegraphics[width=0.45\linewidth]{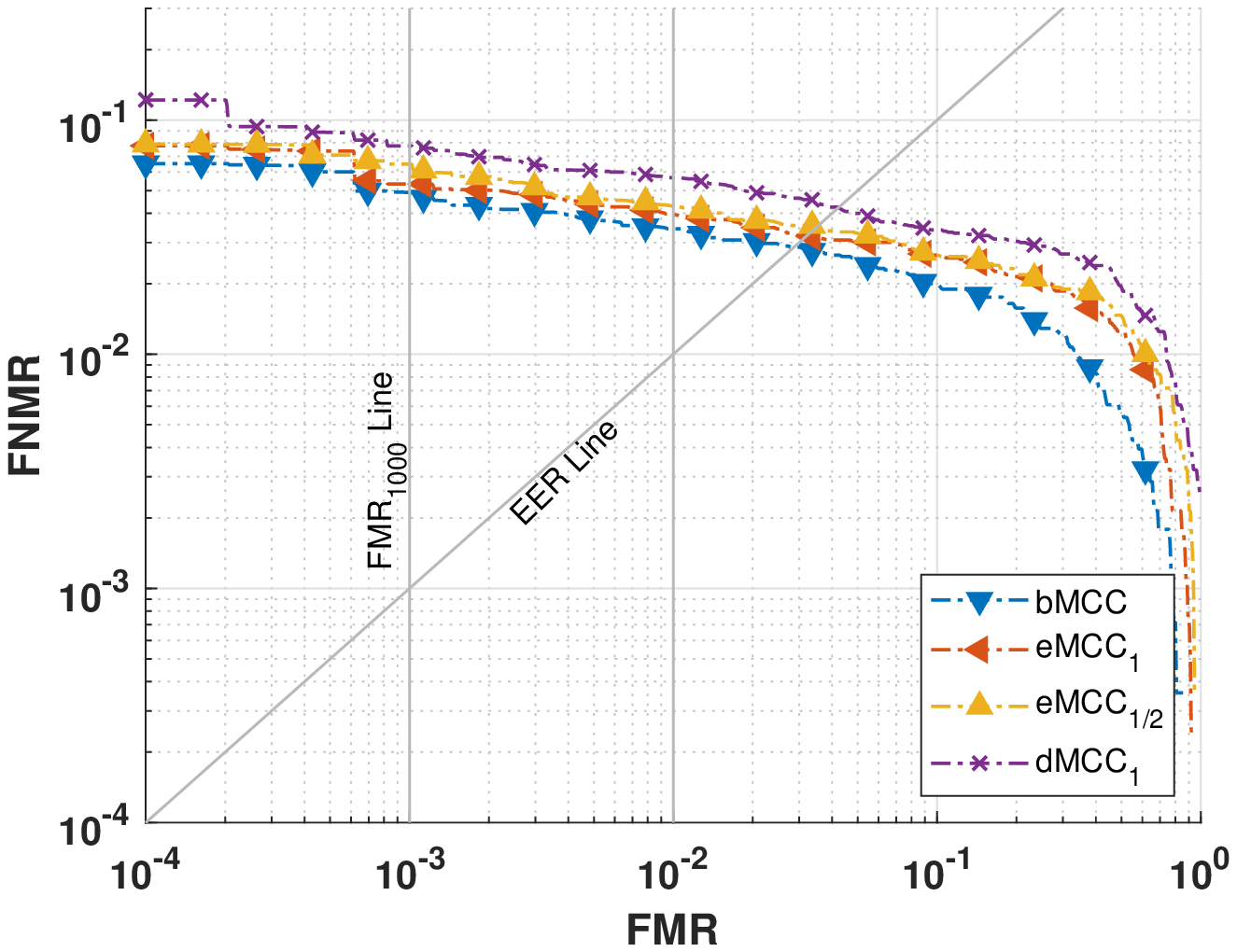}
			}
			\caption{\color{color}Comparison of DET curves obtained by bMCC,  eMCC$_1$, eMCC$_{1/2}$ and dMCC$_{1}$ on datasets FVC2002 DB1-DB4.}
			\label{fig:FVC2002_DET_all_4}
		\end{figure}
		
		\begin{table}[htbp]
			\color{color}
			\setlength\tabcolsep{6pt}
			\renewcommand{\arraystretch}{1.5}
			\caption{\color{color}Comparison of the proposed template with bMCC in the case of $N_S=16$, $N_D=5$ and the number of minutiae $n=50$, with $L_c = N_S \times N_S \times N_D$ and $L_b = N_S \times N_S$}
			\label{tab:templateLength_8}
			\centering
			\begin{tabular}{l|c}
				\hline\hline
				&Template length (bits) \\ \hline
				bMCC~\cite{RN159}                     & $n(L_c + L_b) = 76,800$ 					\\ \hline
				eMCC$_1$ with $p = 1 $            & $n(\frac{1}{4}L_c+\frac{1}{8}N_DL_b) = 24,000$ \\  \hline
				eMCC$_{1/2}$ with $p = 1/2 $ & $n(\frac{1}{8}L_c +\frac{1}{16}N_DL_b) = 12,000$ \\  \hline
				dMCC$_{1}$ with $p = 1 $         & $n(\frac{1}{8}L_c +\frac{1}{16}N_DL_b) = 12,000 $ \\  \hline		
				\hline
			\end{tabular}
	\end{table}}

	\subsection{Comparison of the Proposed Lightweight Cancelable Template with State-of-the-art Methods} \label{sec:validation}
	
	In this section, we comprehensively compare the proposed lightweight cancelable template with state-of-the-art methods in four essential aspects: 
	\begin{itemize}
		\item[1)] Template characteristics, including template length, IoT oriented, binary, and cancelable;
		\item[2)] Distributions of matching score;
		\item[3)] DET curves; and
		\item[4)] EER and FMR$_{1000}$ evaluation. 
	\end{itemize}
	Similar to Section \ref{sec:emcc_p}, to reduce the impact of missing and spurious minutiae, the commercial software Verifinger 12.1 is adopted in this experiment for minutia extraction. The LGS algorithm introduced in Section~\ref{sec:matching} is used to perform the template matching.
	The three state-of-the-art templates used as the baseline are summarized as follows: 
	1) the original MCC template~\cite{RN159} (denoted as `MCC'); 
	2) the original binary MCC template~\cite{RN159} (denoted as `bMCC') obtained by binarizing the MCC template; and 
	3) the latest IoT-oriented privacy-preserving template~\cite{RN2761} (denoted as `cMCC') developed upon the MCC template. 
	The experimental results for MCC, bMCC, and cMCC are provided by \cite{RN2761}.
	
	\subsubsection{Comparison of the Template Characteristics}
	Table \ref{tab:templateSummary} compares the template characteristics of the proposed lightweight cancelable template and the baseline (i.e., the aforementioned three state-of-the-art fingerprint templates~MCC, bMCC and cMCC). 
	Compared with MCC and bMCC, the IoT-oriented binary cancelable template cMCC reduces half of the cell value part but does not save the cell validity part. 
	By contrast, the proposed template achieves substantial storage savings in both the cell value part and the cell validity part. 
	\begin{table}[htbp]
		\setlength\tabcolsep{4pt}
		\renewcommand{\arraystretch}{1.5}
		\caption{Template characteristics of the baseline and proposed templates}
		\label{tab:templateSummary}
		\centering
		\begin{tabular}{l|c|c|c|c}
			\hline\hline
			&IoT-oriented &Binary          &Cancelable    & Template length \\ \hline
			MCC~\cite{RN159}    &					    &                      &				      & $n(L_c+L_b)$ 					\\ \hline
			bMCC~\cite{RN159}   &					   &\checkmark &				      & $n(L_c+L_b)$ 					\\ \hline
			cMCC~\cite{RN2761} &\checkmark &\checkmark  &\checkmark & $n(\frac{1}{2}L_c+\frac{1}{2}L_b)$    \\ \hline
			eMCC$_1$   			     &\checkmark &\checkmark  &\checkmark & $n(\frac{1}{4}L_c+\frac{1}{8}N_DL_b)$ \\  \hline
			eMCC$_{2/3}$  		   &\checkmark &\checkmark  &\checkmark & $n(\frac{1}{6}L_c +\frac{1}{12}N_DL_b)$ \\  \hline		
			eMCC$_{1/2}$  		   &\checkmark &\checkmark  &\checkmark & $n(\frac{1}{8}L_c +\frac{1}{16}N_DL_b)$ \\ 
			\hline \hline
		\end{tabular}
	\end{table}

	\subsubsection{Comparison of Matching Score Distributions}
\begin{figure*}[htbp]
		\centering
		\subfloat[MCC on DB1]
		{
			\includegraphics[width=0.16\linewidth]{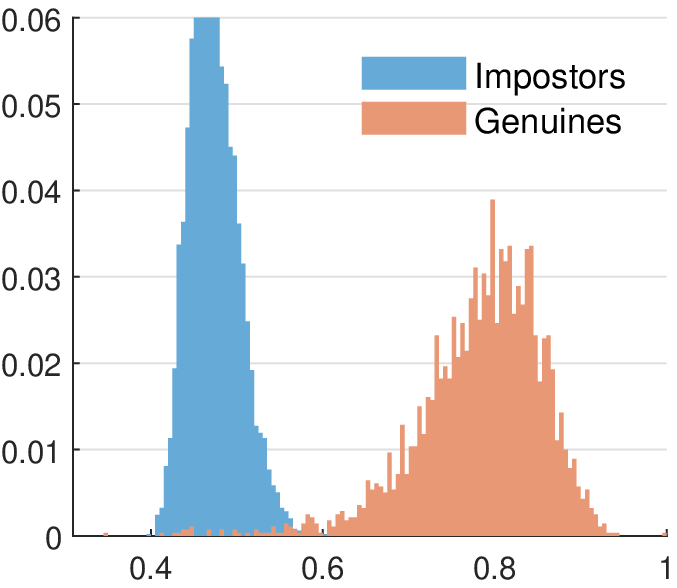}
		}
		\subfloat[bMCC on DB1]
		{
			\includegraphics[width=0.16\linewidth]{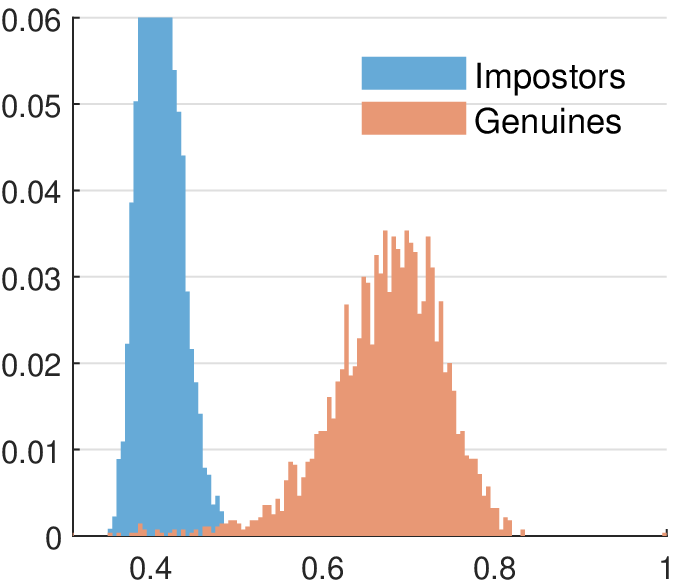}
		}
		\subfloat[cMCC on DB1]
		{
			\includegraphics[width=0.16\linewidth]{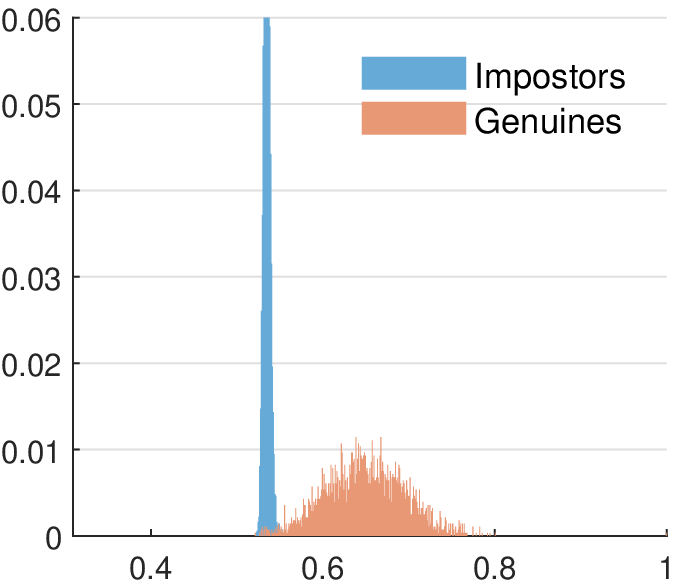}
		}
		\subfloat[eMCC$_1$ on DB1]
		{
			\includegraphics[width=0.16\linewidth]{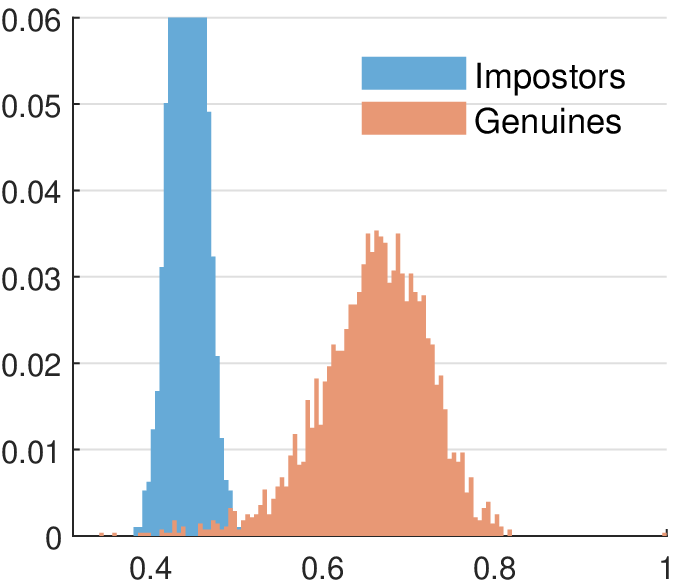}
		}
		\subfloat[eMCC$_{2/3}$ on DB1]
		{
			\includegraphics[width=0.16\linewidth]{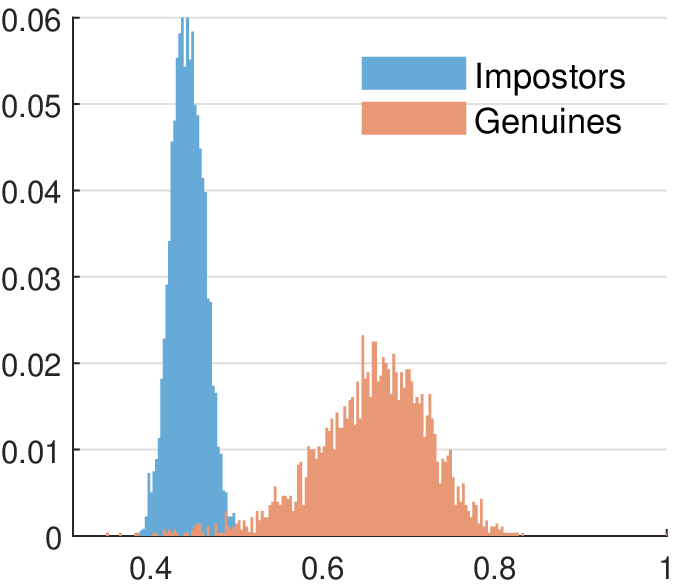}
		}
		\subfloat[eMCC$_{1/2}$ on DB1]
		{
			\includegraphics[width=0.16\linewidth]{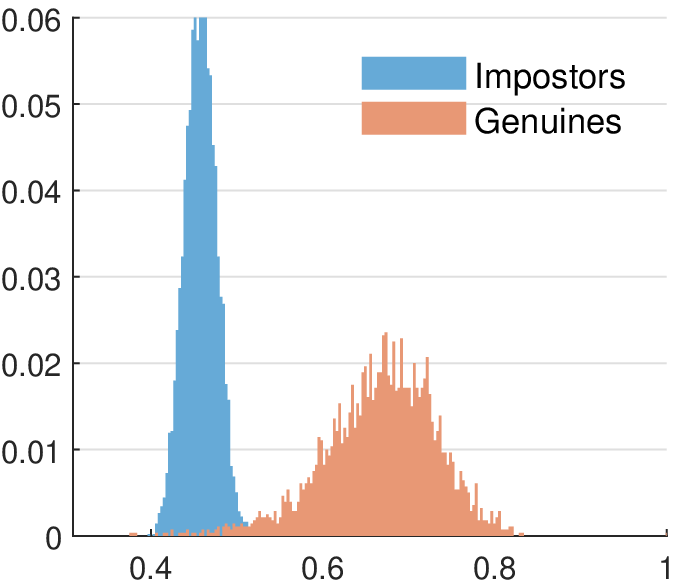}
		}
		\vfil
		
		\subfloat[MCC on DB2]
		{
			\includegraphics[width=0.16\linewidth]{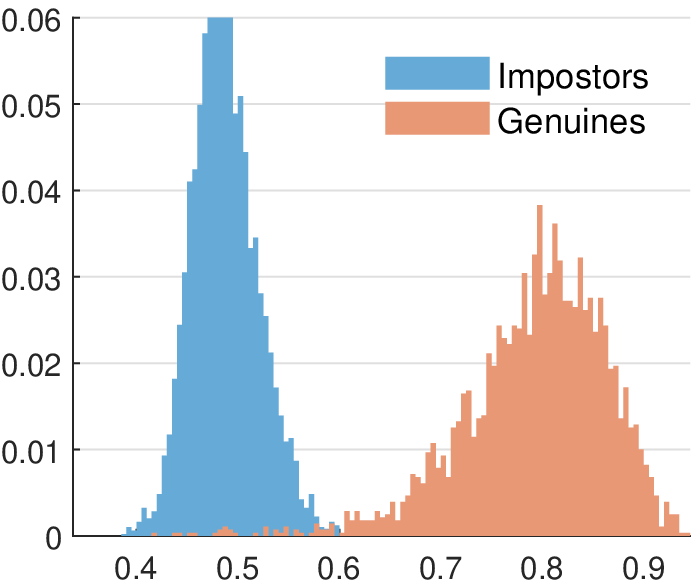}
		}
		\subfloat[bMCC on DB2]
		{
			\includegraphics[width=0.16\linewidth]{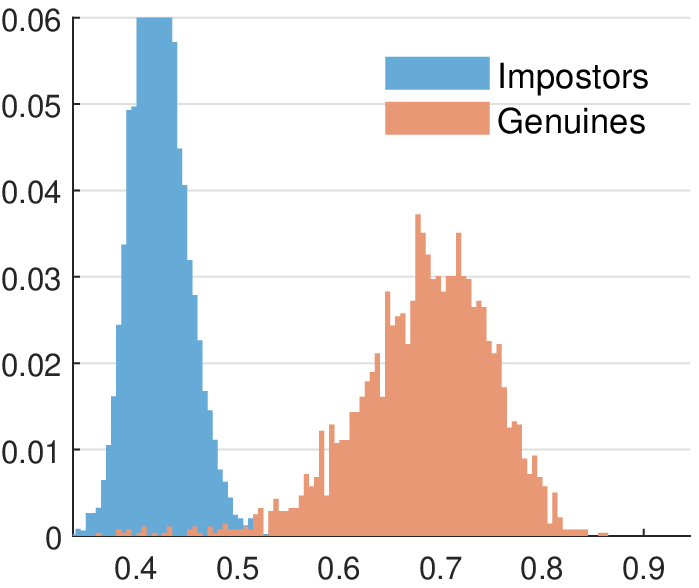}
		}
		\subfloat[cMCC on DB2]
		{
			\includegraphics[width=0.16\linewidth]{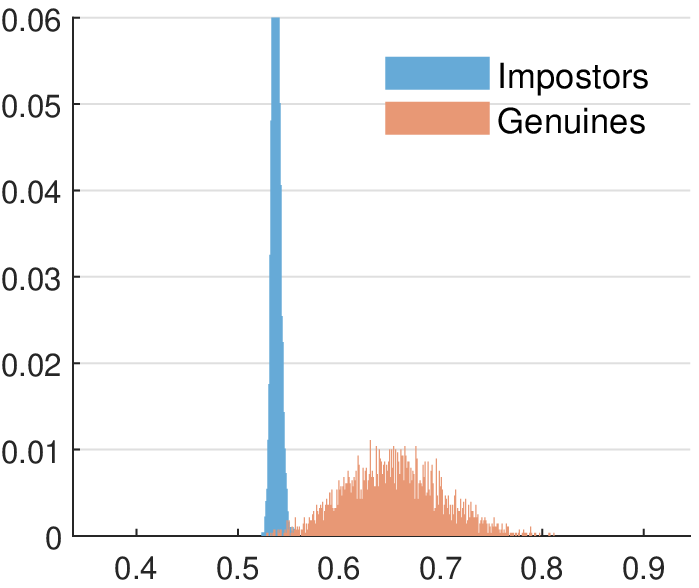}
		}
		\subfloat[eMCC$_1$ on DB2]
		{
			\includegraphics[width=0.16\linewidth]{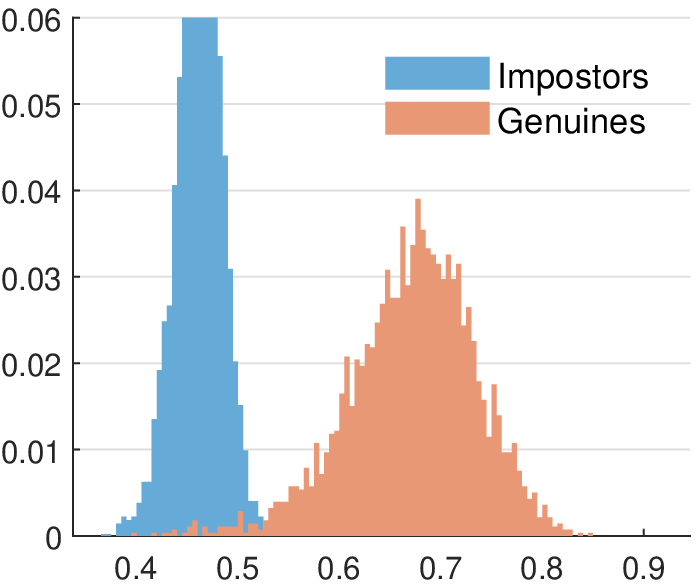}
		}
		\subfloat[eMCC$_{2/3}$ on DB2]
		{
			\includegraphics[width=0.16\linewidth]{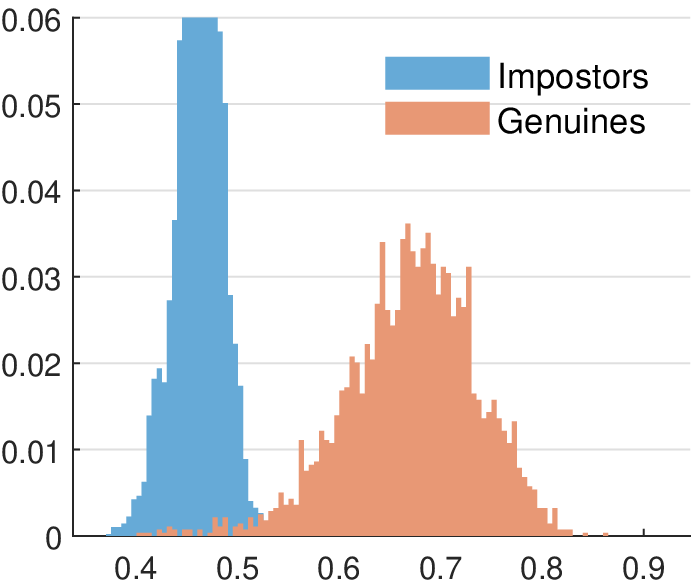}
		}
		\subfloat[eMCC$_{1/2}$ on DB2]
		{
			\includegraphics[width=0.16\linewidth]{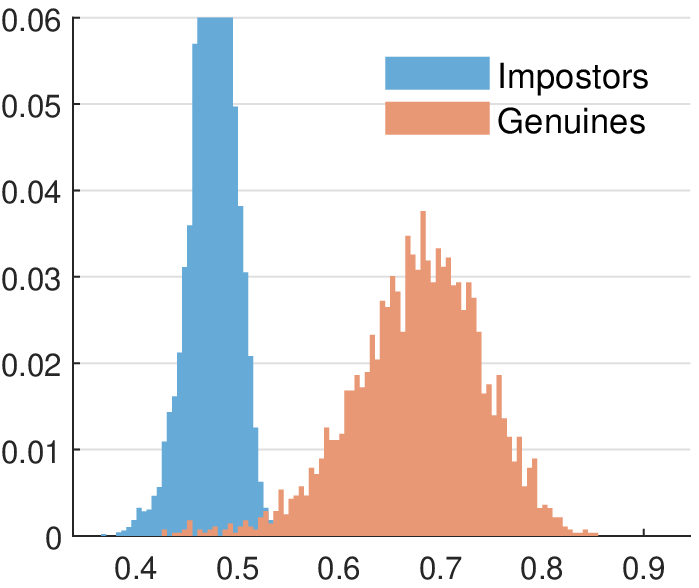}
		}
		\vfil
		
		\subfloat[MCC on DB3]
		{
			\includegraphics[width=0.16\linewidth]{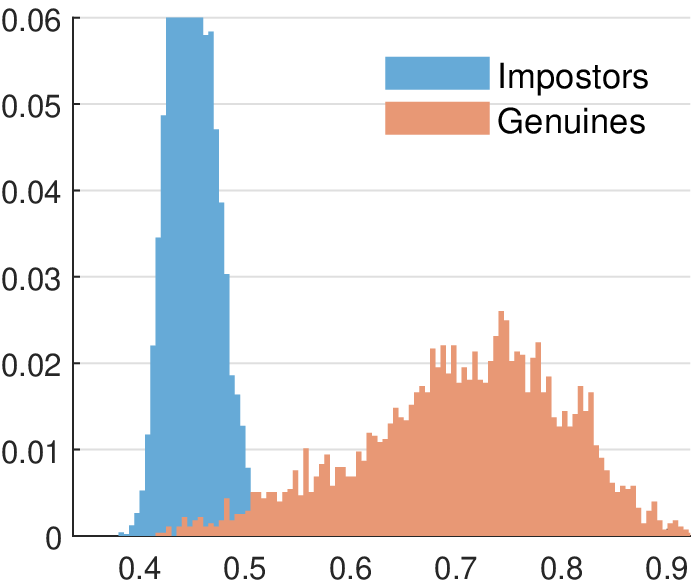}
		}
		\subfloat[bMCC on DB3]
		{
			\includegraphics[width=0.16\linewidth]{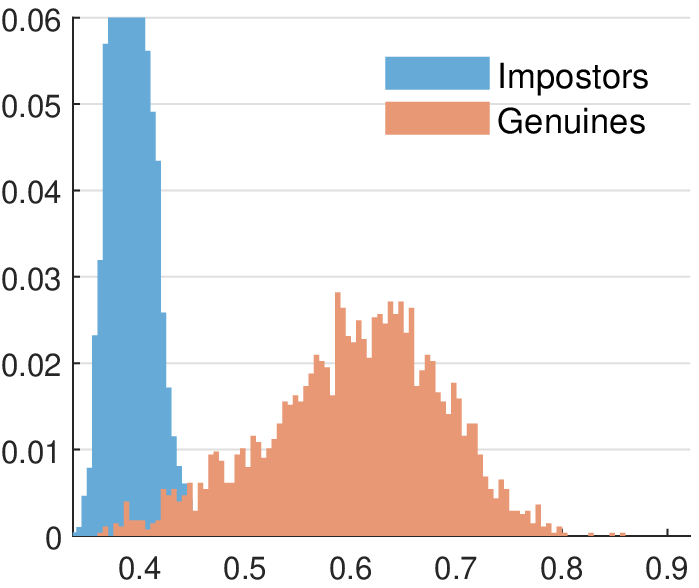}
		}
		\subfloat[cMCC on DB3]
		{
			\includegraphics[width=0.16\linewidth]{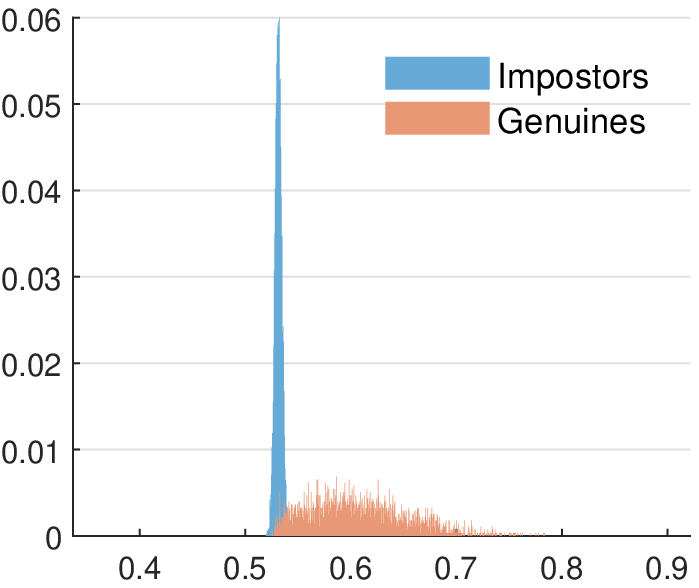}
		}
		\subfloat[eMCC$_1$ on DB3]
		{
			\includegraphics[width=0.16\linewidth]{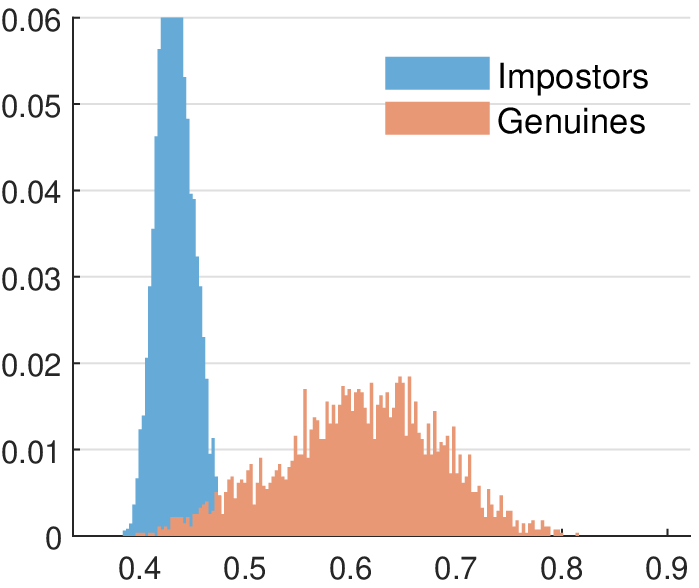}
		}
		\subfloat[eMCC$_{2/3}$ on DB3]
		{
			\includegraphics[width=0.16\linewidth]{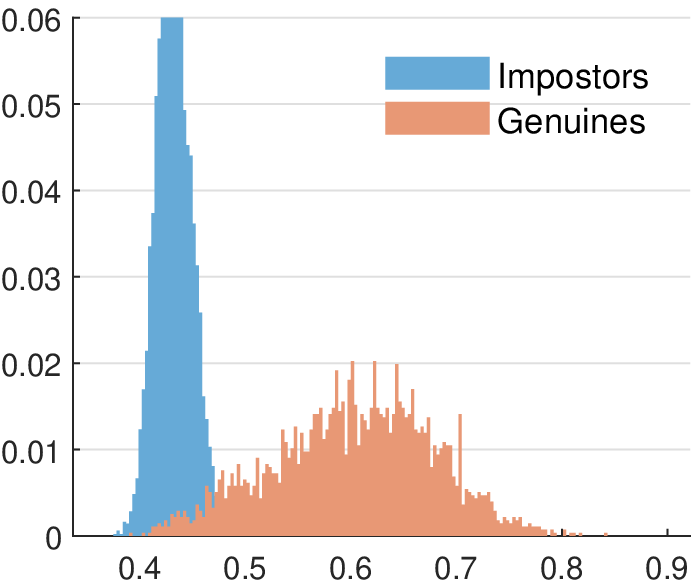}
		}
		\subfloat[eMCC$_{1/2}$ on DB3]
		{
			\includegraphics[width=0.16\linewidth]{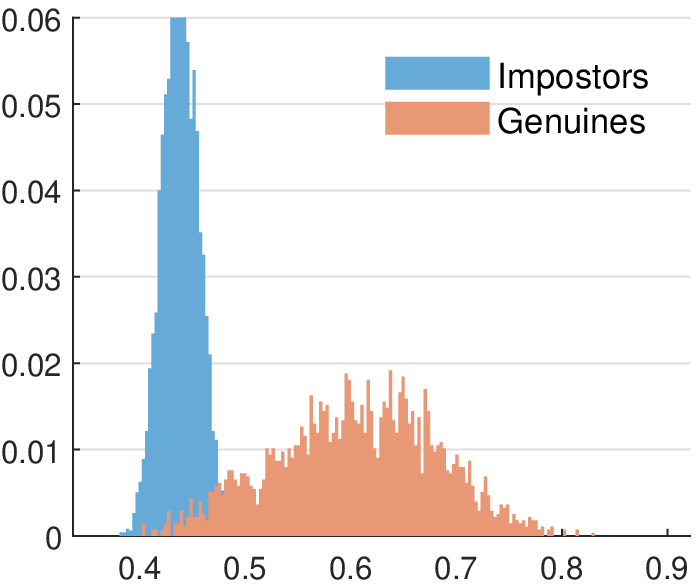}
		}
		\vfil
		
		\subfloat[MCC on DB4]
		{
			\includegraphics[width=0.16\linewidth]{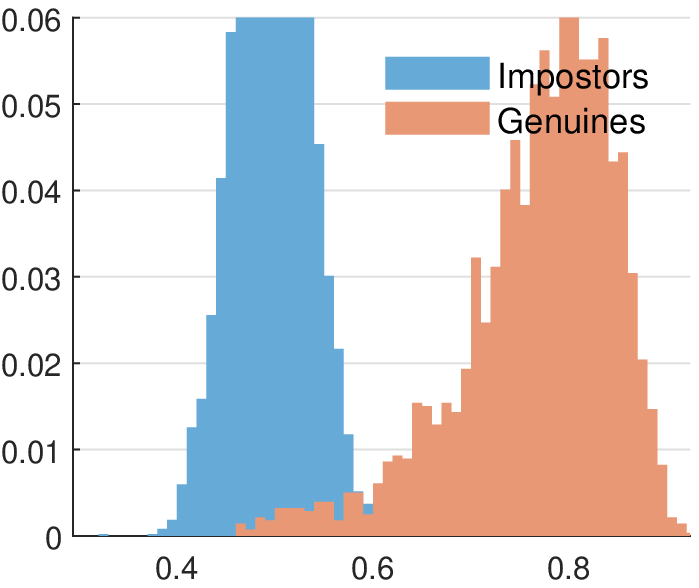}
		}
		\subfloat[bMCC on DB4]
		{
			\includegraphics[width=0.16\linewidth]{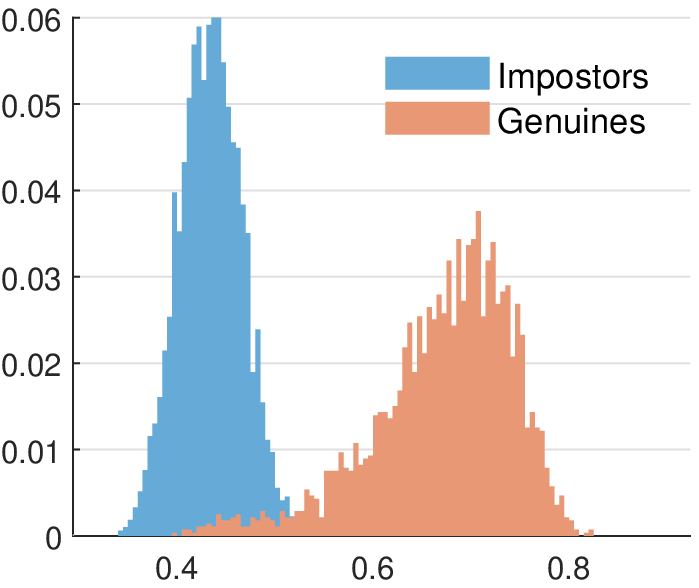}
		}
		\subfloat[cMCC on DB4]
		{
			\includegraphics[width=0.16\linewidth]{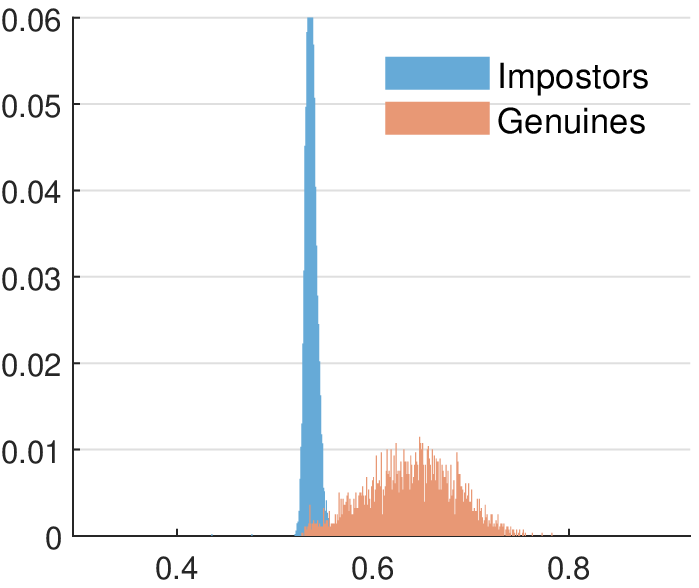}
		}
		\subfloat[eMCC$_1$ on DB4]
		{
			\includegraphics[width=0.16\linewidth]{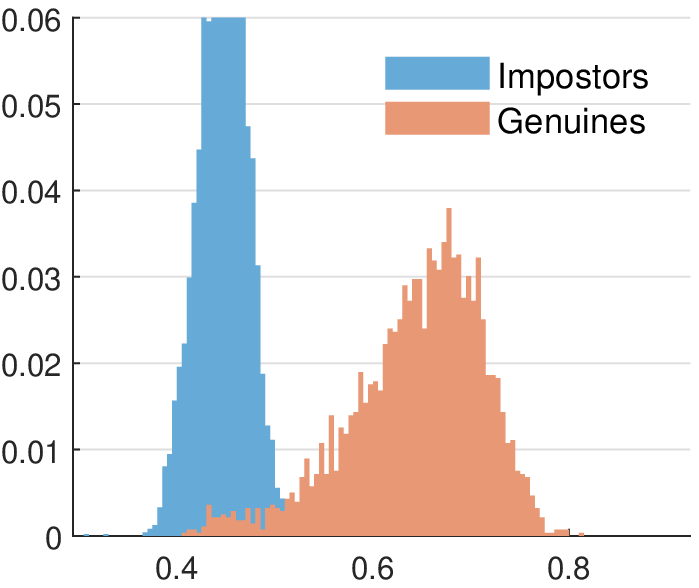}
		}
		\subfloat[eMCC$_{2/3}$ on DB4]
		{
			\includegraphics[width=0.16\linewidth]{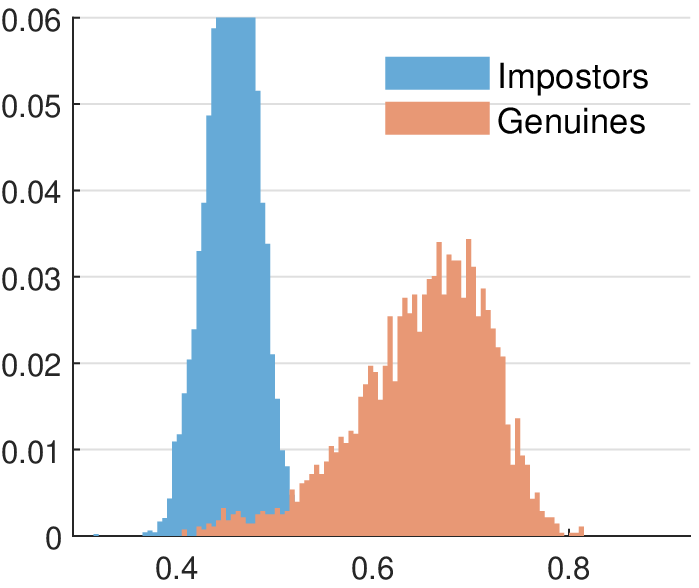}
		}
		\subfloat[eMCC$_{1/2}$ on DB4]
		{
			\includegraphics[width=0.16\linewidth]{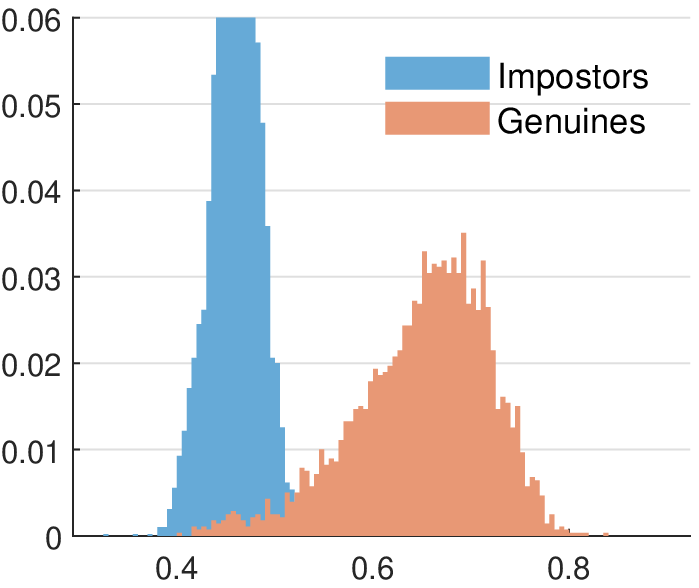}
		}
		
		\caption{Comparison of score distributions by MCC, bMCC, cMCC, eMCC$_1$, eMCC$_{2/3}$ and eMCC$_{1/2}$ evaluated on datasets FVC2002 DB1-DB4. The x-axis is the matching score, and the y-axis is the proportion of scores falling into each score bin.}
		\label{fig:score_distribution_FVC2002}
	\end{figure*}

\begin{figure*}[htbp]
		\centering
		\subfloat[MCC on DB1]
		{
			\includegraphics[width=0.16\linewidth]{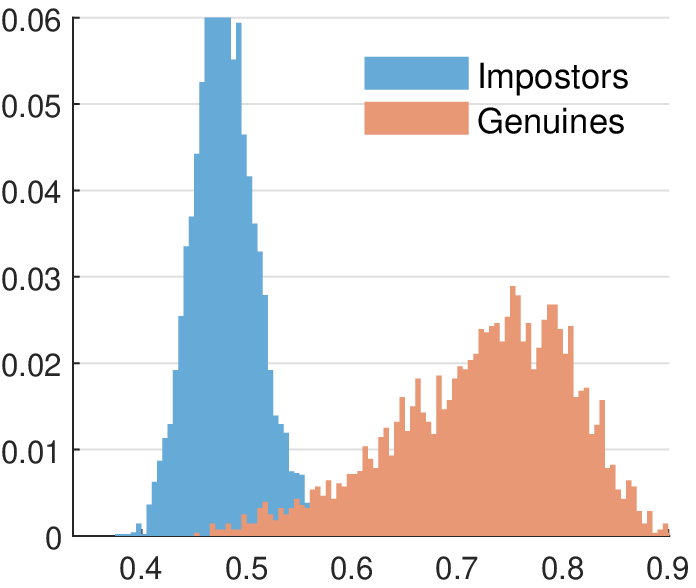}
		}
		\subfloat[bMCC on DB1]
		{
			\includegraphics[width=0.16\linewidth]{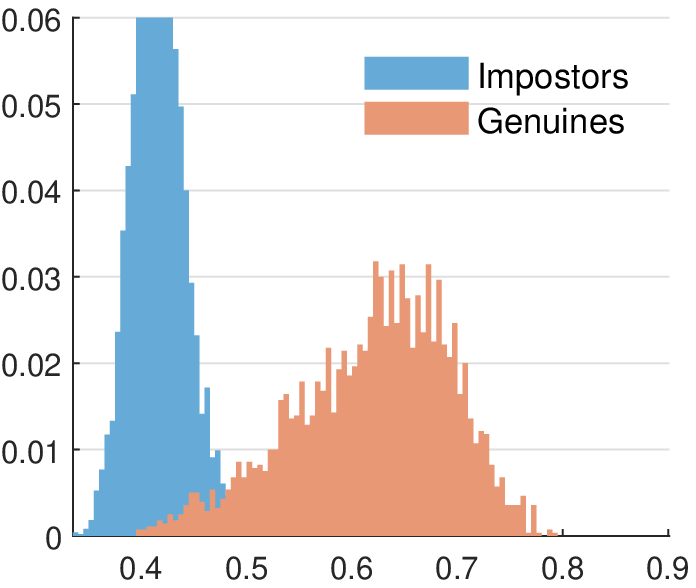}
		}
		\subfloat[cMCC on DB1]
		{
			\includegraphics[width=0.16\linewidth]{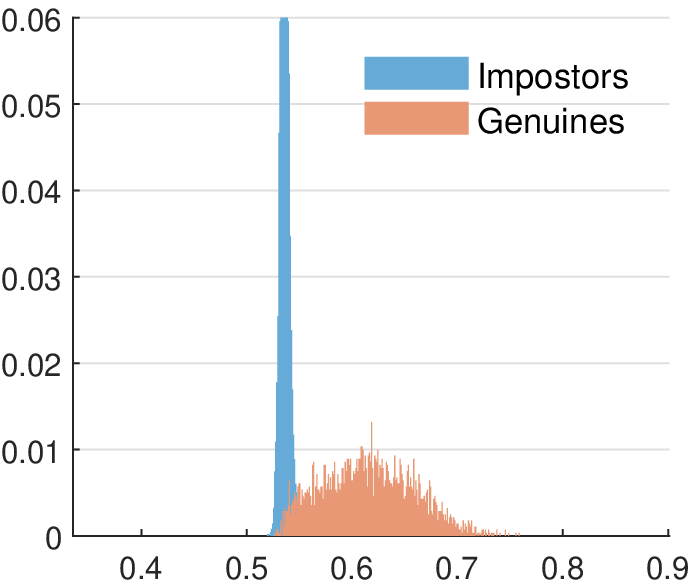}
		}
		\subfloat[eMCC$_1$ on DB1]
		{
			\includegraphics[width=0.16\linewidth]{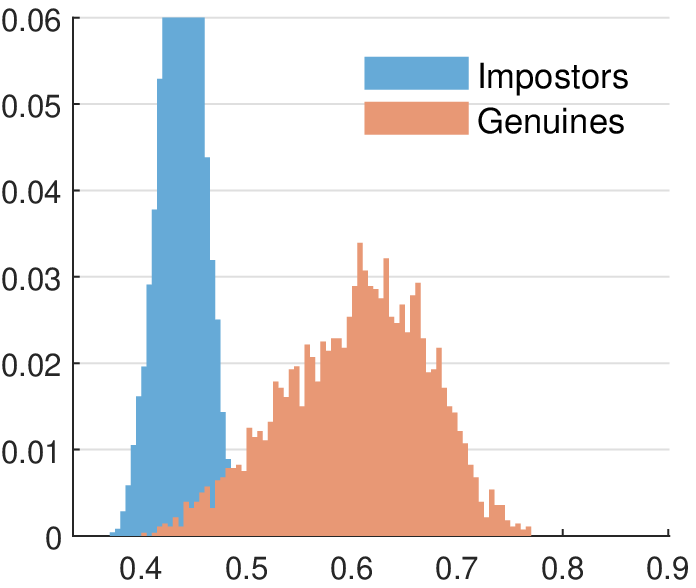}
		}
		\subfloat[eMCC$_{2/3}$ on DB1]
		{
			\includegraphics[width=0.16\linewidth]{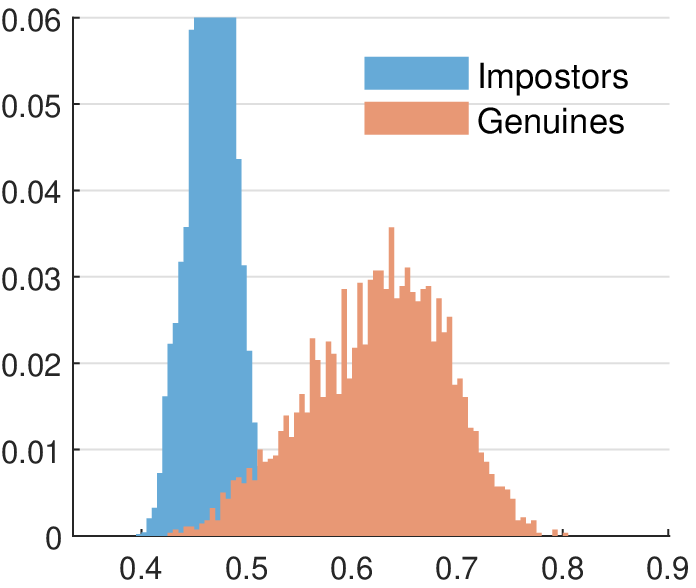}
		}
		\subfloat[eMCC$_{1/2}$ on DB1]
		{
			\includegraphics[width=0.16\linewidth]{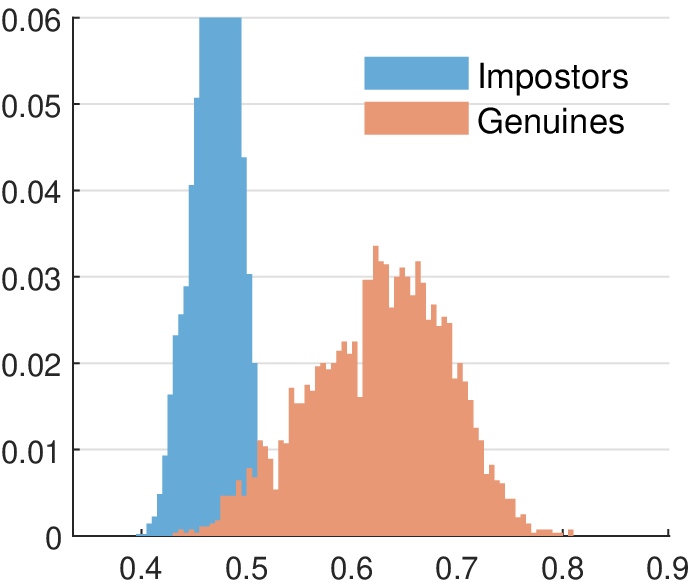}
		}
		\vfil
		
		\subfloat[MCC on DB2]
		{
			\includegraphics[width=0.16\linewidth]{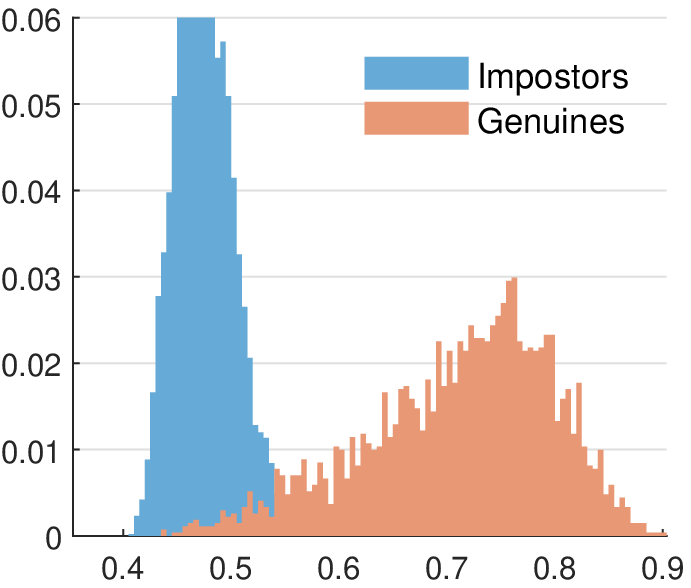}
		}
		\subfloat[bMCC on DB2]
		{
			\includegraphics[width=0.16\linewidth]{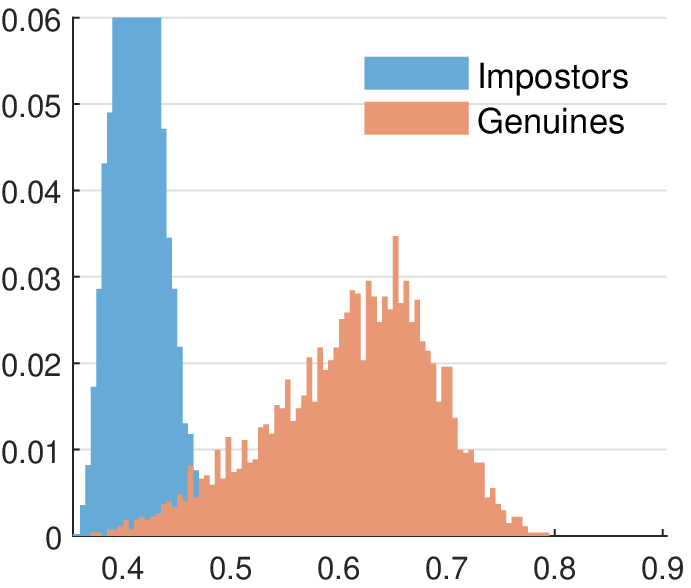}
		}
		\subfloat[cMCC on DB2]
		{
			\includegraphics[width=0.16\linewidth]{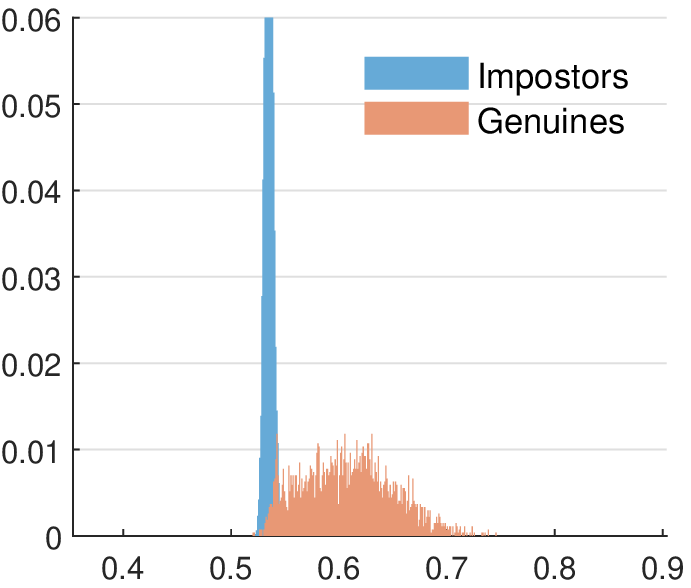}
		}
		\subfloat[eMCC$_1$ on DB2]
		{
			\includegraphics[width=0.16\linewidth]{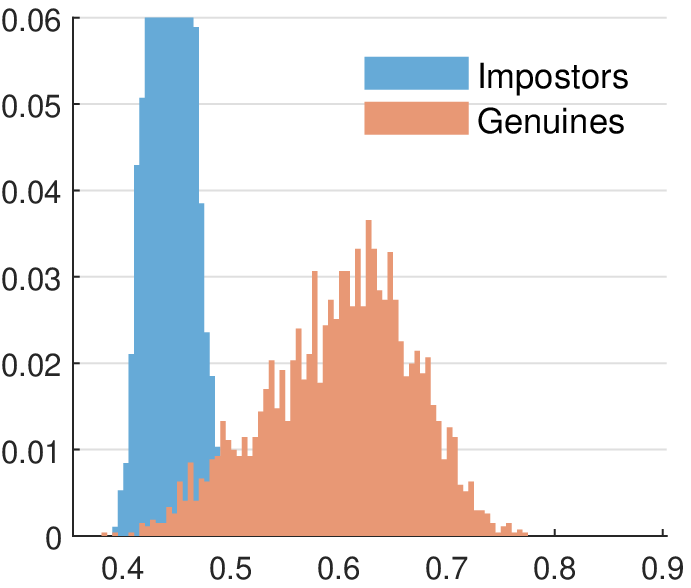}
		}
		\subfloat[eMCC$_{2/3}$ on DB2]
		{
			\includegraphics[width=0.16\linewidth]{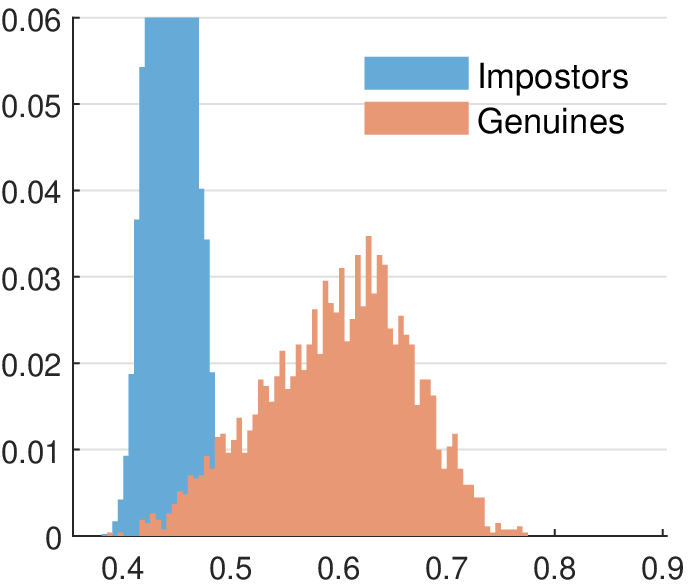}
		}
		\subfloat[eMCC$_{1/2}$ on DB2]
		{
			\includegraphics[width=0.16\linewidth]{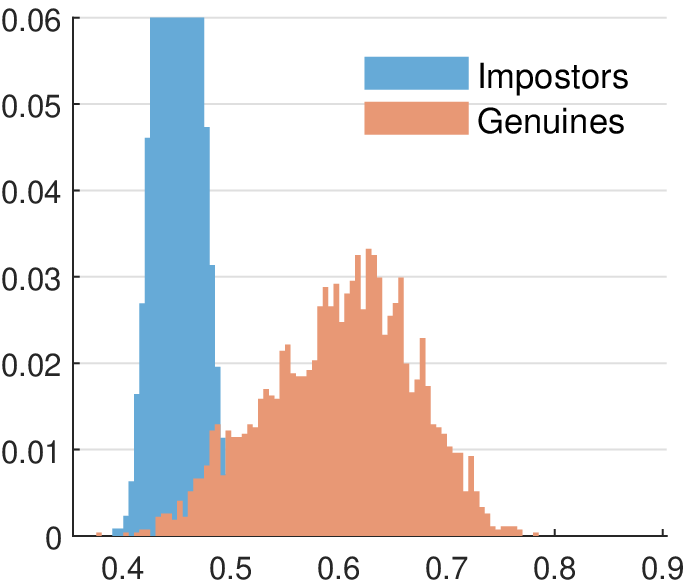}
		}
		\vfil
		
		\subfloat[MCC on DB3]
		{
			\includegraphics[width=0.16\linewidth]{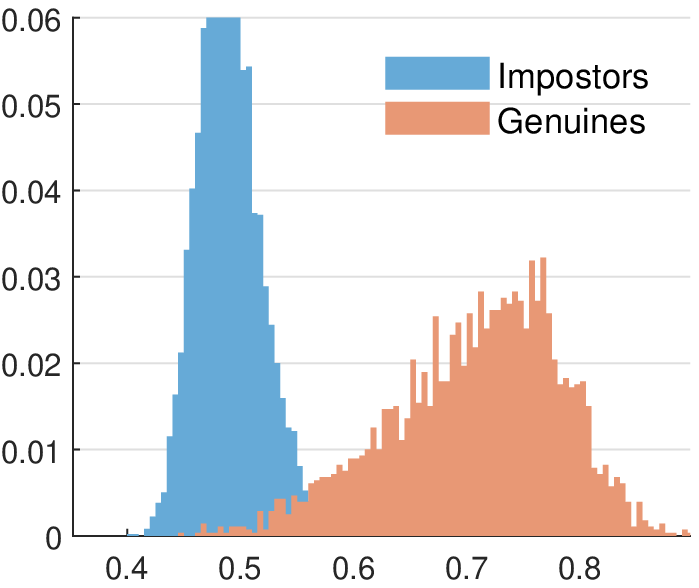}
		}
		\subfloat[bMCC on DB3]
		{
			\includegraphics[width=0.16\linewidth]{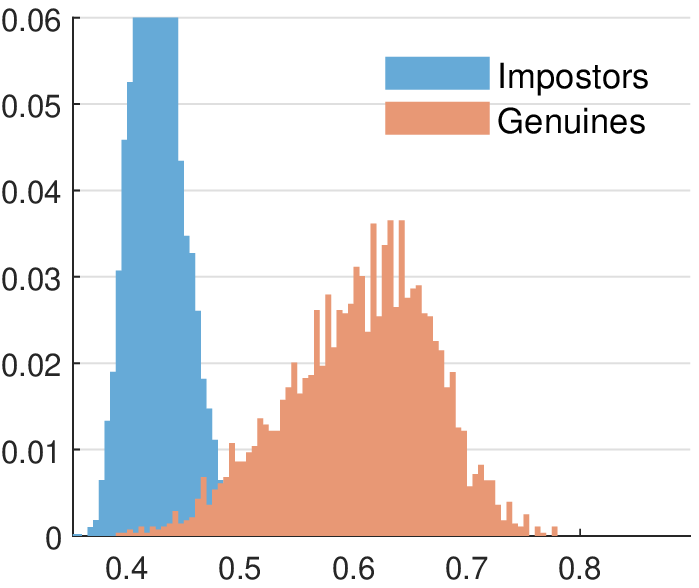}
		}
		\subfloat[cMCC on DB3]
		{
			\includegraphics[width=0.16\linewidth]{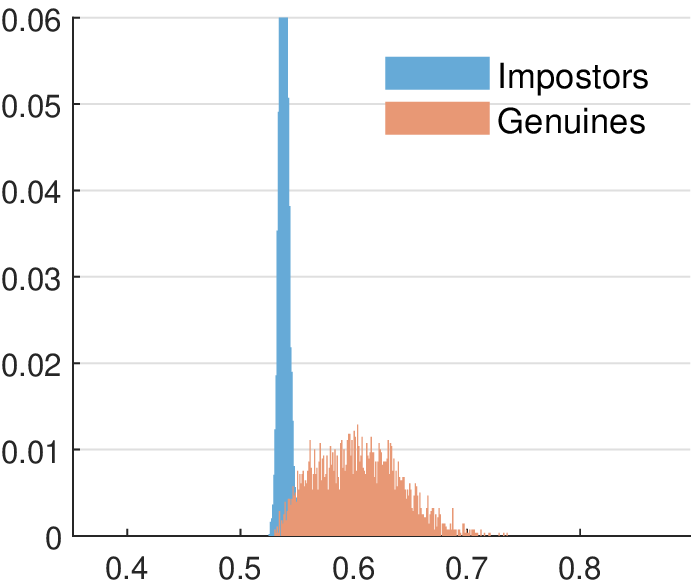}
		}
		\subfloat[eMCC$_1$ on DB3]
		{
			\includegraphics[width=0.16\linewidth]{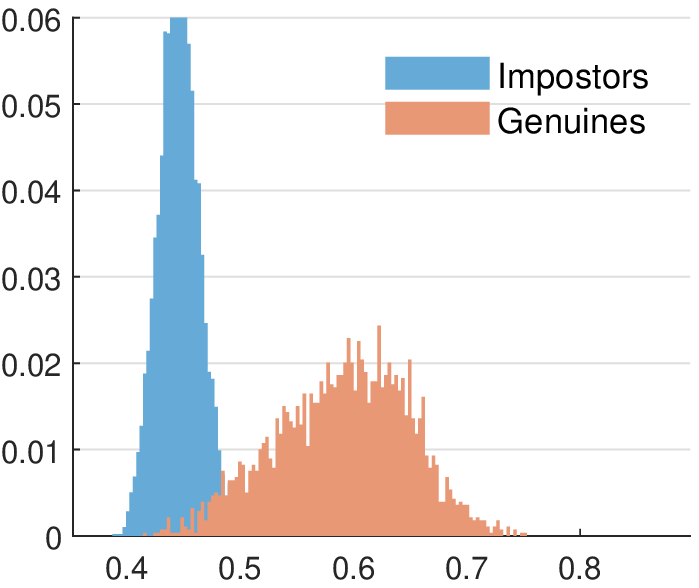}
		}
		\subfloat[eMCC$_{2/3}$ on DB3]
		{
			\includegraphics[width=0.16\linewidth]{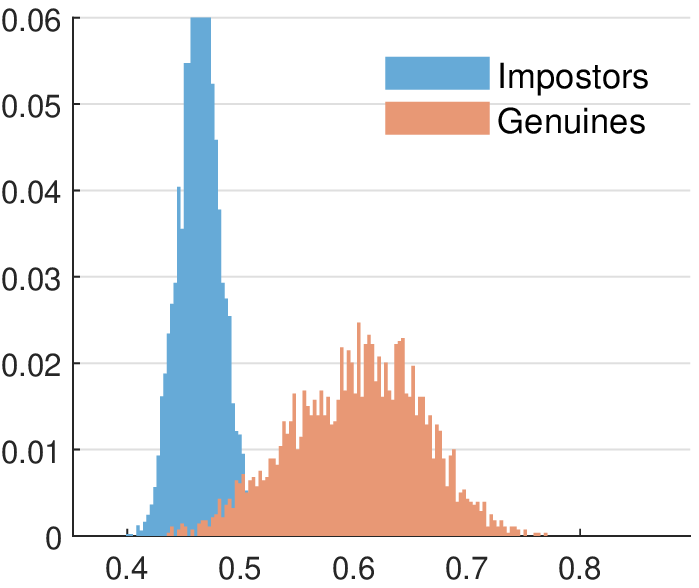}
		}
		\subfloat[eMCC$_{1/2}$ on DB3]
		{
			\includegraphics[width=0.16\linewidth]{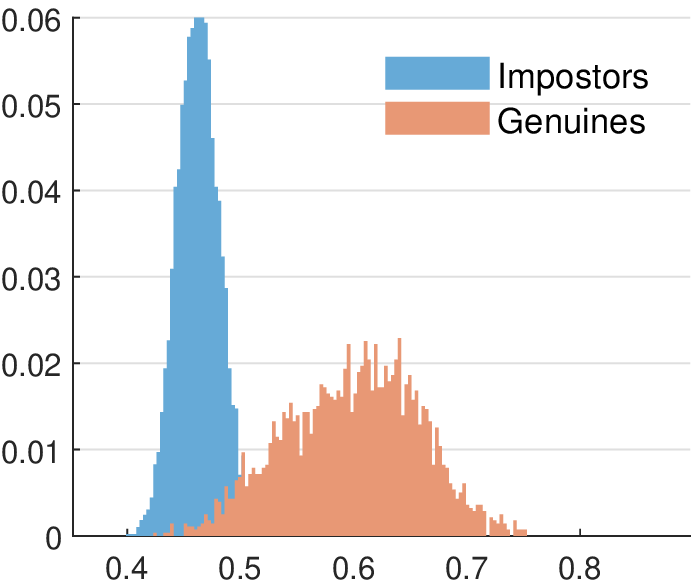}
		}
		\vfil
		
		\subfloat[MCC on DB4]
		{
			\includegraphics[width=0.16\linewidth]{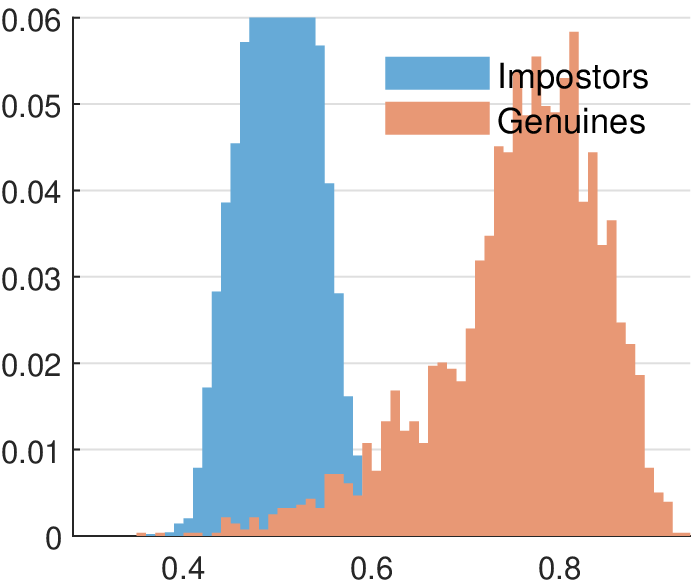}
		}
		\subfloat[bMCC on DB4]
		{
			\includegraphics[width=0.16\linewidth]{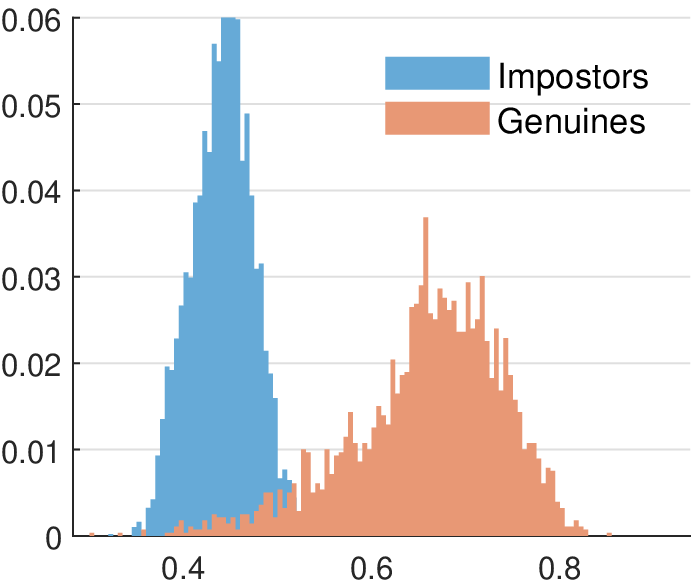}
		}
		\subfloat[cMCC on DB4]
		{
			\includegraphics[width=0.16\linewidth]{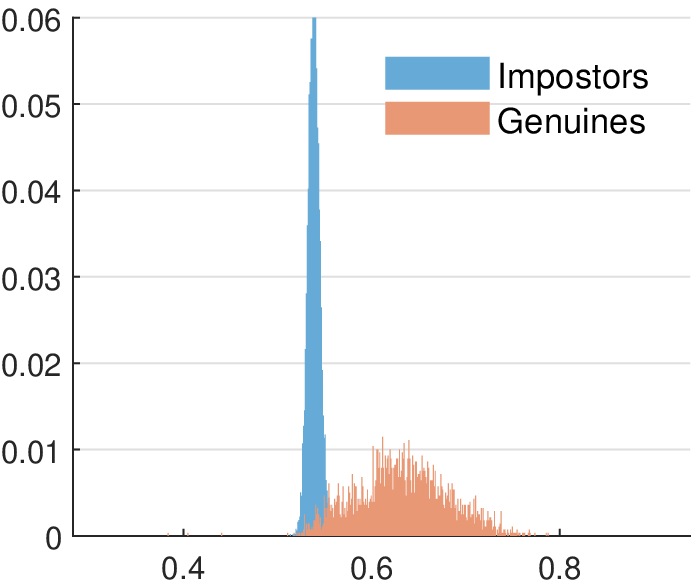}
		}
		\subfloat[eMCC$_1$ on DB4]
		{
			\includegraphics[width=0.16\linewidth]{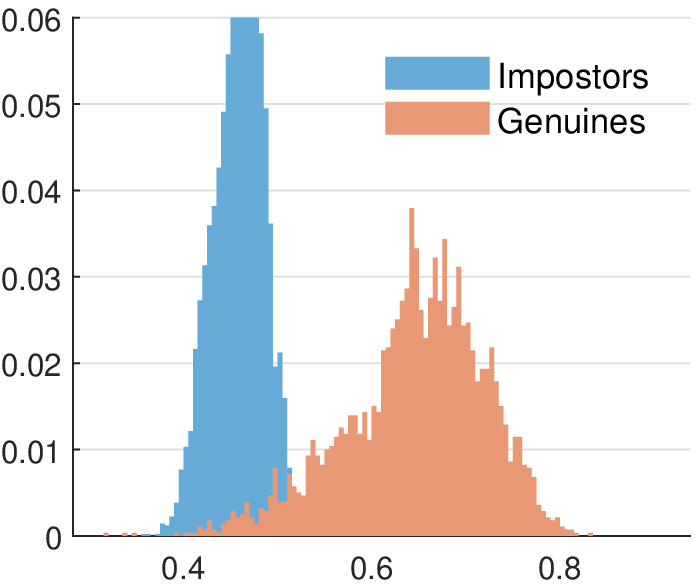}
		}
		\subfloat[eMCC$_{2/3}$ on DB4]
		{
			\includegraphics[width=0.16\linewidth]{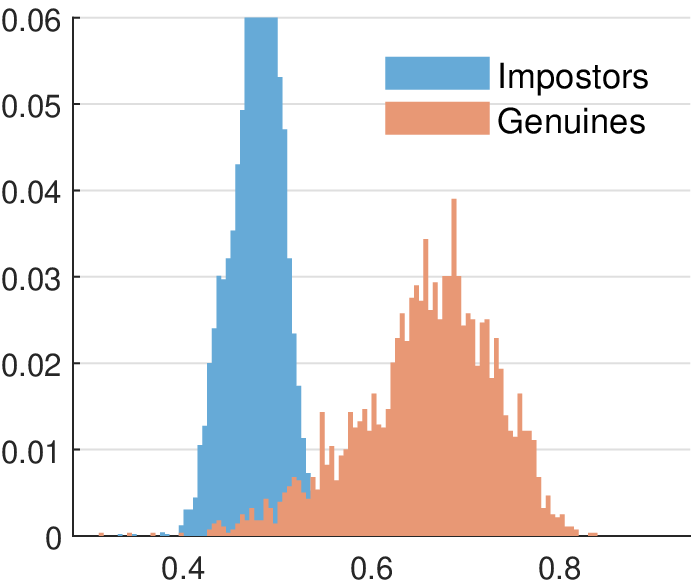}
		}
		\subfloat[eMCC$_{1/2}$ on DB4]
		{
			\includegraphics[width=0.16\linewidth]{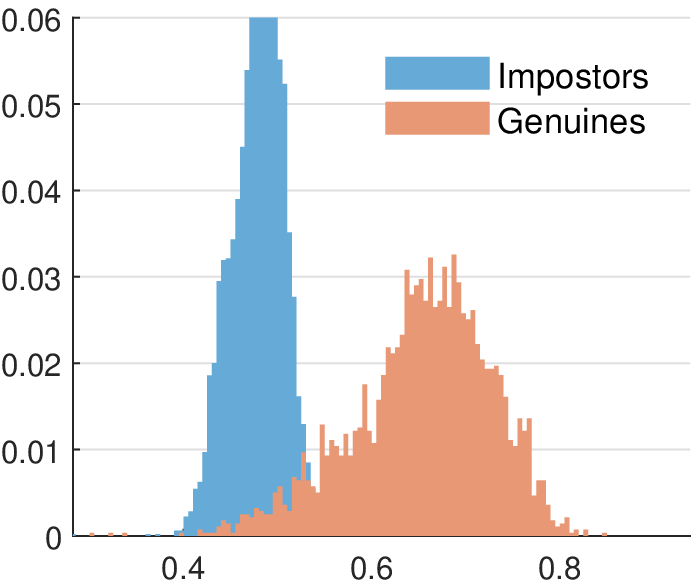}
		}
		\caption{Comparison of score distributions by MCC, bMCC, cMCC, eMCC$_1$, eMCC$_{2/3}$ and eMCC$_{1/2}$ evaluated on datasets FVC2004 DB1-DB4. The x-axis is the matching score, and the y-axis is the proportion of scores falling into each score bin.}
		\label{fig:score_distribution_FVC2004}
	\end{figure*}

	Fig.~\ref{fig:score_distribution_FVC2002} and Fig.~\ref{fig:score_distribution_FVC2004} 
	show the comparison of matching score distributions between MCC, bMCC, cMCC, eMCC$_1$, eMCC$_{2/3}$ and eMCC$_{1/2}$ evaluated on datasets FVC2002 DB1-DB4 and FVC2004 DB1-DB4, respectively. 
	It is clear that the imposter scores of eMCC$_1$, eMCC$_{2/3}$ and eMCC$_{1/2}$ mainly concentrate in the range [0.4, 0.5], while the genuine scores of the three templates mainly concentrated in the range [0.4, 0.8].
	The imposter and genuine scores of MCC are mainly in the range [0.4, 0.6] and [0.4, 0.9], respectively.
	Similarly, the imposter and genuine scores of bMCC are mainly in the range [0.4, 0.5] and [0.5, 0.8], respectively, while cMCC's imposter scores are mainly clustered in the range [5.3, 5.5] and its genuine scores are mainly in the range [0.55, 75].
	In summary, the proposed eMCC$_1$, eMCC$_{2/3}$ and eMCC$_{1/2}$ have similar imposter and genuine score distributions in comparison with MCC and bMCC, but different distributions compared to cMCC.

	\subsubsection{DET Curves}
\begin{figure}[htbp]
		\centering
		\subfloat[DET curves evaluated on DB1]
		{
			\includegraphics[width=0.45\linewidth]{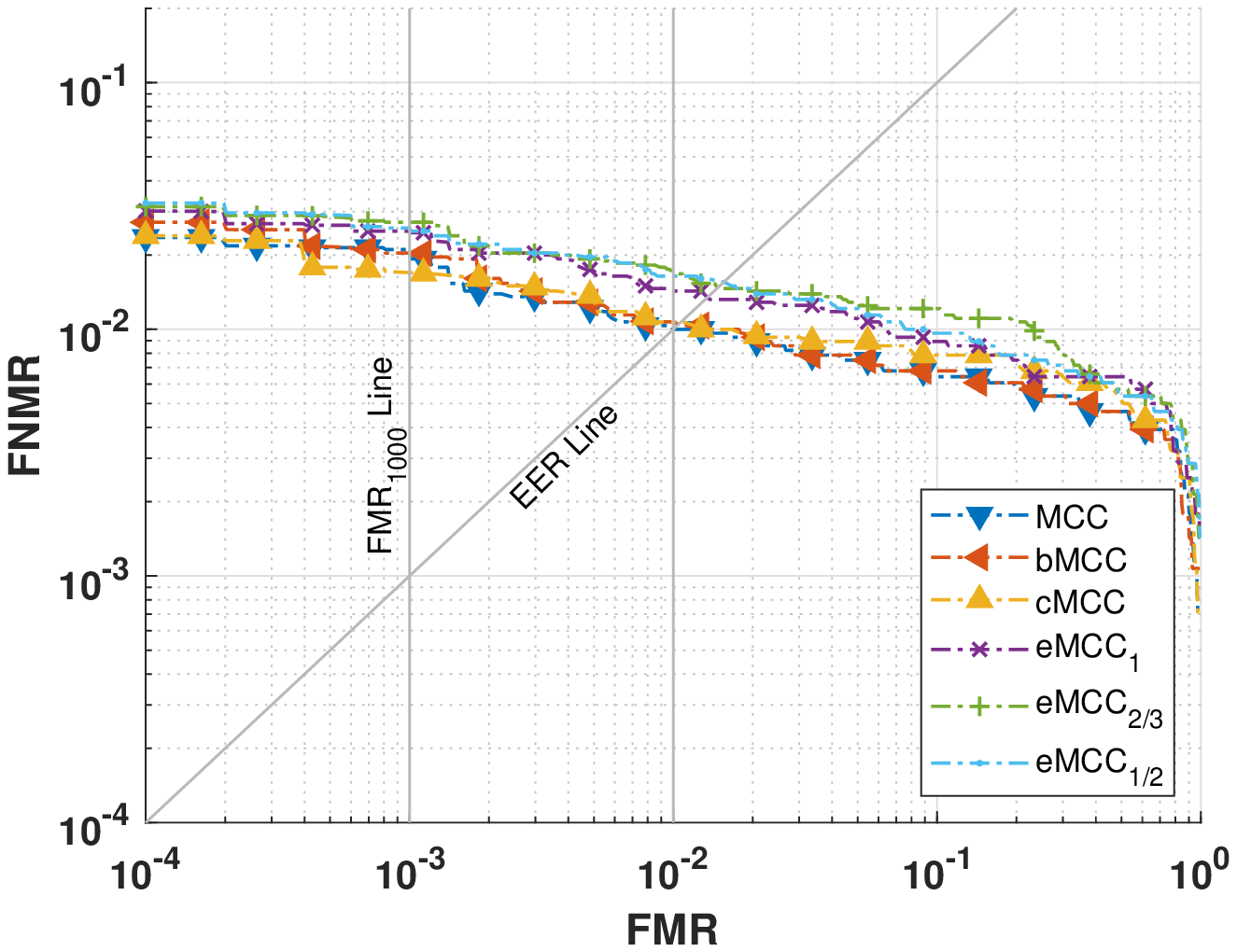}
		}
		\subfloat[DET curves evaluated on DB2]
		{
			\includegraphics[width=0.45\linewidth]{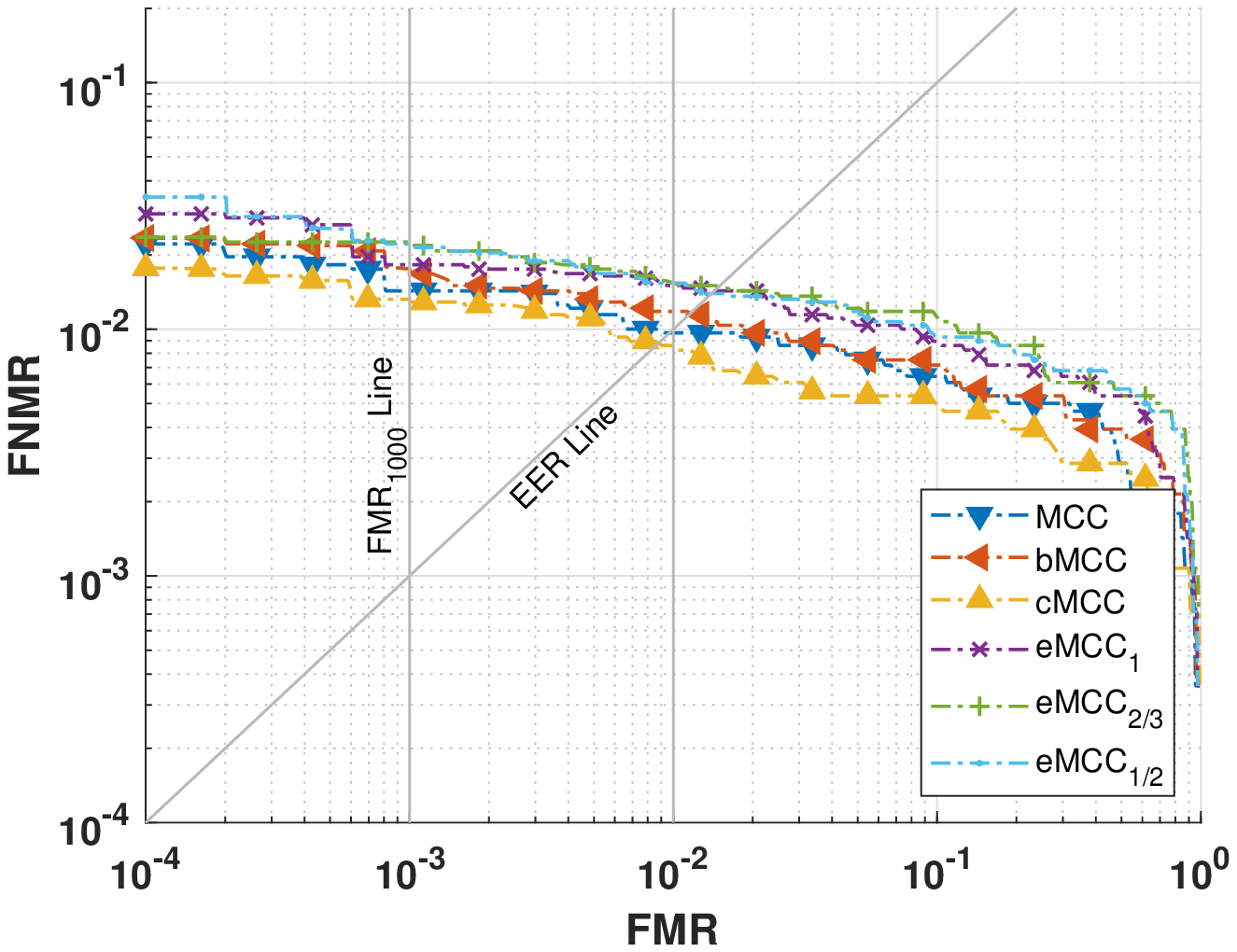}
		}
		\vfil
		\subfloat[DET curves evaluated on DB3]
		{
			\includegraphics[width=0.45\linewidth]{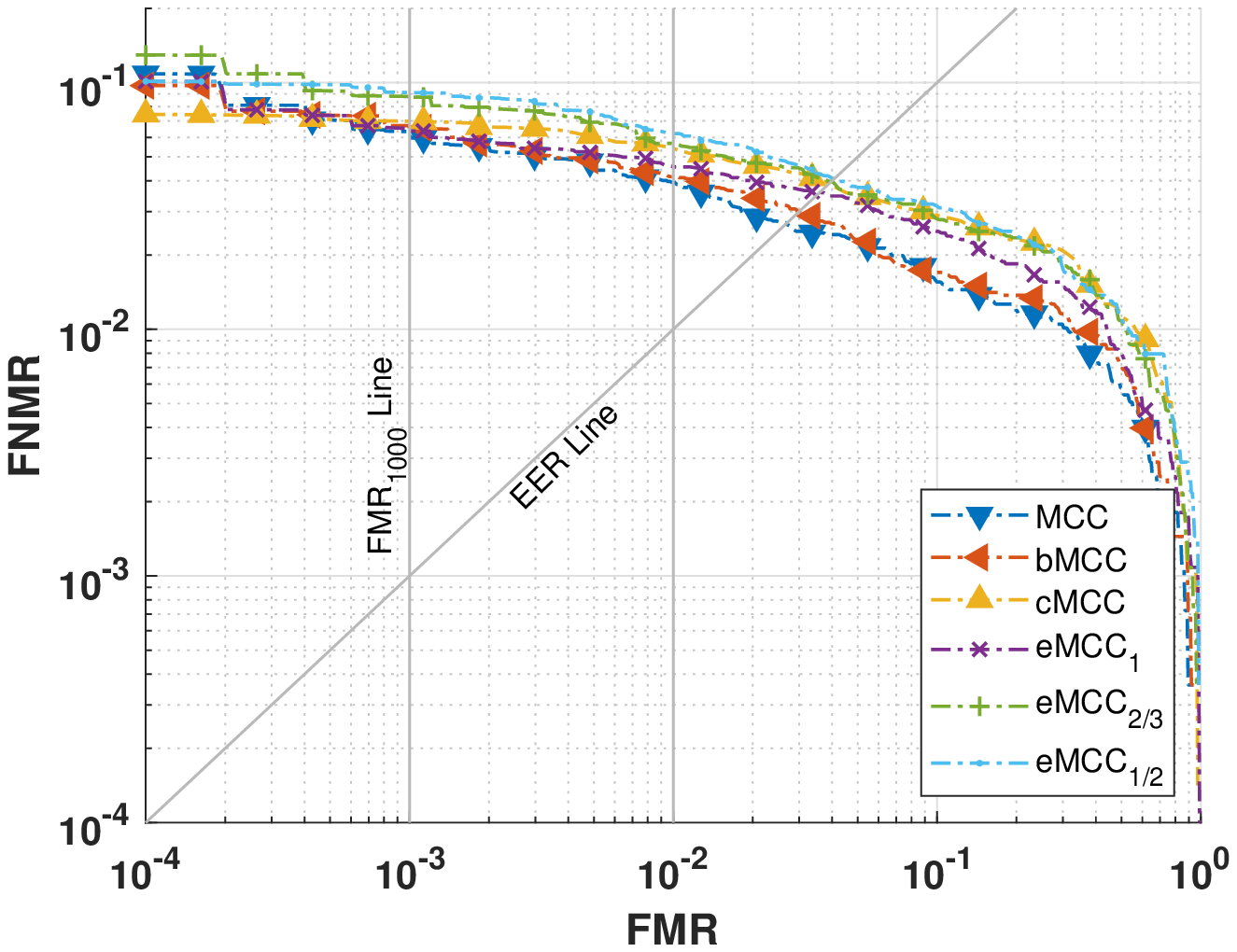}
		}
		\subfloat[DET curves evaluated on DB4]
		{
			\includegraphics[width=0.45\linewidth]{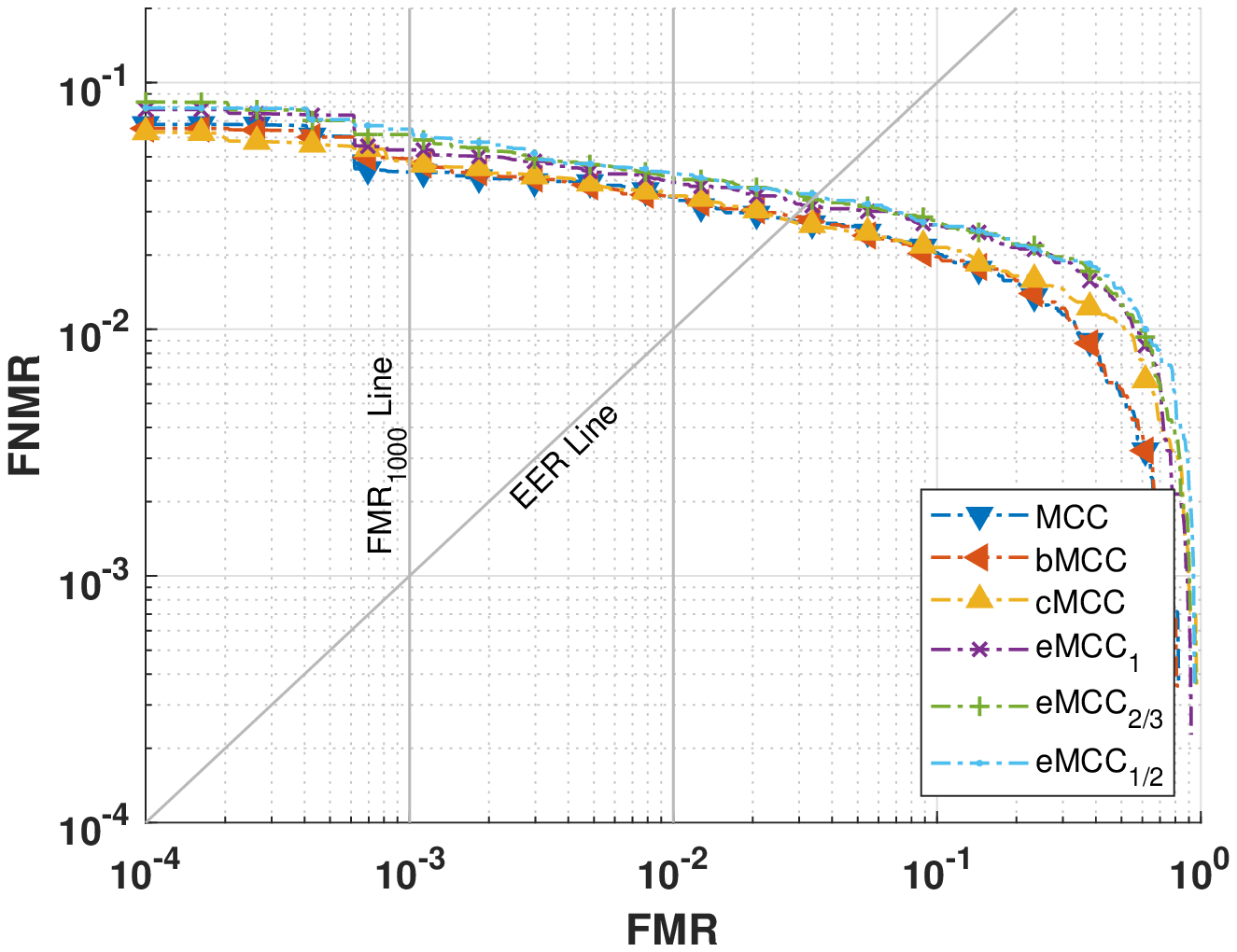}
		}
		\caption{Comparison of DET curves obtained by MCC, bMCC, cMCC, eMCC$_1$, eMCC$_{2/3}$ and eMCC$_{1/2}$ on datasets FVC2002 DB1-DB4.}
		\label{fig:FVC2002_DET_all_6}
	\end{figure}

\begin{figure}[htbp]
		\centering
		\subfloat[DET curves evaluated on DB1]
		{
			\includegraphics[width=0.45\linewidth]{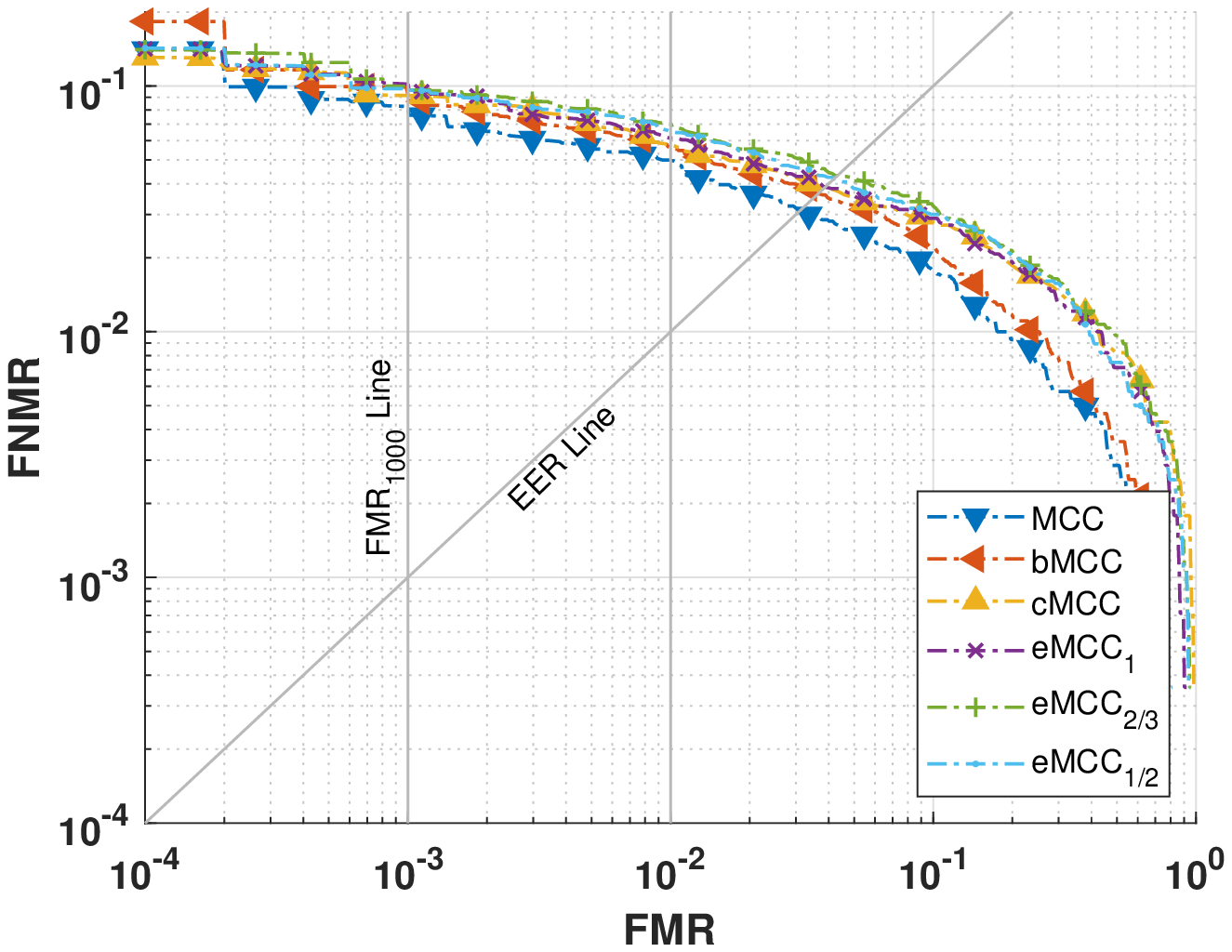}
		}
		\subfloat[DET curves evaluated on DB2]
		{
			\includegraphics[width=0.45\linewidth]{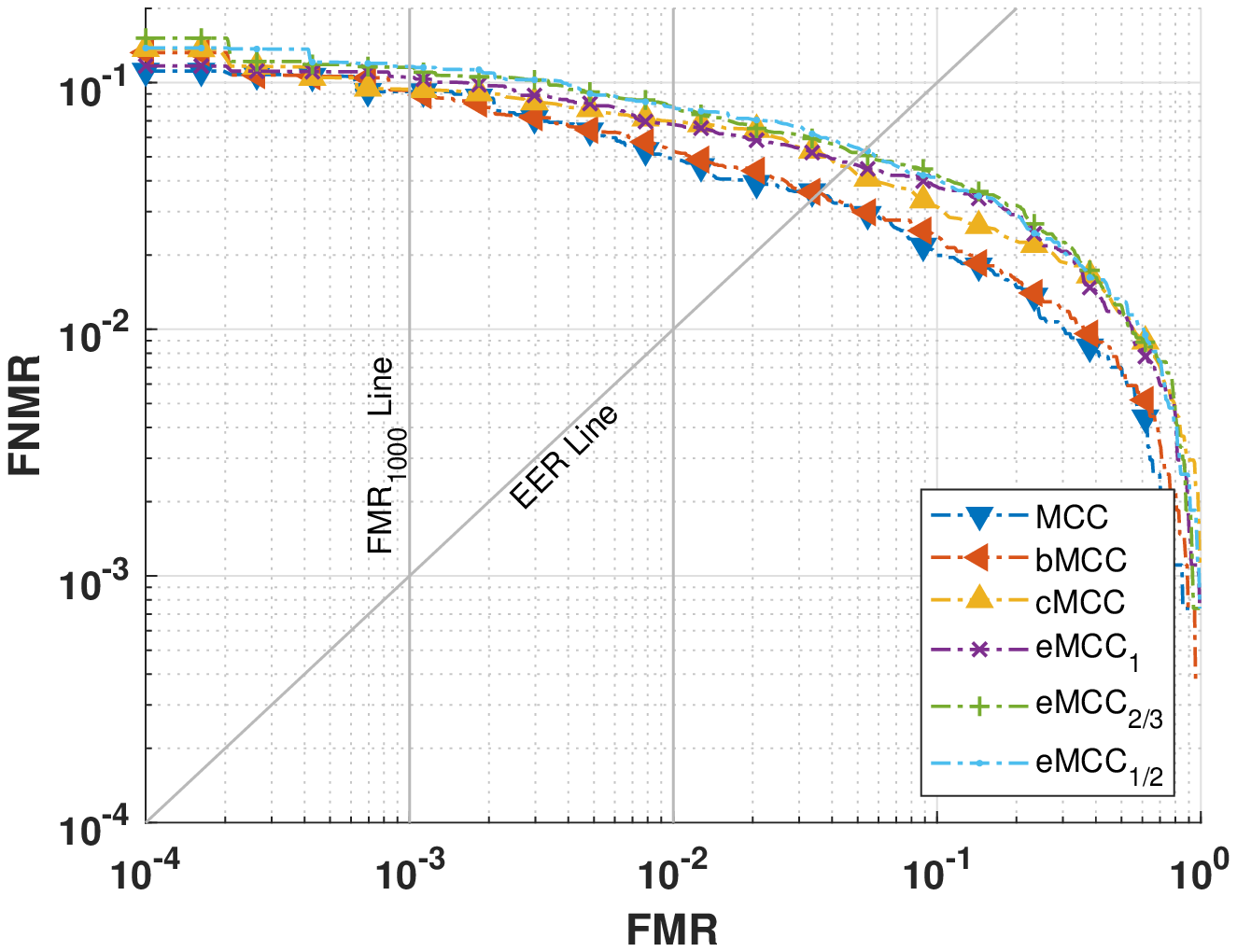}
		}
		\vfil
		\subfloat[DET curves evaluated on DB3]
		{
			\includegraphics[width=0.45\linewidth]{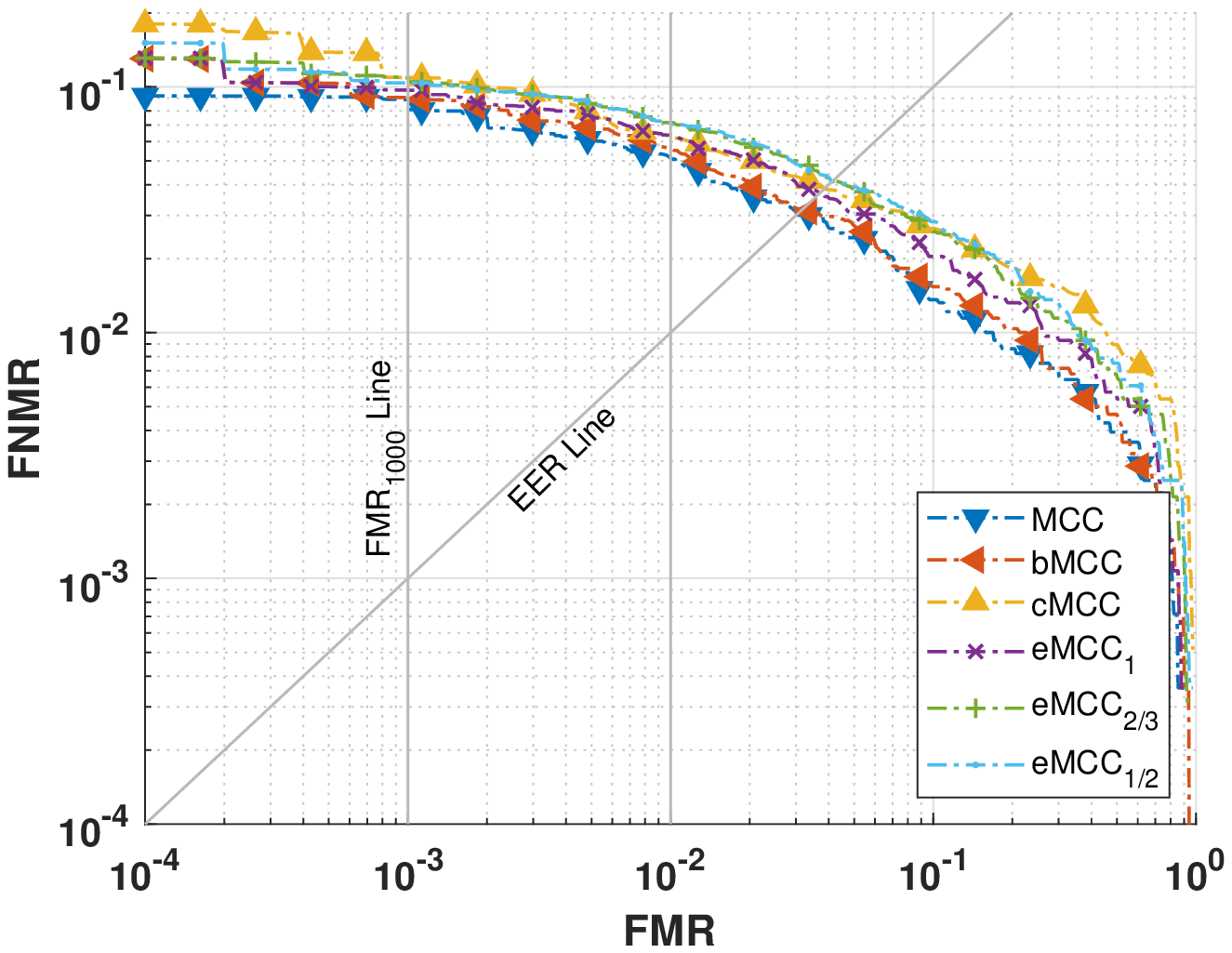}
		}
		\subfloat[DET curves evaluated on DB4]
		{
			\includegraphics[width=0.45\linewidth]{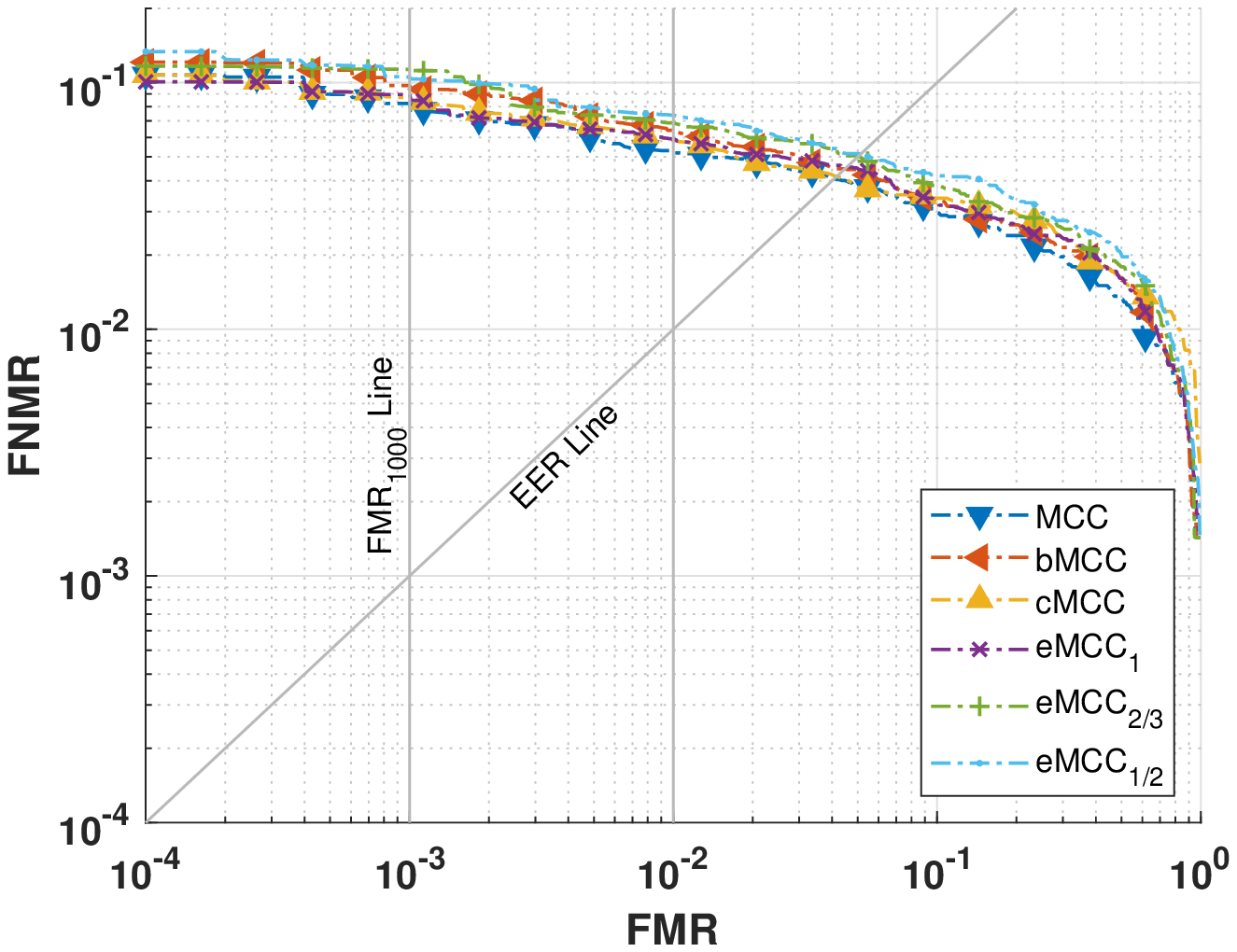}
		}
		\caption{Comparison of DET curves obtained by MCC, bMCC, cMCC, eMCC$_1$, eMCC$_{2/3}$ and eMCC$_{1/2}$ on datasets FVC2004 DB1-DB4.}
		\label{fig:FVC2004_DET_all_6}
	\end{figure}

	Fig.~\ref{fig:FVC2002_DET_all_6} and Fig.~\ref{fig:FVC2004_DET_all_6}
	show the comparison of DET curves obtained by MCC, bMCC, cMCC, eMCC$_1$, eMCC$_{2/3}$, and eMCC$_{1/2}$ on datasets FVC2002 DB1-DB4 and FVC2004 DB1-DB4, respectively. 
	On FVC2002 DB4 and FVC2004 DB4, the proposed eMCC$_1$, eMCC$_{2/3}$ and eMCC$_{1/2}$ show similar DET curves compared to MCC, bMCC and cMCC. 
	On FVC2002 DB1, FVC2002 DB3, FVC2004 DB1 and FVC2004 DB2, for FMR $<10^{-3}$, the proposed eMCC$_1$, eMCC$_{2/3}$, and eMCC$_{1/2}$ achieve close FNMR values compared to MCC, bMCC and cMCC.
	In summary, eMCC$_1$, eMCC$_{2/3}$ and eMCC$_{1/2}$ only have minor deterioration in authentication accuracy, but they make considerable savings in the storage space. This demonstrates the validity of the proposed template.

	\subsubsection{EER and FMR$_{1000}$}
	Table~\ref{tab:EER_FMR1000_emcc_all_6} compares the EER and FMR$_{1000}$ obtained by MCC, bMCC, cMCC, eMCC$_1$, eMCC$_{2/3}$ and eMCC$_{1/2}$ on datasets FVC2002 DB1-DB4 and FVC2004 DB1-DB4. 
	With a sizable reduction on the template length, the proposed eMCC$_1$, eMCC$_{2/3}$ and eMCC$_{1/2}$ achieve relatively close EER on most of these eight datasets, and eMCC$_1$ even outperforms the IoT-oriented template cMCC on FVC2002 DB3 and FVC2004 DB3. 
	Regarding FMR$_{1000}$, the proposed eMCC$_1$, eMCC$_{2/3}$, and eMCC$_{1/2}$ also achieve similar accuracy on most of the eight datasets.
	On FVC2002 DB3,	eMCC$_1$ even performs better than bMCC and cMCC.
	On FVC2004 DB4, eMCC$_1$ has a better FMR$_{1000}$ than bMCC.
	On FVC2004 DB1, eMCC$_{1/2}$ achieves a better FMR$_{1000}$ than bMCC.
	In summary, this demonstrates that with a significantly reduced template length, the proposed lightweight cancelable template shows no degradation in authentication accuracy.
	
	\begin{table*}[htbp]
		\setlength\tabcolsep{2pt}
		\renewcommand{\arraystretch}{1.5}
		\caption{Comparison of verification accuracy in terms of the EER and FMR$_{1000}$ obtained by MCC, bMCC, cMCC, eMCC$_1$, eMCC$_{2/3}$ and eMCC$_{1/2}$ on FVC2002 DB1-DB4 and FVC2004 DB1-DB4.}
		\label{tab:EER_FMR1000_emcc_all_6}
		\centering
		\resizebox{\linewidth}{!}
		{
			\begin{tabular}{c |c||c|c|c|c|c|c||c|c|c|c|c|c} 
				\hline		\hline
				\multicolumn{2}{c||}{\multirow{2}{*}{Dateset}}	&\multicolumn{6}{c||}{\textbf{EER (\%)}}  	          &  \multicolumn{6}{c}{\textbf{FMR$_{1000}$ (\%)}}		\\ \cline{3-14} 
				\multicolumn{2}{c||}{} & MCC & bMCC & cMCC & eMCC$_1$ & eMCC$_{2/3}$ & eMCC$_{1/2}$  &MCC & bMCC & cMCC & eMCC$_1$ & eMCC$_{2/3}$ & eMCC$_{1/2}$	\\ 	\hline \hline
				
				\multirow{4}{*}{\textbf{FVC2002}} 
				& DB1& 1.00 &1.04 &1.07 &1.35 &1.46 &1.55 &2.11 &2.04 &1.70 &2.50 &2.71 &2.57		\\		\cline{2-14}
				& DB2& 0.97	&1.15 &0.86 &1.43 &1.47 &1.40 &1.43 &1.77 &1.32 &1.83 &2.26 &2.22       \\      \cline{2-14}
				& DB3& 2.68	&3.00 &3.90 &3.61 &3.97 &4.04 &6.34 &6.69 &6.98 &6.53 &8.79 &9.12		\\ 		\cline{2-14}
				& DB4& 2.76	&2.86 &2.74 &3.18 &3.33 &3.50 &4.34 &4.92 &4.81 &5.33 &6.15 &6.48		\\ 		\hline		\hline
				
				\multirow{4}{*}{\textbf{FVC2004}} 
				& DB1&3.11  &3.64 &3.85 &3.89 &4.32 &4.21 &8.27 &9.92 &9.15 &10.21 &9.97   &9.75 \\ 	\cline{2-14}
				& DB2&3.51	&3.55 &4.51 &4.72 &5.20 &5.28 &9.20 &9.29 &9.39 &10.52 &11.53 &11.61		\\ 		\cline{2-14}
				& DB3&3.05	&3.15 &3.83 &3.71 &4.15 &4.11 &8.89 &9.06 &9.98 &9.71   &10.92 &10.38		\\ 		\cline{2-14}
				& DB4&4.08	&4.48 &4.15 &4.54 &5.01 &5.12 &8.23 &9.74 &8.69 &8.95   &11.32 &10.55		\\ 		\hline		\hline
			\end{tabular}
		}
	\end{table*}
	
	\subsection{Evaluation in an IoT Prototype System} \label{sec:simulation}
	In this section, we evaluate the proposed template on an IoT prototype system, implemented using the popular open-source software Open Virtual Platforms\texttrademark~\footnote{https://www.ovpworld.org/} (OVP\texttrademark, version 20210408.0) and the RISC-V instruction set architecture.\footnote{https://www.ovpworld.org/dlp/}
	
	\subsubsection{Storage of eMCC Template and Runtime}
	Table \ref{tab:templateLength} shows the comparison of the proposed template with bMCC and cMCC in the case of $N_S=16$, $N_D=5$ and the number of minutiae $n=50$, with $L_c = N_S \times N_S \times N_D$ and $L_b = N_S \times N_S.$
		\begin{table}[htbp]
		\setlength\tabcolsep{6pt}
		\renewcommand{\arraystretch}{1.5}
		\caption{Comparison of the proposed template with bMCC and cMCC in the case of $N_S=16$, $N_D=5$ and the number of minutiae $n=50$, with $L_c = N_S \times N_S \times N_D$ and $L_b = N_S \times N_S$}
		\label{tab:templateLength}
		\centering
		\begin{tabular}{l|c}
			\hline\hline
			&Template length (bits) \\ \hline
			bMCC~\cite{RN159}                     & $n(L_c + L_b) = 76,800$ 					\\ \hline
			cMCC~\cite{RN2761}                   & $n(\frac{1}{2}L_c+\frac{1}{2}L_b) = 38,400$    \\ \hline
			eMCC$_1$ with $p = 1 $            & $n(\frac{1}{4}L_c+\frac{1}{8}N_DL_b) = 24,000$ \\  \hline
			eMCC$_{2/3}$ with $p = 2/3 $ & $n(\frac{1}{6}L_c +\frac{1}{12}N_DL_b) = 16,000 $ \\  \hline		
			eMCC$_{1/2}$ with $p = 1/2 $ & $n(\frac{1}{8}L_c +\frac{1}{16}N_DL_b) = 12,000$ \\ 
			\hline \hline
		\end{tabular}
	\end{table}
	As shown in Table \ref{tab:templateLength}, eMCC$_1$ template requires 24K bits, saving 52.8K bits over bMCC and 14.4K bits over cMCC, while eMCC$_{1/2}$ template only requires 12K bits, saving 64.8K bits over bMCC and 26.4K bits over cMCC. 
	This manifests that the proposed length-flexible template can remarkably reduce template storage space, which is highly beneficial to resource-constrained IoT devices. 
	The prototype system requires about 12K bits, 16K bits and 24K bits for the template storage of eMCC$_{1/2}$, eMCC$_{2/3}$ and eMCC$_{1}$, respectively. 
	Therefore, they are applicable to commercial smart cards (e.g., C5-M.O.S.T. Card Contact Microprocessor Smart Cards CLXSU064KC5 32K Bits~\footnote{https://www.cardlogix.com/product/contact-smart-card-most-c5-microproccessor-cards/} and Atmel AT24C16C Memory Smart Card 16K Bits~\footnote{https://www.cardlogix.com/product/atmel-at24c16c-memory-smart-chip-card-16k/}).

	The average time taken for fingerprint enrollment and verification is measured by evaluating the prototype system on FVC2002 DB1 with eight hundred fingerprints of size $388 \times 374$. 
	The fingerprint enrollment process aims to extract minutiae in the format of ISO/IEC 19794-2 and to generate an eMCC template to be stored, so the original fingerprint image or feature is not stored to prevent privacy leakage.
	The fingerprint verification process sharing the common minutiae extraction and template generation aims to match a query template against the enrolled template.
	The open-source algorithm Mindtct~\cite{RN675} is utilized to implement the minutiae extraction in this simulation experiment.
	The average runtime of the minutiae extraction is around 2,300 milliseconds, which obviously can be  optimized further. Since minutiae extraction is a relatively independent process, it is beyond the scope of this work. 
	The average runtime of generating eMCC$_1$, eMCC$_{2/3}$ and eMCC$_{1/2}$ is approximately 255, 240 and 225 milliseconds, respectively. 
	Compared with minutiae extraction and template generation, template matching is time-efficient, with an average runtime of 70 milliseconds for eMCC$_1$, 55 milliseconds for eMCC$_{2/3}$, and 45 milliseconds for eMCC$_{1/2}$. 
	In summary, slightly depending on the parameter~$p$, the average runtime of the enrollment process varies approximately from 2, 525 milliseconds to 2,555 milliseconds, while the average runtime of the verification process varies approximately from 2570 milliseconds to 2,625 milliseconds. 
	Note that the enrollment and verification procedures require much more time on the time-consuming minutia extraction. Therefore, there is much room for reducing the runtime by either optimizing the minutiae extraction process or integrating a time-saving minutiae extraction method.

	{\color{color}Table \ref{tab:efficiencyComparison} shows the comparison of efficiency of the proposed eMCC$_{1/2}$ with state-of-the-art cryptographic fingerprint authentication methods, namely M1-2021 \cite{RN3220}, M2-2021 \cite{RN2771} and M3-2020 \cite{RN2559}. As shown in Table \ref{tab:efficiencyComparison}, compared with M1-2021 \cite{RN3220} and M2-2021 \cite{RN2771}, even though they are based on cloud computing, the proposed eMCC$_{1/2}$ achieves better efficiency in terms of authentication time, storage space, and communication cost. Compared with the cloud-based method M3-2020 \cite{RN2559}, the proposed eMCC$_{1/2}$ performs better in authentication time. In addition, the proposed eMCC$_{1/2}$ performs much faster than M3-2020 \cite{RN2559}. The proposed eMCC$_{1/2}$ costs about 2,525 ms for a 12,000-bit template, while M3-2020 \cite{RN2559} needs about 123,537 ms for encrypting a 300-bit template. 
	Besides, the proposed eMCC$_{1/2}$ achieves a better EER of 1.4\% than M3-2020 \cite{RN2559} with an EER of 8.25\% on FVC2002 DB2.  
	}
	\begin{table}[htbp]
		\setlength\tabcolsep{6pt}
		\renewcommand{\arraystretch}{1.5}
		\caption{\color{color}Efficiency comparison between the proposed and state-of-the-art cryptographic fingerprint authentication methods}
		\label{tab:efficiencyComparison}
		\centering
		\resizebox{\linewidth}{!}
		{
			\color{color}
			\begin{tabular}{l|c|c|c|c|c}
				\hline\hline
				Methods & \tabincell{c}{Cryptographic\\techniques} &\tabincell{c}{Cloud\\computing} & \tabincell{c}{Authentication\\time (ms)} &\tabincell{c}{Storage\\space (Bytes)} & \tabincell{c}{Communication\\cost (Bytes)}\\ \hline
				M1-2021 \cite{RN3220} & \tabincell{c}{Homomorphic\\encryption}& Yes &$\sim 7,168$ & 20K & 20K \\  \hline
				M2-2021 \cite{RN2771} & \tabincell{c}{Homomorphic\\encryption}& Yes & $\sim 7,413$ &23.125K & 20.0625K\\  \hline
				M3-2020 \cite{RN2559} & \tabincell{c}{Homomorphic\\encryption}& Yes &  $\sim 3,028$ & unavailable & unavailable\\  \hline		
				eMCC$_{1/2}$ & nil & No & $\sim 2,570$ & $\sim 1.5$K & nil \\  \hline	
				\hline
			\end{tabular}
		}
	\end{table}

	\subsubsection{Comparison of DET Curves}
\begin{figure}[htbp]
		\centering
		\subfloat[DET curves evaluated on DB1]
		{
			\includegraphics[width=0.48\linewidth]{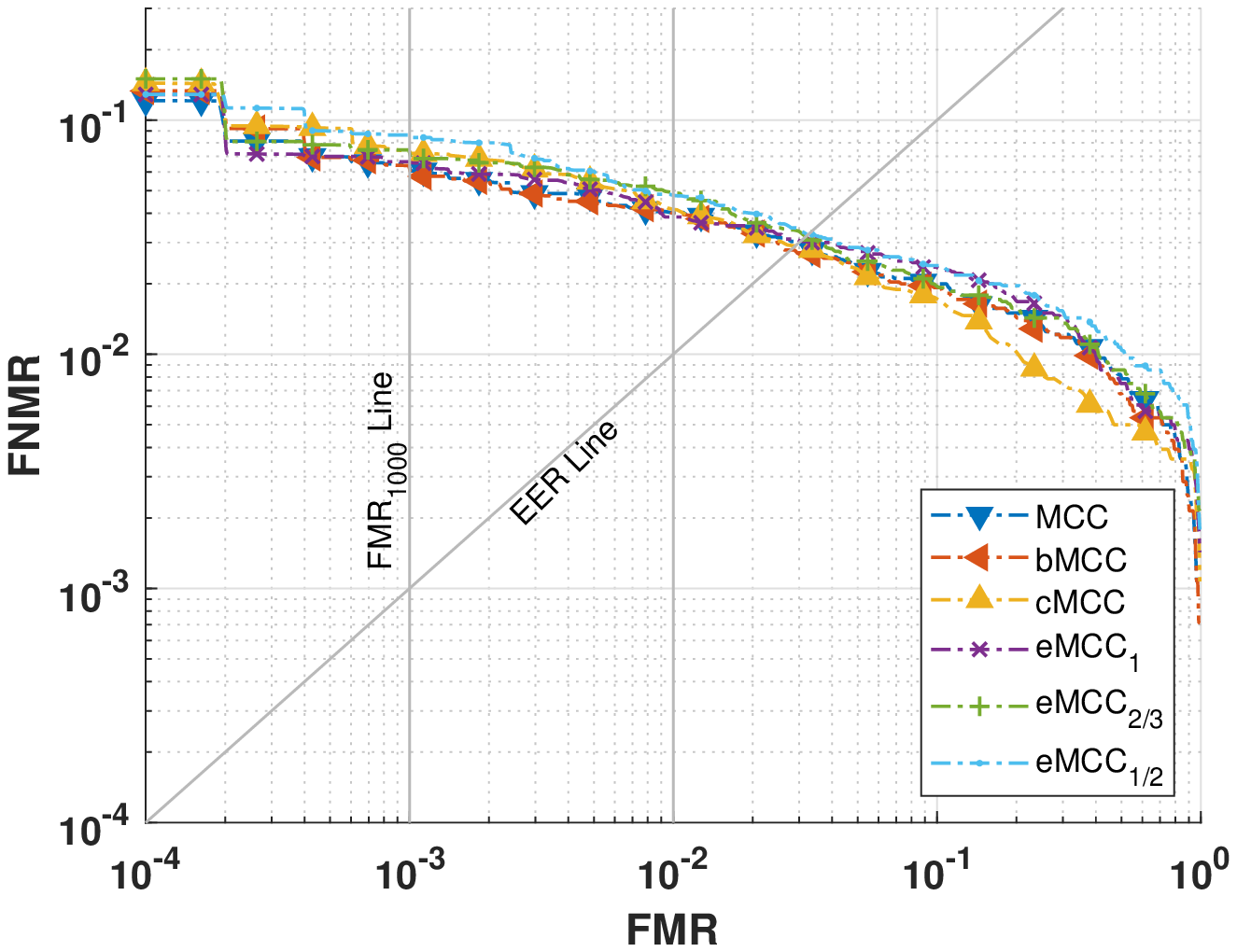}
		}
		\subfloat[DET curves evaluated on DB2]
		{
			\includegraphics[width=0.48\linewidth]{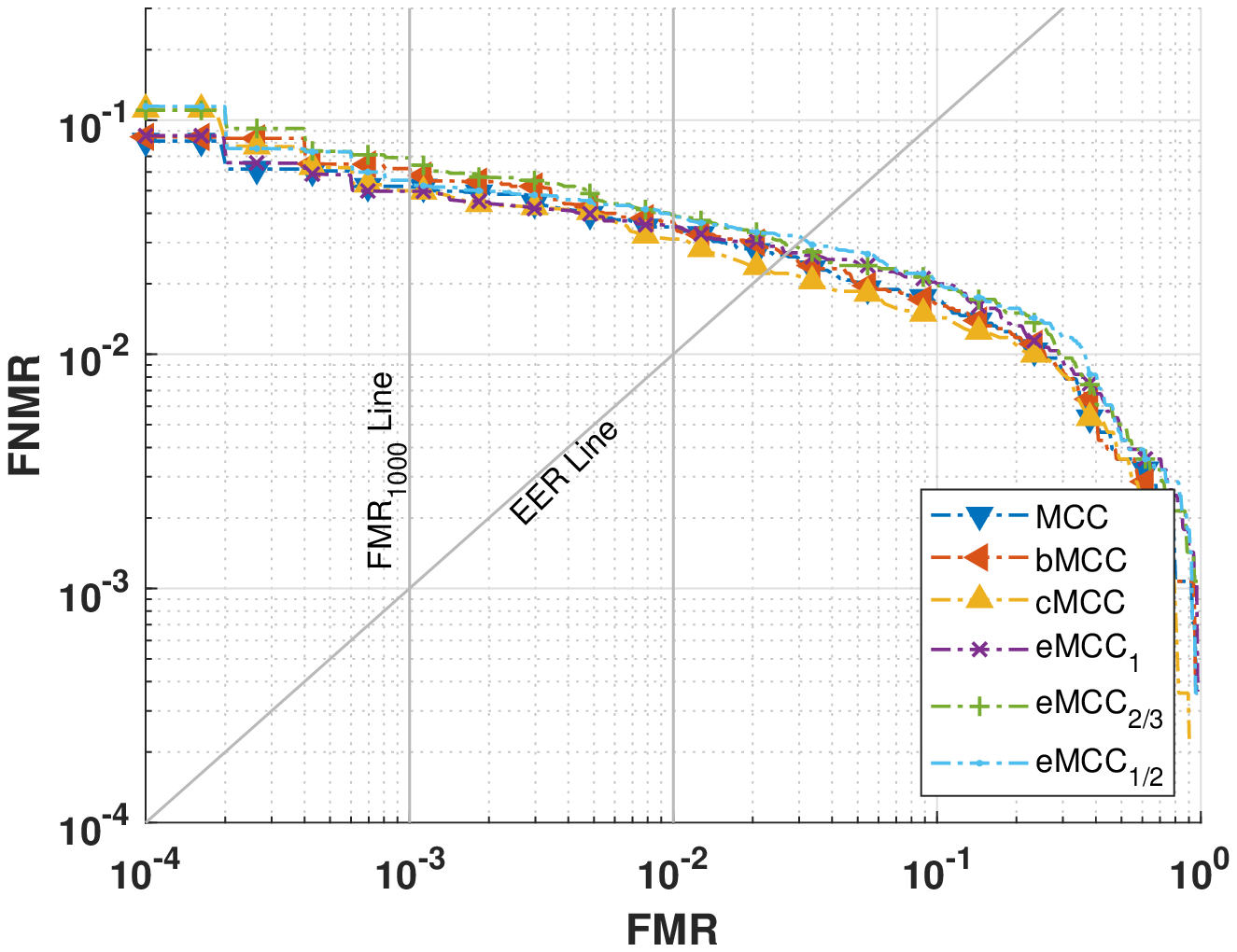}
		}
		\vfil
		\subfloat[DET curves evaluated on DB3]
		{
			\includegraphics[width=0.48\linewidth]{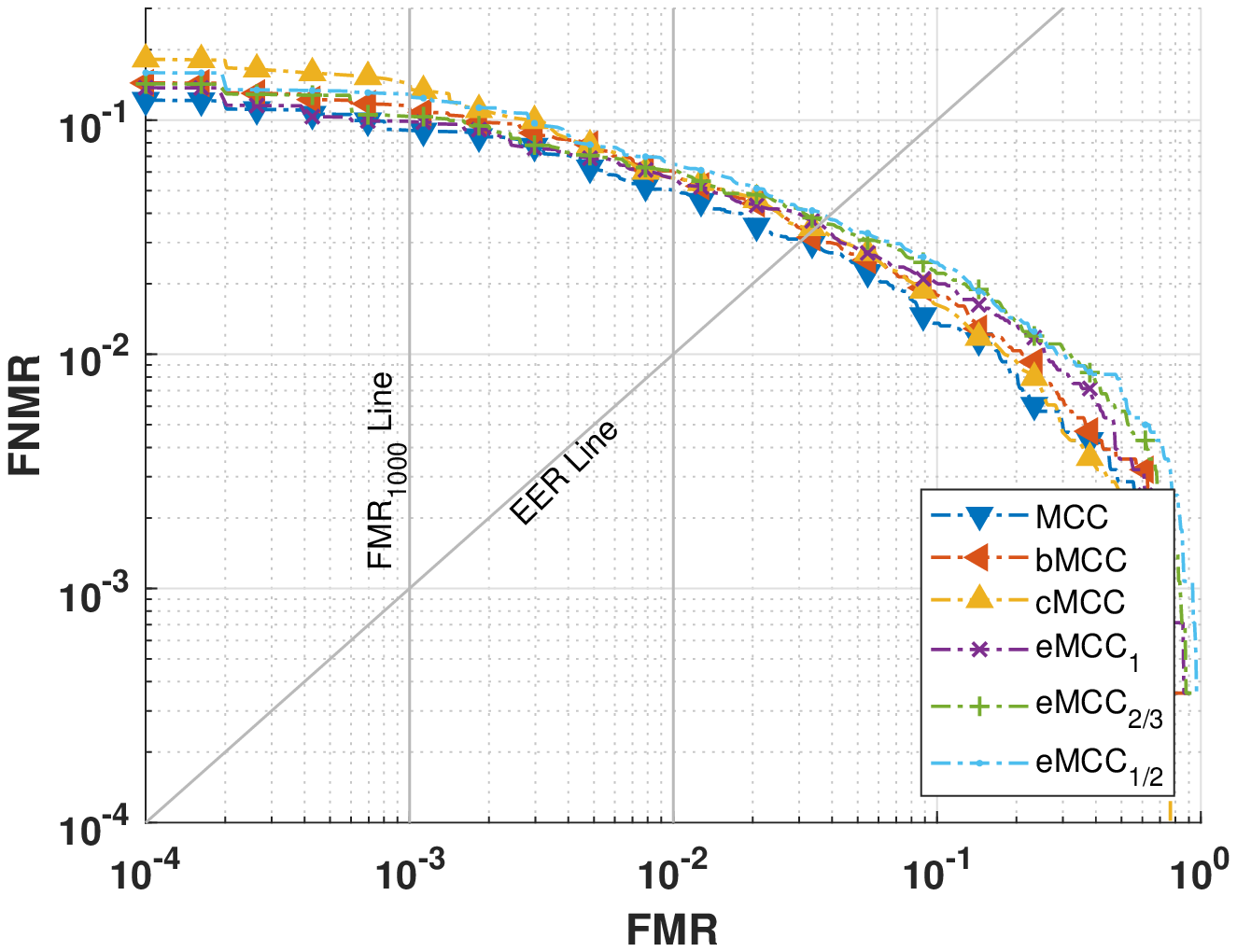}
		}
		\subfloat[DET curves evaluated on DB4]
		{
			\includegraphics[width=0.48\linewidth]{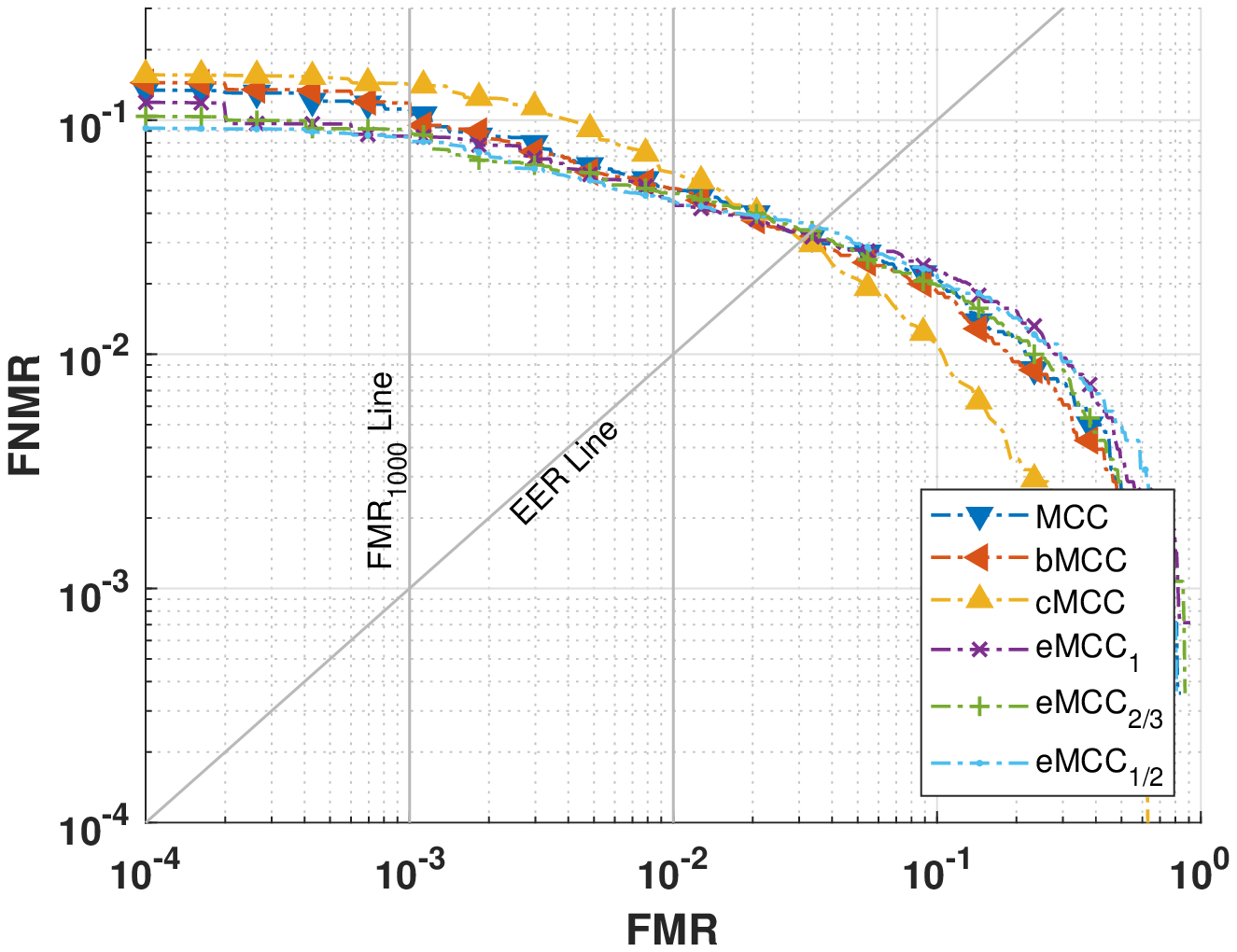}
		}
		\caption{Comparison of DET curves obtained by MCC, bMCC, cMCC, eMCC$_1$, eMCC$_{2/3}$ and eMCC$_{1/2}$ based on the minutiae extraction presented in Section~\ref{sec:extraction} on datasets FVC2002 DB1-DB4, evaluated using the implemented IoT prototype system.}
		\label{fig:DET_FVC2002_IoT}
	\end{figure}

\begin{figure}[htbp]
		\centering
		\subfloat[DET curves evaluated on DB1]
		{
			\includegraphics[width=0.48\linewidth]{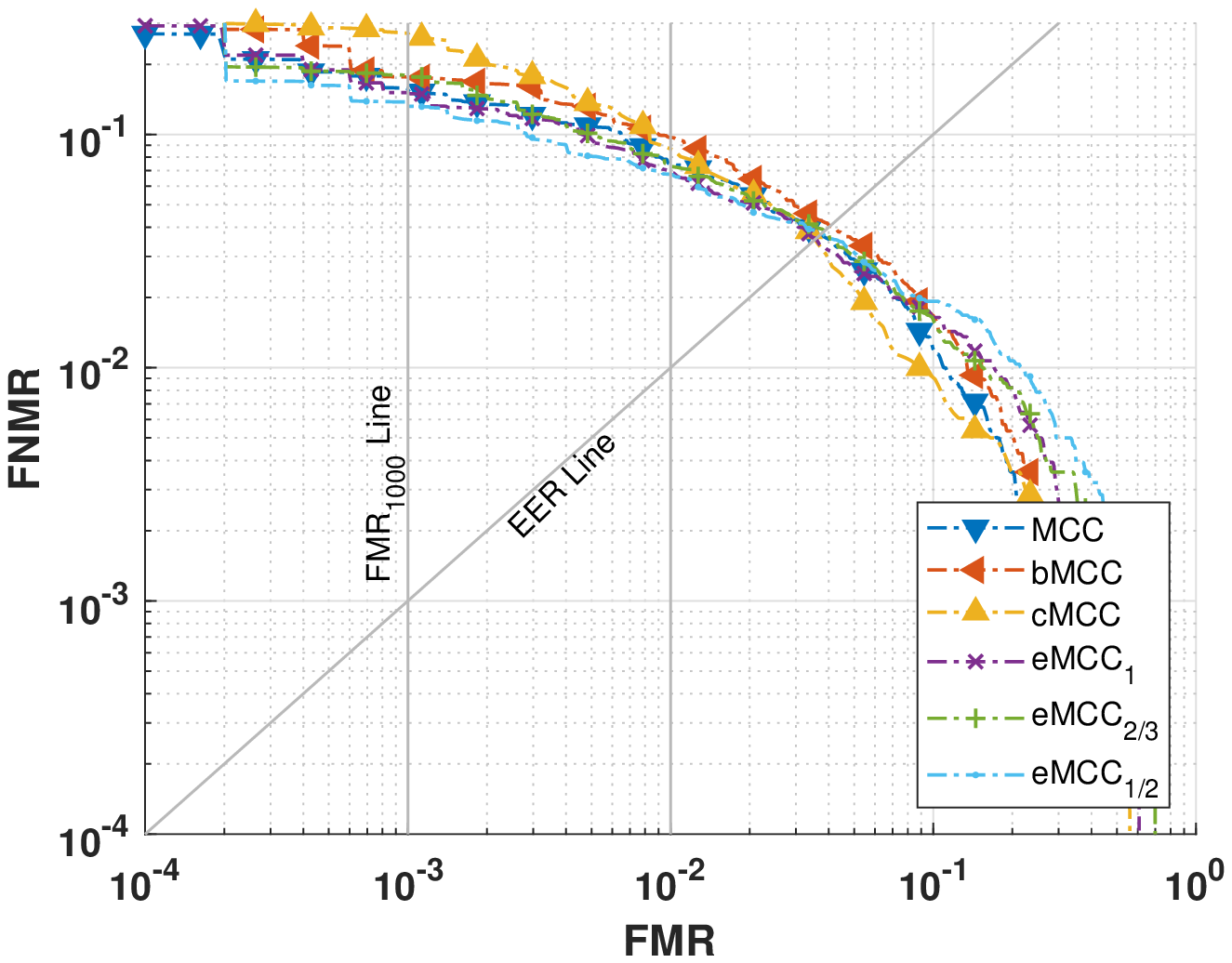}
		}
		\subfloat[DET curves evaluated on DB2]
		{
			\includegraphics[width=0.48\linewidth]{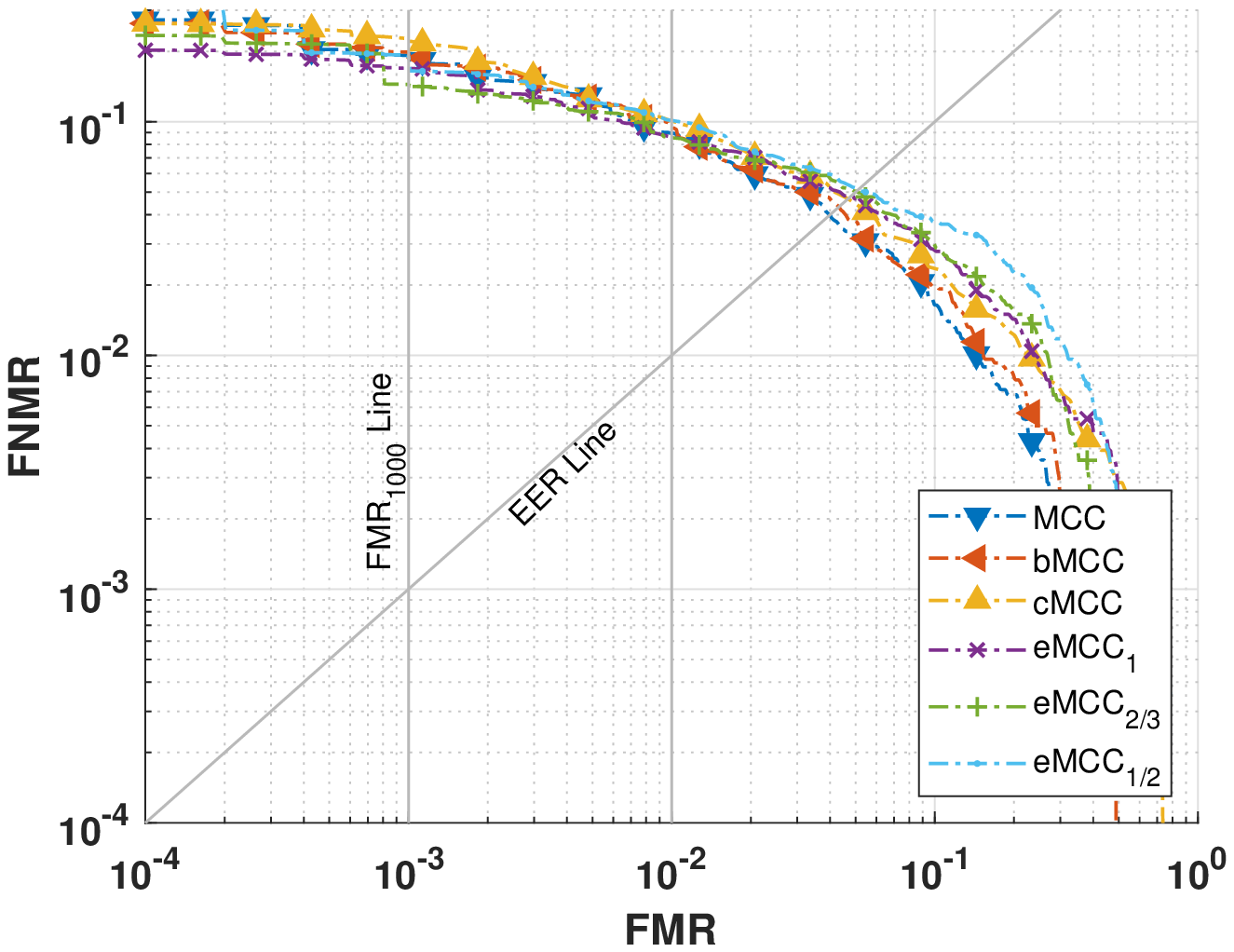}
		}
		\vfil
		\subfloat[DET curves evaluated on DB3]
		{
			\includegraphics[width=0.48\linewidth]{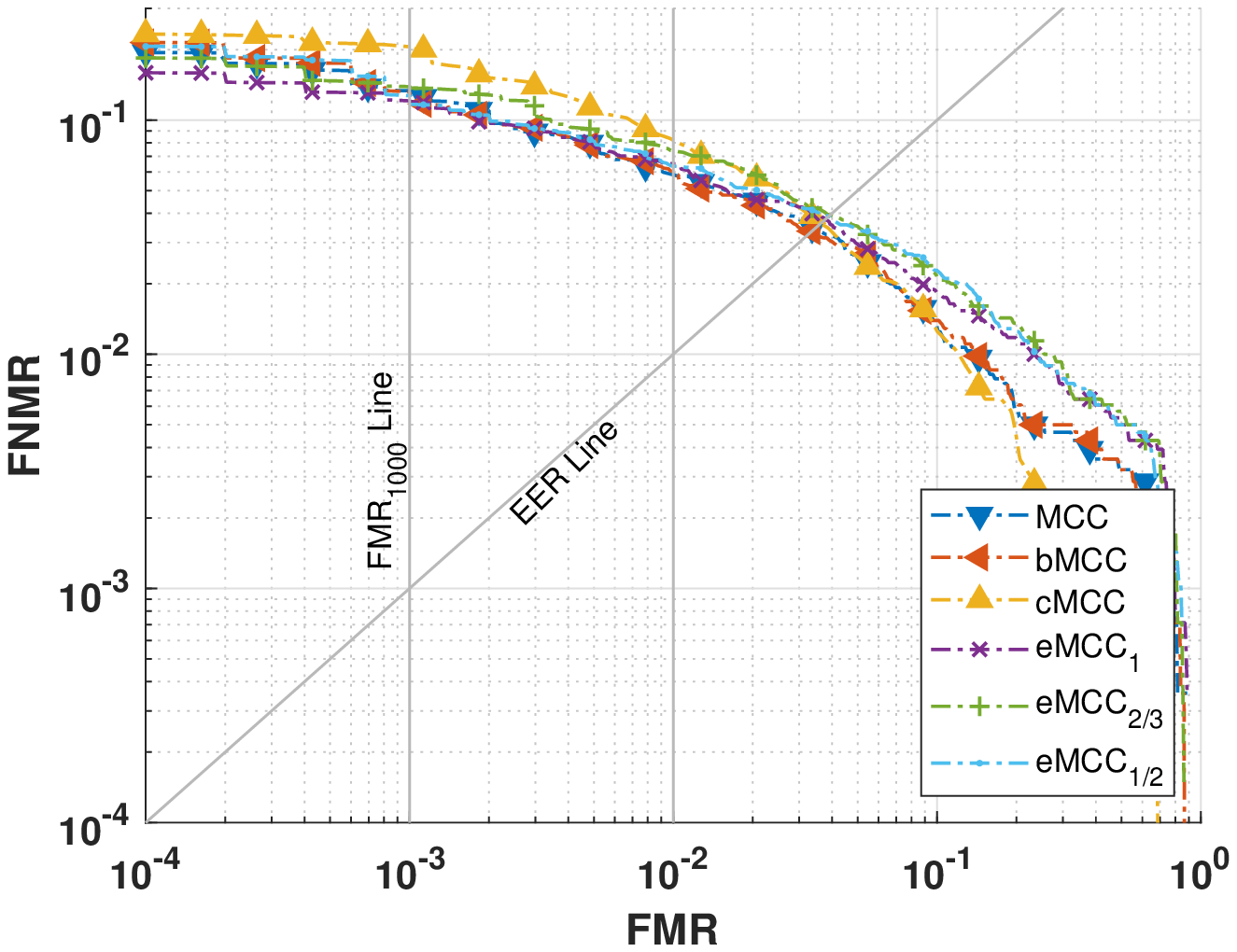}
		}
		\subfloat[DET curves evaluated on DB4]
		{
			\includegraphics[width=0.48\linewidth]{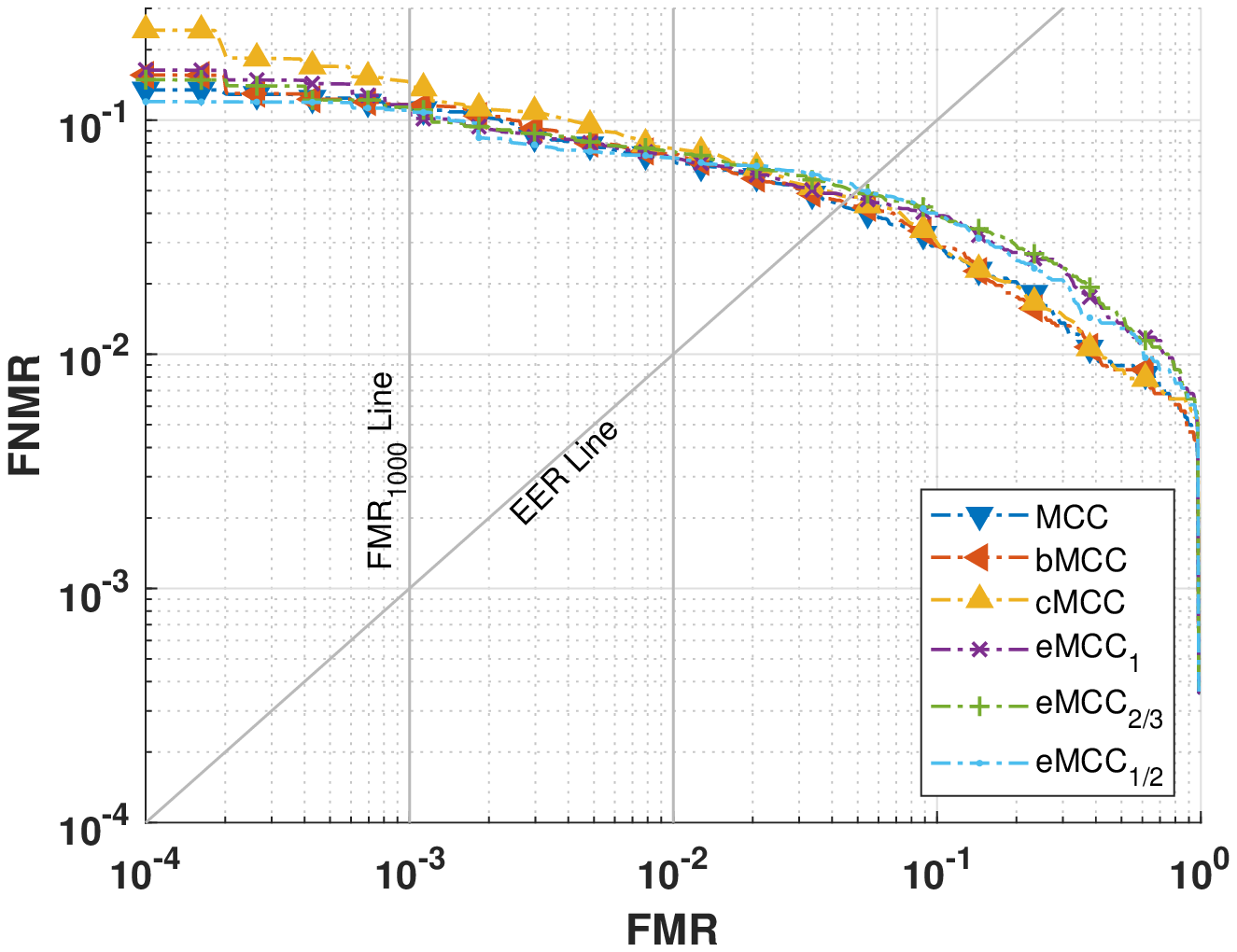}
		}
		\caption{Comparison of DET curves obtained by MCC, bMCC, cMCC, eMCC$_1$, eMCC$_{2/3}$, and eMCC$_{1/2}$ based on the minutiae extraction presented in Section \ref{sec:extraction} on datasets FVC2004 DB1-DB4, evaluated using the implemented IoT prototype system.}
		\label{fig:DET_FVC2004_IoT}
	\end{figure}

	Fig.~\ref{fig:DET_FVC2002_IoT} and Fig.~\ref{fig:DET_FVC2004_IoT} compare the DET curves obtained by MCC, bMCC, cMCC, eMCC$_1$, eMCC$_{2/3}$ and eMCC$_{1/2}$ on datasets FVC2002 and FVC2004, respectively, evaluated using the implemented IoT prototype system. 
	The results for MCC, bMCC, and cMCC are provided by~\cite{RN2761}. The results for eMCC$_1$, eMCC$_{2/3}$, and eMCC$_{1/2}$ are obtained using the implemented IoT system.
	As shown in Fig.~\ref{fig:DET_FVC2002_IoT}, on FVC2002 DB1, FVC2002 DB2, and FVC2002 DB3, the DET curves are similar to each other.  
	On the left side of the FMR$_{1000}$ line on FVC2002 DB1 and FVC2002 DB2, eMCC$_1$ show better DET curves than cMCC. 
	We also observe that on the left side of the FMR$_{1000}$ line on FVC2002 DB3, eMCC$_1$, eMCC$_{2/3}$ and eMCC$_{1/2}$ perform better than cMCC.
	On FVC2002 DB4, it is shown that above the EER line, eMCC$_1$, eMCC$_{2/3}$ and eMCC$_{1/2}$ exhibit better DET curves than the other three templates. 
	Similar experimental results are also observed on FVC2004, as can be seen in Fig.~\ref{fig:DET_FVC2004_IoT}.
	This shows that the authentication accuracy of the proposed template is comparable to that of the state-of-the-art templates.
	
	\subsubsection{EER and FMR$_{1000}$}
	Table~\ref{tab:EER_FMR1000_emcc_all_6_IoT} compares the EER and FMR$_{1000}$ obtained by MCC, bMCC, cMCC, eMCC$_1$, eMCC$_{2/3}$, and eMCC$_{1/2}$ on datasets FVC2002 DB1-DB4 and FVC2004 DB1-DB4, evaluated using the implemented IoT prototype system. 
	As shown in Table~\ref{tab:EER_FMR1000_emcc_all_6_IoT}, compared with MCC, bMCC and cMCC, the proposed templates eMCC$_1$, eMCC$_{2/3}$ and eMCC$_{1/2}$ achieve comparable accuracy in terms of the EER and FMR$_{1000}$. 
	eMCC$_1$ even performs better than MCC on FVC2002 DB4 and FVC2004 DB1. 
	eMCC$_{2/3}$ and eMCC$_{1/2}$ also achieve better EER values than bMCC on FVC2004 DB1.
	Regarding the FMR$_{1000}$, the accuracy of eMCC$_1$, eMCC$_{2/3}$ and eMCC$_{1/2}$ are close to that of MCC, bMCC and cMCC.
	On FVC2002 DB2, eMCC$_{1/2}$ achieves a better FMR$_{1000}$ than bMCC, and eMCC$_{1}$ outperforms MCC and bMCC.
	Similar results are also clearly observed on FVC2004. 
	In summary, comparable authentication accuracy is demonstrated between the proposed and state-of-the-art templates. 
	 
	\begin{table*}[htbp]
		\setlength\tabcolsep{2pt}
		\renewcommand{\arraystretch}{1.5}
		\caption{Comparison of verification accuracy in terms of the EER and FMR$_{1000}$ obtained by MCC, bMCC, cMCC, eMCC$_1$, eMCC$_{2/3}$ and eMCC$_{1/2}$ on FVC2002 DB1-DB4 and FVC2004 DB1-DB4, evaluated using the implemented IoT prototype system.}
		\label{tab:EER_FMR1000_emcc_all_6_IoT}
		\centering
		\resizebox{\linewidth}{!}
		{
			\begin{tabular}{c |c||c|c|c|c|c|c||c|c|c|c|c|c} 
				\hline		\hline
				\multicolumn{2}{c||}{\multirow{2}{*}{Dateset}}	&\multicolumn{6}{c||}{\textbf{EER (\%)}}  	          &  \multicolumn{6}{c}{\textbf{FMR$_{1000}$ (\%)}}		\\ \cline{3-14} 
				\multicolumn{2}{c||}{} & MCC & bMCC & cMCC & eMCC$_1$ & eMCC$_{2/3}$ & eMCC$_{1/2}$  &MCC & bMCC & cMCC & eMCC$_1$ & eMCC$_{2/3}$ & eMCC$_{1/2}$	\\ 	\hline \hline
				
				\multirow{4}{*}{\textbf{FVC2002}} 
				& DB1& 2.96 &2.78 &2.89 &3.03 &3.17 &3.25 &6.50 & 6.41&7.39 &6.64 &7.46 &8.64		\\		\cline{2-14}
				& DB2& 2.64	&2.68&2.29 &2.71 &2.89 &3.09 &5.21 &6.21&5.00 &4.96 &6.90 &5.53       \\      \cline{2-14}
				& DB3& 3.07	&3.29&3.43 &3.57 &3.71 &3.83 &9.07 &11.56&14.48 &9.90 &10.40 &12.92		\\ 		\cline{2-14}
				& DB4& 3.28	&3.18&3.11 &3.25 &3.37 &3.49 &11.14 &11.82&14.32 &8.53 &9.11 &8.47		\\ 		\hline		\hline
				
				\multirow{4}{*}{\textbf{FVC2004}} 
				& DB1&3.70 &4.04 &3.56 &3.57 &3.82 &3.78 &15.81 & 17.66&27.28 &15.11 &17.98  &13.79 \\ 	\cline{2-14}
				& DB2&4.00	&4.32&4.64 &4.82 &5.03 &5.13 &19.26 &19.89&22.78 &16.95 &14.45 &19.29		\\ 		\cline{2-14}
				& DB3&3.42	&3.36&3.63 &3.75 &3.92 &3.89 &13.36 &13.30&20.67 &12.11   &13.91 &12.76		\\ 		\cline{2-14}
				& DB4&4.33	&4.44&4.69 &4.72 &4.90 &5.07 &11.28 &11.67&14.68 &11.63   &11.35 &11.02		\\ 		\hline		\hline
			\end{tabular}
		}
	\end{table*}
	
	\subsection{Authentication Accuracy and Security Analysis}\label{sec:security}
	
	{\color{color}
	\subsubsection{Authentication Accuracy Analysis}\label{sec:accuracyAnalysis}
	In authentication, there are two cases: genuine matching and imposter matching. For the genuine matching, because the query cancelable template and the enrolled cancelable template are processed by the same cancelable system, it is obviously clear that the authentication result of the query cancelable template against the enrolled cancelable template will be the same as the results of the original query and enrolled templates with a high probability. This is also supported by authentication accuracy results in Section \ref{sec:validation} and \ref{sec:simulation}. 
	
	Next we analyze the authentication accuracy for the imposter matching case, 
	As shown in Fig. \ref{fig:nested_difference}, according to the distribution of $\frac{e_i}{2}$ (in Table \ref{tab:encoding}) collected from 18,539 valid MCC feature vectors from five hundred fingerprints, we have the following probabilities: 
	\begin{equation}\nonumber 
		\begin{cases}
			P(-0.2 \leqslant \frac{e_i}{2} \leqslant 0.2) & \approx 0.75,\\
			P(\frac{e_i}{2}\geqslant 0.2) &\approx 0.125,\\
			P(\frac{e_i}{2}\leqslant -0.2) &\approx 0.125.
		\end{cases}
	\end{equation}
	Hence, for an eMCC$_1$ feature vector defined in Eq. \ref{eq:xorVector} with $k=160$, according to the encoding scheme in Table \ref{tab:encoding}, we have the following probabilities:
	\begin{equation}\nonumber 
		\begin{cases}
			P(\hat{e}_i = 00) & \approx 0.75,\\
			P(\hat{e}_i = 10) &\approx 0.125,\\
			P(\hat{e}_i = 01) &\approx 0.125.
		\end{cases}
	\end{equation}
Therefore, for a fake query cancelable template matching against the enrolled cancelable template, the probability that the authentication result is the same as the genuine result is about $0.75^{120}*0.125^{20}*0.125^{20} \approx 7.65\times10^{-52}.$ 
In summary, there is near zero probability for a fake cancelable template to obtain a highly-close authentication accuracy as the genuine query cancelable template. 
		\begin{figure}[htbp]
			\centering
			\includegraphics[width=0.9\linewidth]{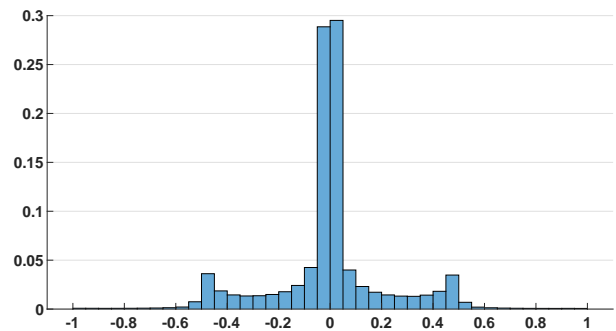}
			\caption{\color{color}Distribution of $\frac{e_i}{2}$ (in Table~\ref{tab:encoding}) collected from 18,539 valid MCC feature vectors from five hundred fingerprints. The x-axis represents the values of $\frac{e_i}{2}$, and the y-axis is the proportion of the values falling into each bin.}
			\label{fig:nested_difference}
		\end{figure}
	}

	\subsubsection{Security Analysis on the Cancelable Template Design} \label{sec:security_4obj}
	The proposed template design meets the four objectives of cancelable biometrics: diversity, revocability, accuracy and non-invertibility. 
	The diversity is guaranteed by the re-indexing scheme, as there exist numerous re-indexing sets, that is $ \frac{L_{c}!}{(L_{c}-l)!}$ (e.g., $L_c$ = 1,280 in the experiments). 
	Regarding the revocability, as the re-indexing process is controlled by a random generator, a completely new template therefore can be easily obtained by choosing a different random seed. 
	The accuracy of the proposed template is comparable to that of the state-of-the-art methods. Especially, when $p = 100\%$, equivalent authentication accuracy is achieved on eight public benchmark datasets FVC2002 DB1-DB4 and FVC2004 DB1-DB4 compared to the state-of-the-art templates. 
	The non-invertibility of the proposed template is guaranteed by the irreversible mapping under two protection mechanisms: 1) the encoding of the nested-difference (Section~\ref{sec:nestedDifference} and Section~\ref{sec:encoding}); and 2) the XOR operation on the encoded vector (Section~\ref{sec:cancelable}).
	As the first protection mechanism, the encoding of the nested-difference utilizes two bits to represent a nested difference of four float values. This process constitutes an infinite-to-one mapping, which is irreversible. Given the encoding bits, there is a near-zero probability to retrieve the original float values, because there exist infinite combinations of float values that can map to the same encoded bits. 
	The second protection mechanism is the XOR Boolean operation on the encoded vector (in Eq. \eqref{eq:encoding2k}). The XOR operation makes it impossible to retrieve the true encoded vector $\bar{\mathbf{e}} $ (in Eq.~\eqref{eq:encoding2k}) from the resultant vector $\hat{\mathbf{e}}$ (in Eq.~\eqref{eq:xorVector}). For example, even with the most lightweight template with $p = 50\%$ for the case of $N_S = 16$ and $N_D=15$, there exist up to $2^{160}$ possible candidate vectors $\bar{\mathbf{e}} $ (in Eq.~\eqref{eq:encoding2k}), which can be used as the input and return the same vector $\hat{\mathbf{e}}$ (in Eq.~\eqref{eq:xorVector}).
	In conclusion, the probability of retrieving the original template from the proposed lightweight cancelable template is almost zero.
	In addition, since the proposed template is revocable, if an enrolled template is compromised, it is easy to issue a completely different template (even of a different length) to ensure the security of the IoT authentication system.
	
	\subsubsection{Security Analysis against Attacks}
	The proposed cancelable template is resistant to attacks via record multiplicity~(ARM), which utilize multiple compromised protected templates to recover the original template.
	This attack can be effectively prevented by the proposed encoding-nested-difference-XOR scheme through two layers of protection.
    The first layer of protection is the XOR operation (Section~\ref{sec:cancelable}). 
    Because the inputs for each bit of the XOR output cannot be uniquely determined, the XOR operation provides computational infeasibility to retrieve the encoded binary feature vector $\bar{\mathbf{e}}$ in Eq.~\eqref{eq:encoding2k} in Section~\ref{sec:encoding}, as shown by the example in Section~\ref{sec:security_4obj}. Taking $C=A ~XOR~ B$ as an example, according to the truth table of the XOR, $C=1$ has two possible inputs: $A=1, B=0$ or $A=0, B=1$. The case of $C=0$ is similar.
    Therefore, even though the adversary acquires multiple compromised protected templates, the encoded binary vectors $\bar{\mathbf{e}}$ in Eq.~\eqref{eq:encoding2k} cannot be uniquely determined.
    The second layer of protection is the encoding operation (Section~\ref{sec:encoding}), where a threshold is defined to binarize the nested difference. Evidently, given the threshold and binarized values, it is of zero probability to restore the nested differences in that infinite combinations of float values can result in the same encoded bits. 
	In summary, the proposed cancelable template is resilient to the ARM. 
	
	The proposed cancelable template is secure against pre-image attacks and optimization-based attacks.
	If the enrolled binary template is compromised, it is not computationally difficult to reconstruct or search for a possible input that can return the same binary vector. 
	However, given that there exist infinite possible candidate inputs, it is of a near-zero probability for the reverted one  to be the genuine fingerprint template.
	In other words, the reverted one cannot be used to generate another legitimate binary template.
	Therefore, these attacks can be effectively prevented by revoking the compromised binary template. 
	In addition, attacks may also be launched through real-world fingerprint datasets. The impact of this attack can be assessed through authentication accuracy (e.g., the EER and FMR$_{1000}$). As demonstrated in Section~\ref{sec:validation} and Section~\ref{sec:simulation}, the proposed template achieves favorable authentication accuracy in terms of the EER and FMR$_{1000}$ in a privacy-preserving IoT environment.

    {\color{color}
    In case the template has been compromised by an adversary, it is infeasible for the attacker to retrieve the input feature due to the non-invertibility of our proposed template. Without this input feature, the attacker cannot launch an attack via the sensor interface which is the normal system interface. It is, however, possible for the attacker to get authenticated if the attacker can inject the compromised template into the matching module after bypassing the sensor and the built-in transformation module. This is exceedingly difficult but possible.  Therefore, it is still difficult to attack a new device even if it has the same compromised template. Our cancelable template design offers further security protection by revoking the compromised template, like the revocation of a password.
    }
\section{Conclusion}\label{sec:conclusion}
	In this paper, we proposed a length-flexible lightweight cancelable fingerprint template design for privacy-preserving authentication systems in resource-constrained IoT applications. 
	The proposed template design consists of two components: 1) length-flexible partial-cancelable feature generation based on the re-indexing scheme; and 2) lightweight cancelable feature generation based on the encoding-nested-difference-XOR scheme. Our template design has a number of benefits to IoT applications, such as flexible feature lengths, lightweight, cancelability and high performance.
	Comprehensive experimental results evaluated on eight benchmark datasets FVC2002 DB1-DB4 and FVC2004 DB1-DB4 demonstrate that the proposed cancelable fingerprint template achieves equivalent authentication performance compared to the state-of-the-art methods, but our design significantly reduces storage space and computational cost.
	More importantly, the proposed length-flexible lightweight cancelable template is suitable for various resource-constrained IoT devices, evidenced by its implementation using a real-world IoT prototype system.
	To the best of our knowledge, it is the first length-flexible lightweight, high-performing cancelable fingerprint template design for resource-constrained IoT applications.
	
	\section*{Acknowledgment}
	
	This research was undertaken with the assistance of resources and services from the National Computational Infrastructure (NCI), which is supported by the Australian Government. 
	This research is supported by ARC Discovery Grants (DP190103660 and DP200103207) and ARC Linkage Grant (LP180100663).

	\ifCLASSOPTIONcaptionsoff
	\newpage
	\fi

	
	
	\bibliographystyle{IEEEtran}
	\bibliography{IoT_2021}

\begin{thebibliography}{10}
\providecommand{\url}[1]{#1}
\csname url@samestyle\endcsname
\providecommand{\newblock}{\relax}
\providecommand{\bibinfo}[2]{#2}
\providecommand{\BIBentrySTDinterwordspacing}{\spaceskip=0pt\relax}
\providecommand{\BIBentryALTinterwordstretchfactor}{4}
\providecommand{\BIBentryALTinterwordspacing}{\spaceskip=\fontdimen2\font plus
\BIBentryALTinterwordstretchfactor\fontdimen3\font minus
  \fontdimen4\font\relax}
\providecommand{\BIBforeignlanguage}[2]{{%
\expandafter\ifx\csname l@#1\endcsname\relax
\typeout{** WARNING: IEEEtran.bst: No hyphenation pattern has been}%
\typeout{** loaded for the language `#1'. Using the pattern for}%
\typeout{** the default language instead.}%
\else
\language=\csname l@#1\endcsname
\fi
#2}}
\providecommand{\BIBdecl}{\relax}
\BIBdecl

\bibitem{Survey_2019}
V.~Hassija, V.~Chamola, V.~Saxena, D.~Jain, P.~Goyal, and B.~Sikdar, ``A survey
  on iot security: application areas, security threats, and solution
  architectures,'' \emph{IEEE Access}, vol.~7, no.~1, pp. 82\,721--82\,743,
  2019.

\bibitem{RN2413}
N.~N. Tran, H.~R. Pota, Q.~N. Tran, and J.~Hu, ``Designing constraint-based
  false data injection attacks against the unbalanced distribution smart
  grids,'' \emph{IEEE Internet of Things Journal}, pp. 1--1, 2021.

\bibitem{IoT-security-2018}
M.~Khan and K.~Salah, ``Iot security: Review, blockchain solutions, and open
  challenges,'' \emph{Future Generation Computer Systems}, vol.~82, pp.
  395--411, 2018.

\bibitem{Demysifying}
N.~Neshenko, E.~Bou-Harb, J.~Crichigno, G.~Kaddoum, and N.~Ghani,
  ``Demystifying iot security: An exhaustive survey on iot vulnerabilities and
  a first empirical look on internet-scale iot exploitations,'' \emph{IEEE
  Communications Surveys Tutorials}, vol.~21, no.~3, pp. 2702--2733, 2019.

\bibitem{riad2019sensitive}
K.~Riad, R.~Hamza, and H.~Yan, ``Sensitive and energetic iot access control for
  managing cloud electronic health records,'' \emph{IEEE Access}, vol.~7, pp.
  86\,384--86\,393, 2019.

\bibitem{pinno2020controlchain}
O.~J.~A. Pinno, A.~R.~A. Gr{\'e}gio, and L.~C. De~Bona, ``Controlchain: A new
  stage on the iot access control authorization,'' \emph{Concurrency and
  Computation: Practice and Experience}, vol.~32, no.~12, p. e5238, 2020.

\bibitem{RN2422}
A.~K. Jain, K.~Nandakumar, and A.~Ross, ``50 years of biometric research:
  Accomplishments, challenges, and opportunities,'' \emph{Pattern Recognition
  Letters}, vol.~79, pp. 80--105, 2016.

\bibitem{RN1145}
X.~F. Yin, Y.~M. Zhu, and J.~K. Hu, ``Contactless fingerprint recognition based
  on global minutia topology and loose genetic algorithm,'' \emph{IEEE
  Transactions on Information Forensics and Security}, vol.~15, pp. 28--41,
  2020.

\bibitem{obaidat2019biometric}
M.~S. Obaidat, S.~P. Rana, T.~Maitra, D.~Giri, and S.~Dutta, ``Biometric
  security and internet of things (iot),'' in \emph{Biometric-Based Physical
  and Cybersecurity Systems}, 2019, pp. 477--509.

\bibitem{ren2016biometrics}
C.-x. Ren, Y.-b. Gong, F.~Hao, X.-y. Cai, and Y.-x. Wu, ``When biometrics meet
  iot: a survey,'' in \emph{International Asia Conference on Industrial
  Engineering and Management Innovation}, 2016, pp. 635--643.

\bibitem{jiang2021enhancing}
X.~Jiang, X.~Liu, J.~Fan, X.~Ye, C.~Dai, E.~A. Clancy, D.~Farina, and W.~Chen,
  ``Enhancing iot saecurity via cancelable hd-semg-based biometric
  authentication password, encoded by gesture,'' \emph{IEEE Internet of Things
  Journal}, 2021.

\bibitem{RN2557}
T.~Kim, Y.~Oh, and H.~Kim, ``Efficient privacy-preserving fingerprint-based
  authentication system using fully homomorphic encryption,'' \emph{Security
  and Communication Networks}, vol. 2020, p. 4195852, 2020.

\bibitem{RN2558}
M.~Barni, T.~Bianchi, D.~Catalano, M.~D. Raimondo, R.~D. Labati, P.~Failla
  \emph{et~al.}, ``A privacy-compliant fingerprint recognition system based on
  homomorphic encryption and fingercode templates,'' in \emph{IEEE
  International Conference on Biometrics: Theory, Applications and Systems},
  2010, pp. 1--7.

\bibitem{RN2560}
Z.~Bohan, W.~Xu, Z.~Kaili, and Z.~Xueyuan, ``Encryption node design in internet
  of things based on fingerprint features and cc2530,'' in \emph{IEEE
  International Conference on Green Computing and Communications and IEEE
  Internet of Things and IEEE Cyber, Physical and Social Computing}, 2013, pp.
  1454--1457.

\bibitem{RN1102}
N.~K. Ratha, S.~Chikkerur, J.~H. Connell, and R.~M. Bolle, ``Generating
  cancelable fingerprint templates,'' \emph{IEEE Transactions on Pattern
  Analysis and Machine Intelligence}, vol.~29, no.~4, pp. 561--572, 2007.

\bibitem{RN2561}
N.~K. Ratha, J.~H. Connell, and R.~M. Bolle, ``Enhancing security and privacy
  in biometrics-based authentication systems,'' \emph{IBM Systems Journal},
  vol.~40, pp. 614--634, 2001.

\bibitem{punithavathi2019partial}
P.~Punithavathi and S.~Geetha, ``Partial dct-based cancelable biometric
  authentication with security and privacy preservation for iot applications,''
  \emph{Multimedia Tools and Applications}, vol.~78, no.~18, pp.
  25\,487--25\,514, 2019.

\bibitem{kaur2020privacy}
H.~Kaur and P.~Khanna, ``Privacy preserving remote multi-server biometric
  authentication using cancelable biometrics and secret sharing,'' \emph{Future
  Generation Computer Systems}, vol. 102, pp. 30--41, 2020.

\bibitem{jiang2020cancelable}
X.~Jiang, K.~Xu, X.~Liu, C.~Dai, D.~A. Clifton, E.~A. Clancy, M.~Akay, and
  W.~Chen, ``Cancelable hd-semg-based biometrics for cross-application
  discrepant personal identification,'' \emph{IEEE Journal of Biomedical and
  Health Informatics}, vol.~25, no.~4, pp. 1070--1079, 2020.

\bibitem{patel2015cancelable}
V.~M. Patel, N.~K. Ratha, and R.~Chellappa, ``Cancelable biometrics: A
  review,'' \emph{IEEE Signal Processing Magazine}, vol.~32, no.~5, pp. 54--65,
  2015.

\bibitem{RN2519}
Q.~N. Tran and J.~Hu, ``A multi-filter fingerprint matching framework for
  cancelable template design,'' \emph{IEEE Transactions on Information
  Forensics and Security}, vol.~16, no.~1, pp. 2926--2940, 2021.

\bibitem{RN159}
R.~Cappelli, M.~Ferrara, and D.~Maltoni, ``Minutia cylinder-code: A new
  representation and matching technique for fingerprint recognition,''
  \emph{IEEE Transactions on Pattern Analysis and Machine Intelligence},
  vol.~32, no.~12, pp. 2128--2141, 2010.

\bibitem{RN2562}
D.~Maio, D.~Maltoni, R.~Cappelli, J.~L. Wayman, and A.~K. Jain, ``{FVC}2002:
  Second fingerprint verification competition,'' in \emph{Object Recognition
  Supported by User Interaction for Service Robots}, vol.~3, 2002, pp. 811--814
  vol.813.

\bibitem{RN2022}
D.~Maio, D.~Maltoni, R.~Cappelli, J.~L. Wayman, and A.~K. Jain, ``{FVC}2004:
  Third fingerprint verification competition,'' in \emph{International
  Conference on Biometric Authentication}, 2004, pp. 1--7.

\bibitem{RN1242}
K.~Xi and J.~Hu, \emph{Bio-cryptography}.\hskip 1em plus 0.5em minus
  0.4em\relax Springer, 2010, ch.~7, pp. 129--157.

\bibitem{RN2559}
W.~Yang, S.~Wang, K.~Yu, J.~J. Kang, and M.~N. Johnstone, ``Secure fingerprint
  authentication with homomorphic encryption,'' in \emph{Digital Image
  Computing: Techniques and Applications}, 2020, pp. 1--6.

\bibitem{RN2768}
M.~S. Azzaz, C.~Tanougast, A.~Maali, and M.~Benssalah, ``An efficient and
  lightweight multi‐scroll chaos‐based hardware solution for protecting
  fingerprint biometric templates,'' \emph{International Journal of
  Communication Systems}, vol.~33, no.~10, p. e4211, 2020.

\bibitem{RN2771}
Y.~Liu, T.~Zhou, Z.~Yue, W.~Liu, L.~Y. Han, Q.~Li \emph{et~al.}, ``Secure and
  efficient online fingerprint authentication scheme based on cloud
  computing,'' \emph{IEEE Transactions on Cloud Computing}, pp. 1--1, 2021.

\bibitem{RN2582}
J.~B. Kho, J.~Kim, I.-J. Kim, and A.~B. Teoh, ``Cancelable fingerprint template
  design with randomized non-negative least squares,'' \emph{Pattern
  Recognition}, vol.~91, pp. 245--260, 2019.

\bibitem{RN2772}
L.~Wu, L.~Meng, S.~Zhao, X.~Wei, and H.~Wang, ``Privacy-preserving cancelable
  biometric authentication based on rdm and ecc,'' \emph{IEEE Access}, vol.~9,
  pp. 90\,989--91\,000, 2021.

\bibitem{RN2775}
I.~Kavati, A.~M. Reddy, E.~S. Babu, K.~S. Reddy, and R.~S. Cheruku, ``Design of
  a fingerprint template protection scheme using elliptical structures,''
  \emph{ICT Express}, 2021.

\bibitem{RN2728}
A.~Bedari, S.~Wang, and W.~Yang, ``Design of cancelable mcc-based fingerprint
  templates using dyno-key model,'' \emph{Pattern Recognition}, vol. 119,
  108074, 2021.

\bibitem{RN2761}
X.~Yin, S.~Wang, M.~Shahzad, and J.~Hu, ``An iot-oriented privacy-preserving
  fingerprint authentication system,'' \emph{IEEE Internet of Things Journal},
  pp. 1--1, 2021.

\bibitem{RN2774}
M.~J. Lee, A.~B.~J. Teoh, A.~Uhl, S.-N. Liang, and Z.~Jin, ``A tokenless
  cancellable scheme for multimodal biometric systems,'' \emph{Computers \&
  Security}, p. 102350, 2021.

\bibitem{RN134}
F.~Benhammadi and K.~B. Bey, ``Embedded fingerprint matching on smart card,''
  \emph{International Journal of Pattern Recognition and Artificial
  Intelligence}, vol.~27, no.~2, p. 1350006, 2013.

\bibitem{RN2551}
P.~Punithavathi, S.~Geetha, M.~Karuppiah, S.~K.~H. Islam, M.~M. Hassan, and
  K.-K.~R. Choo, ``A lightweight machine learning-based authentication
  framework for smart iot devices,'' \emph{Information Sciences}, vol. 484,
  no.~1, pp. 255--268, 2019.

\bibitem{RN2510}
C.~T. Pang, W.-Y. Yau, R.~Mueller, and L.~Yih, \emph{Biometric
  system-on-card}.\hskip 1em plus 0.5em minus 0.4em\relax Springer US, 2014,
  pp. 1--6.

\bibitem{RN2563}
K.~Habib, A.~Torjusen, and W.~Leister, ``A novel authentication framework based
  on biometric and radio fingerprinting for the iot in ehealth,'' in
  \emph{International Conference on Smart Systems, Devices, and Technologies},
  2014, pp. 32--37.

\bibitem{RN2777}
M.~Golec, S.~S. Gill, R.~Bahsoon, and O.~Rana, ``Biosec: A biometric
  authentication framework for secure and private communication among edge
  devices in iot and industry 4.0,'' \emph{IEEE Consumer Electronics Magazine},
  pp. 1--1, 2020.

\bibitem{RN2779}
M.~Sabri, M.-S. Moin, and F.~Razzazi, ``A new framework for match on card and
  match on host quality based multimodal biometric authentication,''
  \emph{Journal of Signal Processing Systems}, vol.~91, no.~2, pp. 163--177,
  2019.

\bibitem{RN2778}
R.~Kumar, ``Internet of things for the prevention of black hole using
  fingerprint authentication and genetic algorithm optimization,''
  \emph{International Journal of Computer Network and Information Security},
  vol.~10, no.~8, p.~17, 2018.

\bibitem{RN675}
C.~I. Watson, M.~D. Garris, E.~Tabassi, C.~L. Wilson, R.~M. Mccabe, S.~Janet
  \emph{et~al.}, ``User's guide to nist biometric image software ({NBIS}),''
  National Institute of Standards and Technology, Tech. Rep., 2007.

\bibitem{RN2538}
R.~Cappelli, M.~Ferrara, D.~Maltoni, and M.~Tistarelli, ``Mcc: A baseline
  algorithm for fingerprint verification in fvc-ongoing,'' in
  \emph{International Conference on Control Automation Robotics \& Vision},
  2010, pp. 19--23.

\bibitem{RN3220}
H.~Zhu, Q.~Wei, X.~Yang, R.~Lu, and H.~Li, ``Efficient and privacy-preserving
  online fingerprint authentication scheme over outsourced data,'' \emph{IEEE
  Transactions on Cloud Computing}, vol.~9, no.~2, pp. 576--586, 2021.

\end{thebibliography}
\end{document}